
\font\fourteenrm=cmr12 scaled 1167
\font\fourteeni=cmmi12 scaled 1167
\font\fourteenit=cmti12 scaled 1167
\font\fourteensy=cmsy10 scaled 1400
\font\fourteensl=cmsl12 scaled 1167
\font\fourteenbf=cmbx12 scaled 1167

\font\fourteencsc=cmcsc10 scaled 1400
\font\fourteentt = cmtt12 scaled 1167
\font\twelverm=cmr12  
\font\twelvei=cmmi12  
\font\twelveit=cmti12 
\font\twelvesy=cmsy10 scaled 1200  
\font\twelvesl=cmsl12 
\font\twelvebf=cmbx12

\font\twelvecsc=cmcsc10 scaled 1200 
\font\twelvett = cmtt12
\font\tenrm=cmr12 scaled 833   
\font\teni=cmmi12    scaled 833

\font\tensy=cmsy10 

\font\tenbf=cmbx12   scaled 833

\font\eightrm=cmr12 scaled 667    \font\sixrm=cmr12 scaled 500    

\font\eighti=cmmi12 scaled 667    \font\sixi=cmmi12 scaled 500

\font\eightsy=cmsy10  scaled 800  \font\sixsy=cmsy10 scaled 600

\font\eightbf=cmbx12  scaled 667

\let\sc=\sevenrm
\let\csc=\tencsc
\def\fourteenpoint{\def\rm{\fam0\fourteenrm}
    \textfont0=\fourteenrm \scriptfont0=\tenrm \scriptscriptfont0=\eightrm
    \textfont1=\fourteeni  \scriptfont1=\teni  \scriptscriptfont1=\eighti
    \textfont2=\fourteensy \scriptfont2=\tensy \scriptscriptfont2=\eightsy
    \textfont\itfam=\fourteenit        \def\it{\fam\itfam\fourteenit}
    \textfont\slfam=\fourteensl        \def\sl{\fam\slfam\fourteensl}
    \textfont\bffam=\fourteenbf \scriptfont\bffam=\tenbf
      \scriptscriptfont\bffam=\eightbf   \def\bf{\fam\bffam\fourteenbf}
      \def\tt{\fam\ttfam\fourteentt}
    \normalbaselineskip=16pt
    \setbox\strutbox=\hbox{\vrule height20pt depth6.5pt width0pt}
    \let\sc=\tenrm  \let\csc=\fourteencsc  \let\er=\tenrm
    \normalbaselines\rm}
\def\twelvepointother{\def\rm{\fam0\twelverm}
    \textfont0=\twelverm \scriptfont0=\eightrm \scriptscriptfont0=\sixrm
    \textfont1=\twelvei  \scriptfont1=\eighti  \scriptscriptfont1=\sixi
    \textfont2=\twelvesy \scriptfont2=\eightsy \scriptscriptfont2=\sixsy
    \textfont\itfam=\twelveit          \def\it{\fam\itfam\twelveit}
    \textfont\slfam=\twelvesl          \def\sl{\fam\slfam\twelvesl}
    \textfont\bffam=\twelvebf   \scriptfont\bffam=\eightbf   
      \scriptscriptfont\bffam=\eightbf   \def\bf{\fam\bffam\twelvebf}
      \def\tt{\fam\ttfam\twelvett}
    \normalbaselineskip=14pt
    \setbox\strutbox=\hbox{\vrule height15pt depth5pt width0pt}
    \let\sc=\eightrm  \let\csc=\twelvecsc  \let\er=\eightrm
    \normalbaselines\rm}
%

\def\axr{{\cal R}}

\def\margrel#1{\par\vskip-1.0\baselineskip \rlap{\hbox to 7.2 in{\hfil\bf #1}}}

\def\unb#1{{\underbar{#1}}}
 
%
\input epsf
\def\figembedps[#1,#2]#3{\midinsert\parindent=0pt
    \vbox{\epsfxsize=#2\centerline{\epsfbox{#1}}}
    {#3\break\vskip 3 mm}
    \def\par{\endgraf\endinsert}}

%
\twelvepointother
\baselineskip = 1.1\baselineskip



\centerline{\fourteenpoint\bf Thermal Properties, Sizes, and Size 
Distribution of}
\centerline{\fourteenpoint\bf Jupiter-Family Cometary Nuclei}

\bigskip
\centerline{%
Y. R. Fern\'andez$^a$, 
M. S. Kelley$^b$,
P. L. Lamy$^c$,
I. Toth$^{c,d}$,
O. Groussin$^c$,
C. M. Lisse$^e$,}
\centerline{%
M. F. A'Hearn$^b$,
J. M. Bauer$^f$,
H. Campins$^a$,
A. Fitzsimmons$^g$,
J. Licandro$^h$,
S. C. Lowry$^{i}$,}
\centerline{%
K. J. Meech$^j$,
J. Pittichov\'a$^f$,
W. T. Reach$^k$,
C. Snodgrass$^{l}$,
H. A. Weaver$^e$}
\bigskip

{\noindent $^a$ Dept. of Physics, Univ. of Central Florida,
	4000 Central Florida Blvd., Orlando, FL 32816-2385, U.S.A.
	+1-407-823-6939, {\tt yan@ucf.edu}} \par
{\noindent $^b$ Dept. of Astronomy, Univ. of Maryland,
	College Park, MD 20742-2421, U.S.A.} \par
{\noindent $^c$ Aix Marseille Universit\'e, CNRS, 
Laboratoire d'Astrophysique de Marseille 
UMR 7326, 13388, Marseille, France} \par
{\noindent $^d$ Konkoly Observatory MTA CSFK CSI, 
PO Box 67, H-1525 Budapest, Hungary} \par
{\noindent $^e$ Applied Physics Laboratory, Johns
Hopkins Univ., 11100 Johns Hopkins Rd., 
Laurel, MD 20723, U.S.A.} \par
{\noindent $^f$ Jet Propulsion Laboratory, California
Institute of Technology, 4800
Oak Grove Drive, Pasadena, CA 91109, U.S.A.} \par
{\noindent $^g$ Astrophysics Research Centre,
School of Physics and Astronomy, Queen's Univ., Belfast
BT7 1NN, U.K.} \par
{\noindent $^h$ Instituto de Astrof\'\i sica de Canarias,
c/Via Lactea s/n, 38200 La Laguna, Spain} \par
{\noindent $^i$ School of Physical Sciences,
Univ. of Kent, Canterbury CT2 7NH, U.K.} \par
{\noindent $^j$ NASA Astrobiology Institute,
Institute for Astronomy, Univ. of
Hawai`i, 2680 Woodlawn Dr., Honolulu, HI 96822, U.S.A.} \par
{\noindent $^k$ Stratospheric Observatory for
Infrared Astronomy, NASA/Ames Research
Center, Moffet Field, CA 94035, U.S.A.} \par
{\noindent $^l$ Max-Planck-Institut f\"ur Sonnensystemforschung,
Max-Plank-Str. 2, 37191 Katlenburg-Lindau, Germany} \par

\bigskip
\centerline{\bf To appear in the journal {\sl Icarus}.}
\centerline{\bf Accepted July 13, 2013.}
\centerline{\bf DOI: 10.1016/j.icarus.2013.07.021}
\bigskip
\centerline{arXiv preprint version 1 - July 23, 2013}

\vfill
\eject


\noindent{\bf Abstract} \par

We present results from SEPPCoN, an on-going Survey of the
Ensemble Physical Properties of Cometary Nuclei.
In this report we discuss mid-infrared
measurements of the thermal emission from
89 nuclei of Jupiter-family comets (JFCs). 
All data were obtained in 2006 and 2007 using
imaging capabilities of the Spitzer Space Telescope. 
The comets were typically 4 to 5 AU from the Sun
when observed and most showed only a point-source with
little or no extended emission from dust. 
For those comets showing dust, we used
image processing to photometrically extract
the nuclei. For all 89 comets, we present
new effective radii, and for 57 comets we present
beaming parameters. Thus our survey provides
the largest compilation of radiometrically-derived
physical properties of nuclei to date.
We have five main conclusions:
(a) The average beaming parameter of the JFC population
is $1.03\pm0.11$, consistent with unity; coupled with the
large distance of the nuclei from the Sun, 
this indicates that most nuclei have
Tempel 1-like thermal inertia. Only two of the 57 nuclei
had outlying values (in a statistical sense) of infrared beaming.
(b) The known JFC population is not complete even
at 3 km radius, and even for comets that
approach to $\sim$2 AU from the Sun and so ought to be
be more discoverable. Several
recently-discovered comets in our survey have
small perihelia and large (above $\sim$2 km)
radii. 
(c) With our radii, we derive an independent estimate
of the JFC nuclear cumulative
size distribution (CSD), and we find that
it has
a power-law slope of around $-1.9$, with the exact
value depending on the bounds in radius. 
(d) This power-law
is close to that derived by
others from visible-wavelength observations that assume
a fixed geometric albedo,
suggesting that there is no strong
dependence of geometric albedo with radius.
(e) The observed CSD shows a hint of structure
with an excess of comets with radii 3 to 6 km.
(f) Our CSD is consistent with the idea that
the intrinsic size distribution
of the JFC population is not a simple power-law and
lacks many sub-kilometer objects.

\vfill
\eject


\noindent{\bf 1. Introduction} \par

While comets are the most-observable, most-pristine
remnants leftover from the era of Solar System formation,
it is still necessary to understand their subsequent evolution
if we are to fully take advantage of what comets can tell us
about the compositional and thermophysical conditions
of the protoplanetary disk. Jupiter-family
comets (JFCs) -- also known as ecliptic comets (Levison 1996) --
are remnants of the icy planetesimals that existed
during the era of planet formation. 
There is dynamical evidence that today's JFCs
come to the inner
Solar System from the scattered disk of objects in the trans-Neptunian 
region (e.g. Duncan et al. 2004),
although the precise origin location of the JFCs is a
subject of ongoing dynamical and compositional study (A'Hearn et al. 2012).
The relationships between JFCs and trans-Neptunian 
objects (TNOs),
and in particular the scattered disk objects (SDOs), are reviewed
in detail elsewhere (e.g. by Lowry et al. (2008)),
but suffice to mention here that the objects themselves
(as JFCs and as SDOs)
have suffered evolutionary processes such
as collisions and devolatilization since their formation
(see e.g. Farinella and Davis 1996,
Kenyon et al. 2008, 
Meech and Svore\v n 2004). The properties of today's
JFC population reflect this history. 
Thus our overarching goal is to use the JFC population
to understand cometary evolution. 

To this end,
our team initiated SEPPCoN -- a ``Survey of the
Ensemble Physical Properties of Cometary Nuclei" --
in 2006 to build up a statistically-significant
database of nuclear radii, albedos, shapes,
and thermal parameters. SEPPCoN is different
from previous surveys in that it makes use
of mid-infrared observations of thermal emission
in conjunction with
non-simultaneous 
visible-wavelength observations
to obtain a clearer picture of the thermophysical
properties of cometary nuclei.
While we aim to understand individual comets
as much as possible, the greater goal is
to infer the ensemble properties of nuclei.

One ensemble property that has been discussed extensively
recently is the JFC size distribution. It is
important to establish the character of the
distribution since its shape will depend
not only on the original distribution of object properties
in the source region but also on
the comets' mass loss since reaching the inner Solar System. 
The mass loss occurs not only through
the `typical' cometary activity but also by the
shedding of small (cm-size) and 
large (meter-size and up) fragments
that over time can radically alter a comet's shape, size, and spin
(Boehnhardt 2004, Chen and Jewitt 1994,
Fern\'andez 2009, Reach et al. 2007). 

In the recent past, there have been several
studies of the size distribution of active nuclei
(e.g. 
Lowry et al. 2003,
Weissman and Lowry 2003,
Lamy et al. 2004,
Meech et al. 2004,
Tancredi et al. 2006,
Fern\'andez and Morbidelli 2006,
Snodgrass et al. 2011
Weiler et al. 2011, 
Fern\'andez and Sosa 2012) and
of dormant or extinct
comets (e.g. 
Alvarez-Candal and Licandro 2006,
Whitman et al. 2006). All
these studies have made
use of visible-wavelength magnitudes
of objects. Since the magnitude is  
proportional to the product of the geometric
albedo and the square of the effective radius,
these studies included an assumption about albedo.
With SEPPCoN's Spitzer observations, we have created an entirely new
and independent estimate of the JFC size
distribution without the need to make a critical albedo
assumption, since for low-albedo objects
the mid-IR flux is largely independent of it.

Another important ensemble property to address
is the physical structure of the nuclei. The structure of
a nucleus today is a result of both the original cometesimal
aggregation process and the subsequent evolution due to
collisions and mass loss.
Traditionally it has been difficult to investigate via
remote sensing 
all the details of a nucleus's structure. 
The Deep Impact impactor experiment on comet
9P/Tempel 1 provided much insight, showing
that, for example, its nucleus must be a highly-porous
object (Richardson et al. 2007, Ernst and Schultz 2007).
One important quantity
that can be investigated remotely is the thermal inertia, i.e.
the geometric mean of the thermal conductivity and
the volumetric heat capacity, or in other words the
thermal memory of a nucleus's surface.
For example, the Deep Impact spacecraft
was able to obtain
thermal mapping of 9P/Tempel 1's surface, revealing
that the thermal inertia must be low
(Groussin et al. 2007),
and that in fact there is likely severe
thermal decoupling between surface ice and
surrounding rock (Sunshine et al. 2006). 
Deep Impact obtained similar data on
comet 103P/Hartley 2, revealing again
a low-thermal inertia nucleus (Groussin et al. 2013). 
These results are consistent with studies
of other cometary nuclei that have been done
using ground-based
mid-IR observations (e.g. Campins et al. 1995,
Fern\'andez 1999, Fern\'andez et al. 2006). 
The survey we present here 
represents the first systematic investigation
of thermal inertia across a large number -- dozens -- of comets.

Some results on individual nuclei in our survey 
have appeared already (Groussin et al. 2009,
Licandro et al. 2009). In this 
paper we
present results for all of the survey's sample,
and also present the distributions of JFC nucleus
size and thermal behavior. 
In \S 2, we discuss the observations and reduction,
and in \S 3, we describe the data analysis techniques.
We show the distributions in \S 4 and discuss
some interpretations. This paper focusses
on the nuclei; a companion paper by
Kelley et al. (2013) discusses the observed extended emission
from dust and the properties of the grains.

\bigskip
\noindent{\bf 2. Observations and Reduction} \par

Our program was approved for observation 
with the Spitzer Space Telescope (Werner et al. 2004)
during
General Observer Cycle 3, from July 2006 to July 2007,
as program \#30908. 
We used the Multiband Imaging Photometer
for Spitzer (MIPS; Rieke et al. 2004) and
the InfraRed Spectrograph (IRS; Houck et al. 2004).

SEPPCoN was designed to study a significant fraction
of the known JFC population so as to overcome
any biases from small-number statistics. In early 2006, 
when the survey was proposed, there were just over
300 known JFCs, so our goal was to have a sample
size logarithmically within half an order of magnitude of
this number. For our purposes, 
we simplistically defined a JFC
as any short-period comet with Tisserand 
invariant $T_J$ between
2 and 3, excluding active Centaurs. (One
Centaur with perihelion $q$ slightly
above the $5.2$ AU cutoff was inadvertently
included in the sample however.) Our choice
of targets was constrained by other
conditions:
First, the comet had to be greater than 4 AU from the Sun
when it would be visible by Spitzer in Cycle 3, since
we wanted to minimize the contamination by coma.
Second,
the nucleus had to have an expected brightness
that would require no more than one hour of clock time
to detect with MIPS or 2.5 hours of clock time
to detect with IRS. Our desired minimum detection 
threshold was 5$\sigma$. 
Third, we excluded comets that were likely to have
very poor ephemerides with position uncertainties greater
than a few arcminutes.
Fourth, we excluded most comets that had already had
their nucleus characterized by other mid-IR
observations or by spacecraft visits.
While the second and third criteria certainly biased our sample
selection, they were necessary to create a feasible sample. Still,
it is important to acknowledge that the criteria mean that we would
tend to avoid observing relatively small (sub-km) nuclei, exactly
the nuclei that would provide the most help in finding the true
size distribution.

With these parameters, we found 100 targets,
about one-third of the total then known. 
Currently (at time of writing) there are
over 450 known\footnote{$^1$}{See 
{\tt http://www.physics.ucf.edu/$\sim$yfernandez/cometlist.html} for
a list.}, but our survey still represents
a significant fraction, and  
is still so far 
the largest study of JFC physical properties
in the infrared. 
The target list is given in Table 1, along with
the geometry of each target and the clock time
duration of each observation (``astronomical
observation request", or AOR, in Spitzer jargon).
Orbital properties of the targets can be found
in the companion paper by Kelley et al. (2013)
and so are not reproduced here.

\margrel{Table 1}

Each comet
was observed with either IRS or MIPS. IRS observations
made use of peak-up (PU) imaging mode in both
the ``blue" and ``red" channels, corresponding
to monochromatic wavelengths of 15.77 and 22.33 $\mu$m
(after color correction).
Those channels throw light onto separate sections
of a 128-by-128 pixel Si:As array. The blue section
is 44-by-31 pixels (80-by-56 arcsec with 1.8-arcsec
wide pixels), and the red section is 46-by-30
pixels (82-by-54 arcsec with same size pixels).
Observations in either array made use of
a five-point dither pattern to improve image
quality. Integration times of individual
exposures were 14 seconds, so the total integration
time on a particular comet in a single 
channel was $14\times5\times n_{cyc}$
seconds, where $n_{cyc}$ is the number of cycles requested and
which ranged from 1 to 50, depending on how bright
we anticipated the nucleus would be in
the blue or red passbands. During
an AOR, the telescope
was tracked at the comet's rate. 
Having two bands allowed us to identify
comets both by the motion and by the color. 
In all, 64 of the 100 comets were targeted with IRS.

Comets for which the 3$\sigma$ ephemeris uncertainty was
thought to be larger than about 30 arcsec were
observed instead with MIPS, with its larger
field of view. Each such comet was observed
twice with MIPS (once with the comet
centered in the field and again
later as a ``shadow" observation)
so that we could identify the comet
by its motion. We used MIPS's 24 $\mu$m channel
(monochromatic wavelength 23.68 $\mu$m after
color correction), which 
makes use of a 128-by-128 pixel Si:As array
with a field of view of 5.4 arcmin (with 2.55-arcsec
wide pixels). Observations with MIPS used
a 14-point dither pattern to improve image
quality. Integration times of individual exposures
were 10 seconds, so the total integration
time on a particular comet was $10\times14\times n_{cyc}$
seconds, where $n_{cyc}$ is the number of cycles requested,
ranging from 1 to 18, depending on how bright
we anticipated the nucleus would be in
the passband. The telescope
tracked the comet's apparent motion during an AOR.
In all, 36 of the 100 comets were targeted with MIPS.

Once the observations were taken, the individual
exposures were processed
by the Spitzer Science Center's own pipeline. For
IRS data, this was version S17.0.0, and for MIPS
data, this was version S17. We then used
these data products to perform further processing,
i.e. extra flat-fielding and stacking of all
images into the comet's reference frame. Our resulting
images are shown in Figs. 1 and 2, for IRS and MIPS observations,
respectively.

\margrel{Fig. 1}

Not all 100 comets in our sample were detected. Comets 43P, 54P, 120P,
141P, 240P/2002 X2, 243P/2003 S2, and P/2003 S1
were not detected by IRS, so in total 57 out
of the 64 targets were seen. 
Comets 203P/1999 WJ$_7$, 
244P/2000 Y3, P/1998 VS$_{24}$, and
P/2004 T1 were not detected by MIPS, so in total
32 out of the 36 targets were seen.
Thus, out of our original 100 targets, we have detections of 89 of them,
a good success rate.

\margrel{Fig. 2}

Among the 11 other comets, there are a few reasons for
their non-detection. Comet 141P was not seen because
the comet was outside the field of view; Spitzer
was using an incorrect ephemeris
(possibly because the comet had split in the 1990s
and the ephemeris was for the original comet, not one of its
fragments).
Comet 240P had not yet been recovered
when the observations occurred, but by using the
newer, two-apparition orbit, we see that the comet
would not have been in the IRS field of view.
Comets 203P, 243P, and 244P 
had likewise not been recovered, but the two-apparition orbit
indicates that Spitzer
was pointed in the correct directions. 
Presumably these nuclei were too faint to be seen, and so we obtained
upper-limit photometry.
Comets 43P, 54P, and 120P -- 
already having multi-apparition orbits 
by the time of the Spitzer observations -- are likewise
simply too faint to be detected, so we obtained
upper-limit photometry for them as well. That leaves
the three single-apparition comets -- P/1998 VS$_{24}$,
P/2003 S1, and P/2004 T1 -- that (at time of writing)
have yet to be recovered. The reason for the non-detection
could be either faintness or incorrect ephemeris;
we provide upper limits for these objects on the
assumption that the former reason is correct.

\bigskip
\noindent{\bf 3. Analysis} \par

\noindent{\sl 3.1. Photometric Nucleus Extraction} \par

A notable property of many of the comets in Figs. 1 
and 2 is the extended emission due to dust despite
the great heliocentric distance of the comets at the time. 
Of the 57 detected IRS comets, 34 were bare but 23 had
discernible dust. Of the 32 detected MIPS comets, 20 were
bare but 12 had discernible dust. 
The properties of the dust from these 35 comets with dust
are discussed
in the companion paper by Kelley et al. (2013). For this 
paper, where we focus on the properties of the nuclei,
we performed
image processing techniques to photometrically
extract the point-source nucleus out of each image.
We describe this procedure below.

One technique that we employed
was a ``coma-fitting technique", which 
was developed by
several of our team members and has been in successful
use for many years with many images of comets
(e.g. Lamy et al. 1996, 1998a, 2004,
Lisse et al. 1999,
Groussin et al. 2004, 2009,
Stansberry et al. 2004).
In particular, this technique allowed us
to correctly predict the nuclear
radii of comets 19P/Borrelly (Lamy et al. 1998b),
81P/Wild 2 (Fern\'andez 1999), 
9P/Tempel 1 (Fern\'andez et al. 2003; Lisse et al. 2005), and
103P/Hartley 2 (Lisse et al. 2009) before
their spacecraft flybys. 

The technique uses 
the radial profiles of surface-brightness in the parts of
the coma and/or tail away from the comet's head
to extrapolate the dust's contribution to
the flux in the pixels that are within the head. 
The radial profiles are assumed to follow power-laws. 
Ideally, after accounting for
the dust's flux, a point-source remains, which
is interpreted as having the flux from the nucleus. 
The technique is described in detail by (e.g.)
Lamy et al. (2004) and Lisse et al. (2009).

This method does assume
that the dust's photometric behavior is consistent
all the way in to the coma-nucleus boundary. This
assumption is most easily confirmed when the
spatial resolution is good, so that one can
actually see what the dust is doing closer
and closer to the
nucleus. In this regard,
the Spitzer data are not optimal since the pixels
are large and the comets are far away. For example,
for $\Delta=4$ AU,
one IRS pixel covers 5200 km and one MIPS pixel
covers 7400 km. However, the method also works
well when there is high contrast between
the point-source nucleus and the dust. Fortunately
this is the case 
for many of our dusty images in Figs. 1 and 2 --
there is often such a strong point-source within the dust
that even the first Airy ring shows up clearly.
This gives us confidence that our
extractions yield realistic estimates of the nuclear flux.
In later sections of this paper we will
mention instances where our results are consistent
with the assumption that our extractions have been successful.

The other processing technique that was applied
is related to the coma-fitting technique, but
attacks the problem by fitting
and subtracting a range of scaled
point-spread functions (PSFs) 
from the image. The criterion for quality is
that the net image must still
have a plausibly-shaped dust coma
and tail; for example, the resulting
dust cannot have
a sharp drop in brightness near the photocenter. 
As before, the photometry of the subtracted
PSF is interpreted as the nucleus's flux. 
This approach has the advantage
of not relying on a single power-law to describe
the dust's brightness as a function of cometocentric
distance, which is useful when there are strong
tails and/or comae whose grains have already
deviated from radial motion due to radiation pressure.

Examples of this latter technique are shown in Figs. 3 and 4,
which show the 23 IRS comets with extended emission and
the 12 MIPS comets with extended emission, respectively.
(For the other 54 comets -- 34 IRS and 20 MIPS -- the 
extraction of the point-source was trivial.) 
The six images in each row pertain to a single comet.  The
first and fourth columns show the comet itself;
for IRS images, these are the ``blue" and ``red" PU images,
whereas for MIPS images, these are the images from the two
visits, though we note that in nearly all cases the sky-subtracted
image is displayed, i.e. the ``visit1" image is the image
from the first visit minus the sidereally-aligned image
from the second visit, and the ``visit2" image is
the image from the second visit minus the sidereally-aligned
image from the first visit. This ``shadow" subtraction 
technique does a great job of
cleaning up the sky background near the comet. 

\margrel{Fig. 3}

The second and fifth columns of Fig. 3 show the best-fitting PSF,
which has been scaled to the best overall flux, aligned
to the best location, and
pixelized to match the IRS or MIPS pixel scale. 
The third and sixth columns show the subtraction 
of the previous two images, i.e. the residuals. In other words this shows
the comet with no nucleus, only dust. We note that
for a few images it was also helpful to remove a planar
sky-background.
As mentioned, in many cases, the point-source nucleus
is quite strong, and is easy to separate. For a few such
cases, the amount of remaining dust has very low surface brightness and
is hard to see in an image (e.g., 6P). However for other
comets, the point-source is rather faint and/or the
flux from dust dominates the total flux from the comet.
In those cases, the fractional uncertainty on the nucleus's flux
is naturally higher since it was more difficult to determine
a ``best" fit.

\margrel{Fig. 4}

A more quantitative demonstration of the technique is shown
in Fig. 5, where we have taken 10 comets from our sample to 
show the effects of removing the PSF. Each comet has two panels;
the comet's name and the particular band or visit is written
in the panel. 
The first five pairs of panels show comets
observed with IRS; the next five pairs of panels
are from comets observed with MIPS.
Each panel shows line cuts in the x-direction
through the image, best-fitting PSF, and the residual. 
If the comet's centroid was in pixel ($x,y$) = ($x_c,y_c$), then the
main plot shows (in grey) line cuts through the image at $y=y_c$,
$y=y_c+1$, and $y=y_c-1$. The main plot also shows 
(in dashed black) line cuts through the best-fitting PSF at those same pixels.
The plots below and to the right of the main plot show the
residuals. The plot below shows (in grey) line cuts through
the residuals at those same $y$ pixels; the plot to the
right shows (in grey) line cuts through the residuals
at $x=x_c$, $x=x_c+1$, and $x=x_c-1$. 

The first IRS comet and the first MIPS comet in Fig. 5
both have almost all their flux coming
from the nucleus; hence the residuals straddle zero. 
The other eight comets have at least some dust in the residual
images and correspond to comets shown in Figs. 3 and 4. 
The second IRS and second MIPS comets have 
faint comae and faint point-sources. 
The third IRS and third MIPS comets have bright comae but
still faint point-sources. 
The fourth IRS and fourth MIPS comets have
faint comae but bright point-sources. 
The fifth IRS and fifth MIPS comets have
bright comae and bright point-sources. Thus this figure
gives a flavor for our ability to extract point-sources
from comets of various signal levels and nucleus-to-coma
contrasts.  

\margrel{Fig. 5}

As a summary of the various strengths of the point-sources
within our comet sample, we list in Table 2 each nucleus's fractional
contribution to the overall comet flux (within
a 3-pixel radius circular aperture).
We note again that
a very sizable majority of our sample showed
all or nearly all nucleus (zero or almost zero dust).
The 35 (out of the 89) nuclei with discernible dust  
are underlined in the table. We do note that a few
of the comets with some dust (e.g. 121P) have nuclei
that contribute about 100\% of the flux within
the aperture anyway; these comets either are (a) faint,
so that the dust is visible but quite diffuse, or (b) 
only showing dust in tails rather than in comae. 
We also note that some comets without dust (e.g. 79P, 127P, 228P)
appear to have fluxes that contribute less than 100\%
or even more than 100\% of the flux. 
This is not a contradiction; these
comets were faint and the flux is obtained from the
equivalent best-fitting PSF rather than from the image
itself. 

\margrel{Table 2}

As a self-consistency check we performed
the image processing twice, with two subsets
of our survey team doing the analysis entirely
independently, using completely different codes.
In particular, one group predominantly used the
coma-fitting technique, with the other using most often
the point-source fitting-and-subtraction approach.
The final nucleus photometry is reported in Table 3,
and those numbers are derived from
the combination of these analyses. 
We note that each team used the same images for their
work -- mosaics created from the output of the Spitzer pipeline. 
In almost all cases, the
two groups yielded consistent results. The two
independent approaches agreed very well with a 6\% 
mean difference in
the IRS photometry 
and 2\% mean difference in the MIPS photometry.
Errors reported in Table 3 account for the differences
in the results from the two techniques. 
In ten cases, one analysis yielded clearly spurious
results (evident e.g. by a very unlikely color temperature)
while the other did not, and so for those comets we used only the
more physically plausible result. 
The reason for the spurious results was the fact that
(a) some comets had unusually-shaped dust emission not conducive
to characterization by power-laws, and (b) some comets had noisy
sky and/or Galactic/extragalactic background sources near the head.

\margrel{Table 3}

Before leaving this topic, it is important to emphasize
a universal problem that affects most studies of nuclei --
not just the study presented here but those already in the literature as well.
Jewitt (1991) gives a list of criteria for deciding when
one is observing a bare nucleus, one of which is that
the ``the image should appear unresolved." But as Jewitt (1991)
goes on to discuss, while this is a necessary condition, it is not
a sufficient one, and unresolved coma can hide within
the seeing disk or PSF. Indeed there is good evidence
that this phenomenon occurs frequently with
comet 2P/Encke (Fern\'andez et al. 2000, 2005b).
Nonetheless, if there is a lack of
evidence to the contrary, most workers assume
that a point-source is identical to the nucleus even though
there is often no way to tell for a given comet if this
assumption is true or false. Many of our images presented
here (Figs. 1 and 2) show point-sources,
and one may hope that the assumption is true, but there
is no telling what unexpected dust flux there is within the volume of space
covered by the central pixel. Our other images show 
extractable point-sources embedded in extended dust emission, and while we
can make more assumptions about the dust's behavior, there is
again no telling what is really going on within the hundreds
or thousands of kilometers within the central pixel. So the general
reliability of nucleus photometry will always be suspect at that
level if there are no corroborating data that can 
independently establish the population of dust in
the near-nucleus environment. On the other hand, 
our success at successfully predicting the effective
radii of nuclei that are then confirmed with spacecraft flybys
(as mentioned at the beginning of this section)
gives us confidence that in many cases the point-source photometry
is indeed identical to the nucleus's photometry.

\medskip
\noindent{\sl 3.2. Thermal Modeling} \par

The conversion of mid-infrared photometry to physical
properties requires the use of a thermal model.
For this work we used the Near-Earth Asteroid
Thermal Model (NEATM), created by Harris (1998)
and based on the Standard Thermal Model (STM; Lebofsky et al. 1986,
Lebofsky and Spencer 1989). We have described
our modeling technique in many other works
(e.g. Lamy et al. 2004, Fern\'andez et al. 2005a, 2006, 2009)
and we only describe here the details that are
particular to the present case.
The basic premise of the thermal model is that the
thermal inertia is very low and so the temperature map of an object's
sunlit surface mainly depends on the instantaneous
local zenith angle of the Sun,
$\theta$. The temperature is maximized at the subsolar point and
falls to zero at the terminator 
as $\root 4 \of {\cos\theta}$. In other words the
night-side retains little thermal memory and so does not normally
contribute much to
the thermal emission.

The thermal model requires some assumptions: we assumed
that (a) the nucleus was spherical, 
(b) the emissivity was 0.95,
(c) the visible-wavelength geometric albedo was 0.04,
and (d) the infrared phase darkening
follows the behavior described by Harris (1998), i.e. is
proportional to the fraction of sunlit hemisphere that
is on the Spitzer-facing side. Of these four items,
the last three are not likely to influence our results 
significantly. Emissivity of rock is unlikely to be
lower than 0.9. The thermal emission 
is proportional to one minus the Bond albedo, so
as long as the geometric albedo is low the emission
is hardly influenced by the actual value. 
Our observations occurred at low phase angles 
(9 to 17 degrees) and so our views of the nuclei
are almost at ``full" phase, requiring relatively
small corrections. The sphericity assumption
is the one most likely to skew our results, since
for example Brown (1985),
Lagerros (1996), and Howell et al. (2012)  
describe how radiometric diameters
of asteroids can be miscalculated with spherical
models due to shape effects. However currently there
is no obvious way to correct for such an effect
without better data on the specific objects,
so our analysis remains for spherical bodies and
their effective radii.  

Usually the crucial parameter in the modeling
is $\eta$, the beaming parameter. This parameter
is a proxy for thermal inertia and surface
roughness, and traditionally
has described the
amount of infrared `beaming' that a body has.
Objects with zero thermal inertia and no significant
topography will have $\eta=1$. 
Of course many real small bodies
have some night-side emission and/or can have non-isotropic
emission due to significant surface features
such as deep and/or large craters.
The former effect lowers the surface
temperatures and raises $\eta$ above 1.
The latter effect can make more of their infrared emission 
`beam' out
at smaller opening angles and reduces $\eta$ below 1. 
The STM originally assumed a constant value
for $\eta$ but the NEATM lets $\eta$ float
as a parameter to be derived. It is important
to either measure $\eta$ or have a solid assumption
for it if one is to obtain radii that are close to reality
(Delbo et al. 2007, Harris et al. 2011, Mainzer et al. 2011),
as demonstrated from cases where spacecraft visits have
tested the analyses of remote observations 
(e.g. Lisse et al. 2005, 2009).

Of the 100 comets in the survey, 
57 were detected by
IRS in both wavelengths, 32 comets were detected
at just one wavelength by MIPS, and 11 were not detected
at all. (None of the IRS comets were detected in just
one IRS wavelength.) For the multi-wavelength objects,
we performed
a two-parameter fit that yielded both the 
effective radius $R_N$ and the beaming parameter $\eta$
(similar to what was done
by Groussin et al. (2009) and Licandro et al. (2009)).
In this case, since 
we formally have no degrees of freedom and could fit
a two-parameter model perfectly to two data points,
we could not
use the usual $\chi^2$ statistic to characterize 
a goodness-of-fit. To remedy this, we performed
multiple model fits to a nucleus's photometry, requiring
that a `good' fit be one where the model passes 
within 1$\sigma$ of both photometry points. The resulting
ranges of parameters that yielded such `good' fits
are reported in Table 4. 

\margrel{Table 4}

For the 32 MIPS comets, we have photometry at only
one wavelength, but we have it at two epochs. We
modeled each photometry point individually. 
We performed a one-parameter
fit to find $R_N$ by assuming $\eta$, and
the value we assumed is $\eta=1.03\pm0.11$
(which we explain in \S 4.1 below).
As before, we have no degrees of freedom, so
we used the same criterion as above to describe
a `good' fit.
The resulting effective radii at each epoch
are reported in Table 5. 

Finally, 
we note that for the six comets that we did not detect
but that were within the field of view, and for
the three other undetected comets that could have
been in the field of view, 
we used the 3$\sigma$ upper limits to photometry 
to derive an upper limit to their radii, using
the same assumption for $\eta$. These numbers
are also in Table 5.

\margrel{Table 5}

\bigskip
\noindent{\bf 4. Discussion} \par

\medskip
\noindent{\sl 4.1. Beaming Parameters} \par

\medskip
\noindent{\sl 4.1.1. Means and Distributions} \par

The 57 beaming parameters in Table 4 are represented
graphically in the top panel of Fig. 6. This represents the 
largest collection by far of
measured cometary beaming parameters to date.
Most of the values are near
unity, suggesting that infrared beaming is not
significant, that there is little night-side
emission from the nuclei, and that the thermal inertia 
is low -- e.g. comparable to the values
measured by Groussin et al. (2007, 2013) for Tempel 1
and Hartley 2 from
spacecraft data.

\margrel{Fig. 6}

We calculated the mean beaming parameter, $\bar\eta$,
by using all 57 points and using 
$\chi^2$ minimization -- i.e. by making use of the
error bars to the individual points. The best fit
is $\bar\eta = 1.00\pm0.12$.\footnote{$^2$}{All
errors quoted in this paper are $1\sigma$.}
We note that the
error is the error in the mean, not the standard
deviation. The minimum $\chi^2_\nu$
is 1.47 ($\nu = 56$), 49\% higher
than the expected value. A closer look reveals
that about 30\% of the $\chi^2$ comes
from just two of the 57 comets, 89P and P/2005 JQ$_5$.
These two points have very low and very high values
of $\eta$, respectively, as listed in Table 4. More
importantly, they have relatively small error bars -- so even
though there are comets with even higher $\eta$, these two are
the farthest outlying comets because of the apparent
ostensible high-quality of each measurement. 89P's
$\eta$ is over $3\sigma$ below the mean, 
and P/2005 JQ$_5$'s $\eta$ is over $3\sigma$ above the mean.
With 57 data points, if the beaming parameters were in reality
distributed normally about a mean, then we would
not expect any points to be that many $\sigma$ away; for example
Chauvenet's criterion is 2.6$\sigma$.
It is of course the case that $\eta$ will not be 
perfectly normally distributed, since $\eta$ can physically
only be so low ($\sim$0.7) but can be quite high. However
for the sample size here we can work on the assumption 
that the distribution would be approximately normal. 

We will discuss comets 89P and P/2005 JQ$_5$ later, but
for now, recalculating $\bar\eta$ without those two outliers
yields $\bar\eta = 1.03\pm0.11$ 
with $\chi^2_\nu = 1.13$, very close to the expected value
of $0.99$ for $\nu=54$. We adopt this value
of $\bar\eta$ as our final answer, rather than the $1.00$ value
mentioned above that includes the outliers.

On the top panel of Fig. 6, 
we have drawn a grey rectangle to represent $\bar\eta$ and its error.
Inspection of the points 
shows good clustering near the mean, but 
an alternate way to see this is to look at the bottom panel
of Fig. 6, where we have plotted each point's offset in units
of its own $\sigma$ from $\bar\eta$. The distribution here
is close to what would be expected for a normal distribution.
For 55 points, 
we would expect 37.5 points within
$\pm1\sigma$ of the mean, and we have 36. 
We would expect 15 points between $\pm$1 and $\pm2\sigma$, and
we have 16. We would expect 2.4 points between $\pm$2 and $\pm3\sigma$
and we have 3. Furthermore, in this representation,
89P and P/2005 JQ$_5$ are clearly outliers, being more than 3$\sigma$ off.

We can roughly quantify the closeness of the $\eta$ distribution to a normal
distribution by calculating the skew and kurtosis. Since each 
data point $\eta_i$ ($i=1,..., n$) 
has its own error bar $\sigma_i$, we calculate skew and kurtosis
using $(\eta_i - \bar\eta)/\sigma_i$, instead of 
$(\eta_i - \bar\eta)/\sigma$, i.e. instead of the standard deviation
$\sigma$. We also use the best-fitting $\bar\eta$ described
above instead of the raw mean (sum of the data divided by 
the number of entries).
Using all 57 data points, we
find that the skew is 1.12 and the kurtosis is 3.98.
With only 55 data points (excluding 89P
and P/2005 JQ$_5$), we calculate a skew of 0.97 (little
change) and
a kurtosis of 0.17 (quite smaller). Press et al. (1992) state that 
estimates of the standard deviations of the skew and kurtosis
are $\sqrt{15/n}$ and $\sqrt{96/n}$, respectively. With $n=55$, 
the numbers here suggest that our $\eta$ distribution is perhaps
somewhat skewed to the positive side but not radically so, and
that the removal of the outliers has removed most of the kurtosis.
This is consistent with the assumption stated above that
the $\eta$ distribution is approximately normal.

One may wonder if there is any difference between the
beaming parameters among the comets that showed some dust
and those that did not, perhaps indicating if
there is any overall systematic problem with the nucleus
extraction described in \S 3.1. The two kinds of comets
are distinguished in Fig. 6 with filled and unfilled
circles. Among the 23 IRS comets
that showed dust (i.e., the underlined comets in Table 2
and the open circles in Fig. 6),
we find that $\bar\eta=1.21\pm0.24$ with $\chi^2_\nu=1.08$
and $\nu=22$. For the 32 IRS comets that
did not show dust (excluding 89P and P/2005 JQ$_5$
from the group of 34, 
as justified earlier), we find that
$\bar\eta=1.00\pm0.11$ with $\chi^2_\nu=1.05$ and
$\nu=31$. Both of these mean values are consistent with
each other. Furthermore, a Student-$t$ test comparison of the two
subgroups' means yields a $t$-statistic significance of 47\%.
For the Student-$t$ test, a lower significance implies a
greater likelihood for the means to be significantly different.
In this case, the significance suggests
that the means of the two subgroups are not significantly different. 

Since the error bars on $\eta$ vary widely from comet to comet,
we can instead perform this statistical test 
using the offsets shown in the bottom
half of Fig. 6, rather than using the values themselves. 
In this case, the Student-$t$ test returns a significance
for a difference in the means
of 1.5\%. We can also compare the entire distribution
of values
by using the Kolmogorov-Smirnov test. 
Again, a lower significance of the statistic suggests 
a higher likelihood that the groups are drawn from different
distributions. We find that this test yields a significance
of 3.7\%. 
Thus both the Student-$t$ and Kolmogorov-Smirnov tests
offer a hint that there may be some systematic difference
in the $\eta$ values from comets that showed dust versus 
those from comets that did not. Since this is unlikely to
be a physical effect, it is probably tied to the general problem
in extracting nuclei from dusty comets. However, since neither significance
crosses the $3\sigma$ threshold, and since the overall distribution
of all $\eta$ values together mimics a Gaussian, we proceed
with an analysis that includes all the comets.

\medskip
\noindent{\sl 4.1.2. Trends} \par

To investigate whether $\eta$
showed trends with any observational or orbital parameters,
we compared our 57 beaming parameters to a variety
of other comet-related quantities. The 16 parameters
we tried are listed in Table 6. Using a variety
of statistical tests, we found no significant correlations
at better than the $3\sigma$ level between
$\eta$ and any of these 16. 
We give more detail on the tests below. 
Scatter plots of four of these 16 quantities are shown in 
Fig. 7. The plots are of the four
that were most likely to reveal a trend: 
heliocentric distance $r$,
phase angle $\alpha$, 
perihelion distance $q$,
and time before perihelion $d_{peri}$.
With regard
to $r$ and $d_{peri}$, it is easier for a cooler object to retain
thermal memory, with could be seen
as higher $\eta$.
With regard to $\alpha$, 
such a trend has been reported in near-Earth asteroids
by Delb\'o et al. (2007). With
regard to $q$, perhaps surface properties
are affected by the maximal heating received
(as coma gas daughter species appear to be; A'Hearn et al. 1995).
Unfortunately, Fig. 7 and Table 6 show that 
there are no significant correlations.

\margrel{Table 6}

Our first test
was to try a simple linear regression with $\eta$ as the ordinate and
each of the 16 parameters as the abscissa. 
The 16 values of the slopes to
the fitted lines, $m$, are given
in the second column of Table 6, and indeed for 13 of the 16 parameters
the slopes are within $1\sigma$ of zero. Two others
have $m$ within $1.5\sigma$ of zero. The slope of the fit
to Spitzer-centric
distance, $\Delta$, is non-zero at the $3\sigma$ level, but it
is hard to explain why this would be significant if the fits to
heliocentric distance and perihelion are not.
Also, further tests (explained below) led us to reject
any significance to this slope.

For a second test we calculated the Spearman ``$\rho$"
(Press et al. 1992) to look for a correlation
between $\eta$ and another quantity. This
test first makes two ranked lists of $\eta$
and the independent quantity, and then
calculates $\rho$, a sum-squared difference of the ranks.
This $\rho$ can be compared to the expected value
if there were no correlation.
Table 6 lists values of $Z$, the number of standard
deviations that $\rho$ is away from this expected
value, and $P_Z$, the probability that one
would by chance get this value of $Z$ in uncorrelated
data. Lower probabilities mean a higher likelihood
of a real correlation. For all 16 quantities,
there is no value of $|Z|$ higher than $3\sigma$
and only one quantity is better than $2\sigma$.
In particular, there is no correlation of $\eta$ with $\Delta$
suggested by this test. 

\margrel{Fig. 7}

The strongest correlation is with days from perihelion, $d_{peri}$,
and indeed visual inspection of
Fig. 7 suggests that pre-perihelion nuclei, on average,
have a slightly lower $\eta$ compared to post-perihelion nuclei.
We calculate that 
the best-fitting mean (in the $\chi^2$ sense) of the
pre-perihelion $\eta$ values is $0.94\pm0.12$; and of
the post-perihelion values, $1.14\pm0.14$. (These
numbers exclude the two outliers 89P and P/2005 JQ$_5$.)
So the numbers are consistent with the plot, but 
the difference between the means is only $0.20\pm0.19$, i.e.
not significant. 
Also, if the correlation were real, one might expect
to see a similar correlation with
heliocentric or perihelion distance, but there is no evidence
for this. So while such a correlation would 
suggest a difference in a nucleus's
thermal memory
post-perihelion vs. pre-perihelion, we conclude
that the result is tantalizing but not yet statistically significant. 

The Spearman $\rho$ calculations make no use
of the error bars to $\eta$, only the measured
values themselves. Since for many $\eta$
the error bars are large, it is possible that
the $\rho$ that we calculate is misleading and
just happens to indicate a correlation (with e.g. $d_{peri}$)
because of our luckiness (or unluckiness) in
the values. To investigate this, we performed
our third test, a modification of the second. We ran
a Monte Carlo simulation of our 57 beaming parameters,
generating 100,000 sets of 57 values of $\eta$
based on the error bars assigned to each $\eta$
(assuming a normal distribution with the error bar
representing the sigma).  
For each set, we then reran the Spearman $\rho$
calculations against the 16
quantities in Table 6, generating 100,000 values of $Z$
for each quantity. We then determined what the
average value of $Z$ ($\bar Z$)
is out of those 100,000 so that we could tell if
our particular 57 values of $\eta$ were indeed
fooling us. We also calculated $\cal N$, the fraction of
the 100,000 runs that yielded values of $Z$ that had
better than $3\sigma$ significance. $\cal N$ is simply
another estimate of our likelihood
in being fooled, though a qualitative estimate,
since we did not explore the expectation of $\cal N$ in detail. These
results are given in two columns of Table 6. 
For 15 of the 16 parameters, $\bar Z$ is always within
$1.1\sigma$ and $\cal N$ is under $1\%$. 
Only for the correlation with $d_{peri}$ did
this analysis return a more significant $\bar Z$ than 1$\sigma$,
but the value is not beyond the $3\sigma$ threshold. 

For our fourth and final test, we attempted to pull out 
any correlation with
the 16 quantities by 
dividing the $\eta$ values into two groups -- one 
group characterized by having low values of the 
independent variable, and one having high values -- and
then using a Student-$t$ test to compare the means.
The division between ``low" and ``high" was made at
the median value of the independent variable (except for
$d_{peri}$, where we made the split at $d_{peri}=0$).
A correlation might reveal itself with a significant
difference in the means of the two groups. The results
of these calculations are given in the last column of Table 6.
The significances are high percentages, so there is no statistically
significant difference found with this analysis.

In summary, from Table 6 we conclude that there is
no statistically significant trend
of $\eta$ vs. any of the other quantities.

We also repeated this entire trend analysis using the subset
of the comets that showed
no emission from dust, i.e. the 34 comets mentioned earlier,
just in case the dusty comets were influencing the results.
No statistically significant correlations were found
in this case between the 34 $\eta$ values and any of
the 16 quantities.  

All this implies
that our results for $\eta$
are not strongly dependent on the observational 
and geometric circumstances, although admittedly for some
of the 16 parameters we do not have a wide
range of values for the test to be strong. (For example,
our range in $r$ only covers a change in subsolar temperature 
of $\sim$30 K.)
Nonetheless, at face value the
results suggest that cometary nuclei have little
infrared beaming and thermal memory. Indeed the
results are consistent with the hypothesis that
all JFC nuclei have about the same bulk thermal properties.
This would be a
very useful result because it means that future
snapshot radiometry can be used to derive cometary
radii and cross sections with good confidence without worrying about
a large uncertainty from an otherwise-unmeasured
beaming parameter.

\medskip
\noindent{\sl 4.1.3. Outliers} \par

While there are individual comets in
our survey with ostensibly
elevated or diminished values of $\eta$ -- such as 127P, 173P, 213P, etc. -- 
for the most part they have sufficiently large error bars that we cannot
claim that they are in reality unusual.
As mentioned, the statistics
suggest that perhaps only two of the 57 comets (3.4\%) 
have $\eta$ significantly outside a normal distribution.

About those two comets, one can ask why 89P
and P/2005 JQ$_5$ could have such unusual results.
As shown in Fig. 1, 89P does not appear to have
dust, so we might
expect the nucleus photometry to be robust. The very low
beaming parameter (0.48) could be explained if the red IRS
photometry was measured 
too low, or the blue IRS photometry was measured too high. 
The bright pixels near the comet in the blue image
were not included in the photometry, so that cannot explain it.
There
was a latent image problem in the red image;
if that had been incorrectly removed so that too much
background behind the comet was removed, then we might
obtain an artificially depressed flux. We would
need to inflate the red flux by about 30\% to bring 89P's
$\eta$ to the ensemble average of 1.03, however, and that is more
than we would expect to be able to correct the flux by based
on any improper latency removal.
A 10\% error would raise 89P's $\eta$ only up to 0.65, still quite low.
We currently have no other systemic explanation for why the
$\eta$ is so low. We know of no other published reports
about this comet that would shed light on the issue.

As for P/2005 JQ$_5$, Fig. 1 again shows that there was
no discernible dust. In this case, the
beaming parameter is very high (2.72), which could be
explained if the red
photometry was measured 
too high, or the blue photometry was measured too low. 
Another explanation however could be that the comet's head is
showing almost mostly dust and very little nucleus. At
the comet's heliocentric distance, we would expect grains
to have a blackbody temperature of about 140 K, and the
flux ratio between the red and blue photometry, converted
into a Planck-function temperature, is in fact
consistent with this -- $143\pm7$ K. So it is possible
that this comet's nucleus is even smaller than reported here
and we are simply seeing grains of a near-nucleus coma that
does not extend past the PSF. 
It should be pointed out though that this comet's nucleus
was detected at radar wavelengths by Harmon et al. (2006),
one of the first to be so observed with delay-Doppler
imaging. Their echoes indicate a nucleus of radius $0.7$ km,
about half the effective radius we report in Table 4. Such a size would
provide roughly one-fourth the flux density of what
we report in Table 3, consistent with the idea
that our observations might be dominated by dust.
Interestingly, Harmon et al. (2006) also say that
the nucleus's surface has
``extremely rough surface texture at wavelength (decimeter) scales."
Such surface structure argues against a high value of $\eta$
unless that roughness is smoothed out at $\sim$10-$\mu$m scales.

\medskip
\noindent{\sl 4.2. Radii and Trends} \par

To analyze our 89 (IRS plus MIPS)
new radii, we could just use the values reported
in Tables 4 and 5. However, we prefer to 
make all the radii self-consistent by first recalculating
the radii of the 57 IRS comets using 
an assumed beaming parameter of 
$\bar\eta=1.03\pm0.11$, i.e. the same
value we used for the MIPS comets. 

There are good reasons for this. 
First, we showed in
\S4.1 that 55 of the 57 beaming parameters
are statistically consistent with $\eta=1.03\pm0.11$,
i.e. a single value with small error bar, 
and in fact we also showed that the distribution is
quite similar to a Gaussian. 

Second, at face value, the range of $\eta$ among
the 57 comets 
covers more than 1 dex, from about 0.5 to 6. The low end
is very low and the high end is very high; indeed there
are virtually no small bodies known with such measured values. If the
distribution were telling us something real about
comet parameters, it would suggest that cometary surfaces in general
are very different from what we have come to believe
from spacecraft flybys. Furthermore such radical differences
in infrared beaming, conductivity, and thermal inertia for 
such a small range of radii -- about 1.5 dex -- seem implausible.
We should point out that a 
Spearman rank correlation calculation does show
that nominally there is a trend between the 57 values of
$R_N$ and of $\eta$ that is significant at the $3.8\sigma$ level. 
But, for the two reasons just described, 
we do not attribute this to a real trend in the surface properties.
Using a mean value for all comets is the safer 
conclusion. 

For these reasons we will continue on the assumption
that all nuclei can be studied using $\eta=1.03\pm0.11$.
An example of the benefit of this 
can be seen by considering comet 173P's
result in Table 4. It is certainly not impossible
that this comet really has radius 16 km and
large thermal inertia. But it seems much more likely
that its $\eta$ is high because of random error;
dropping the $\eta$ to 1.03 drops the radius
(since a lower
beaming parameter means the object is hotter overall
and so requires a smaller radius to explain a given
flux density) and so this in a sense debiasses
173P's ostensible radius to make its entry into
a size distribution more appropriate.

The recalculated radii that all use $\bar\eta$ -- for
MIPS comets and IRS comets alike -- are listed
in Table 7. 
The table gives the values that we use
to look for trends with other
quantities and to create the size distribution. 
We note that for the multi-epoch MIPS radii,
we have combined the two results into one value. 
If the standard deviation of
the two radii was higher than the error bars on the
individual measurements, then we assigned the error to be 
the standard
deviation. Otherwise, we assigned the error to be
the average of the errors of the individual measurements.

\margrel{Table 7}

We first look at some trends in the radii.
Figure 8 shows how the radii compare with
perihelion distance $q$. This is important to understand
since it is presumably on average easier to discover the
smaller JFCs that happen to
approach closer to the Sun and Earth. Therefore we may expect
that there would be a trend of greater $R_N$ with
greater $q$. 
From the Spearman $\rho$ value, we find however that the trend is 
not significant, yielding a significance
at only the $2.6\sigma$ level. In fact
any apparent trend is enhanced
perhaps more than it should be by the one comet at
the highest perihelion (5.5 AU); without it, the
significance is $2.4\sigma$. It is also
enhanced by our exclusion of some previously-studied,
relatively-large, small-$q$ JFCs from the survey (see below). 

\margrel{Fig. 8}

Figure 8 also shows that there is little difference in the
radii among comets for which there was no discernible dust
versus those that had discernible dust. 
The only appreciable
difference is in the 3.75- and 4.25-AU bins, where
there are not very many comets anyway. 
A Kolmogorov-Smirnov
test of these two subgroups of radii -- 54 with no dust and
35 with dust -- returns a statistic with 
significance 51.6\%, suggesting
that the two subgroups are indeed drawn from the same distribution
and that there is no significant difference in the radius
distributions. This is an important
point since it gives us more confidence about the
the quality of the nucleus extraction procedure on the
35 comets that showed dust. 

Figures 7 and 8 make it clear that many
comets in our sample lie in a restricted range
of perihelia. Fortunately, our sample is not much
different from the entire {\sl known} JFC population. 
This is shown in Fig. 9, which shows the cumulative
fractional distribution of perihelia among the
89 detected SEPPCoN comets and all 450 JFCs 
that were known at time of writing. The curves match quite closely,
and a Kolmogorov-Smirnov test yields a statistic
with significance of 81\%. 
So while there is certainly a bias in the discovery
rate of JFCs, the biases in SEPPCoN are no different.

\margrel{Fig. 9}

Fig. 8 also shows graphically the relative
number of large, medium, and small nuclei. In particular,
the upper left part of the plot is fairly blank. Why
are there so few nuclei larger than about 3.0 km with
perihelia below 2 AU? This is partly a result of our
excluding from the survey
several comets that have already been characterized. 
In particular, there are nine nuclei we
excluded whose
effective radii have been measured by spacecraft flybys
or by earlier mid-IR photometry. 
Those nine radii are: 
 2.4 km for 2P/Encke (Fern\'andez et al. 1999, Harmon and Nolan 2005),
 3.0 km for 9P/Tempel 1 (A'Hearn et al. 2005),
 5.2 km for 10P/Tempel 2 (A'Hearn et al. 1989),
 2.2 km for 19P/Borrelly (Soderblom et al. 2002),
10.7 km for 28P/Neujmin 1 (Campins et al. 1987),
 4.6 km for 49P/Arend-Rigaux (Campins et al. 1995, Millis et al. 1988),
 2.0 km for 67P/Churyumov-Gerasimenko (Lamy et al. 2008),
 2.1 km for 81P/Wild 2 (Brownlee et al. 2004, Duxbury et al. 2004),
and
 1.0 km for 103P/Hartley 2 (A'Hearn et al. 2011, Lisse et al. 2009).  
All of these comets have $q<2$ AU, and all but one of them
have radii above our SEPPCoN sample median of 1.4 km. So they
would fill in some of the sparser area in Fig. 8.

Yet another interesting feature of Fig. 8 
is apparent if one checks the discovery date
of the largest nuclei in the plot. Of the
12 nuclei with $R_N>3$ km, six (50\%) were discovered
in just the four years before the survey. 
And of the 17 nuclei with $R_N$ between 2 and 3 km,
five (29\%) were discovered as recently. In other words,
had our survey been done, for example, in the 1990s, many
large nuclei would not have been discovered
or available for study.

This point is illustrated in Fig. 10, where
we have plotted perihelion distance and discovery year
and used the symbol size to represent radius. This
plot includes not only our SEPPCoN detections
but also those comets among the nine discussed above that were discovered
after 1970. Many
large nuclei in our sample were discovered in the 2000s. 
This indicates that the census of
the JFC population is likely incomplete at most
radii, certainly even up to 
2 or 3 km. Furthermore, we are probably
incomplete even for comets with low perihelion. Among
those 11 recent comets with nuclei greater than 2 km
are comets 162P, 169P, P/2005 XA$_{54}$ --
three comets with perihelia of only 1.2, 0.6, and 1.8 AU,
respectively. If we divide the 98 nuclei here (89 SEPPCoN
plus nine from the literature) into a pre-2000 and post-2000
group, there are 61 comets in the former and 37 in the latter.
A Kolmogorov-Smirnov test of the radii in those two groups
indicates that they are likely drawn from the same distribution,
since the statistic yields a significance of 68\%.

\margrel{Fig. 10}

\medskip
\noindent{\sl 4.3. Notes about Radii} \par

\noindent{\sl 4.3.1. Previous Studies of Radii} \par

To make a broad comparison with our radii, we
checked for comets that were both in our sample and
in several recent compilations of radii created
by Lamy et al. (2004, hereafter L04;
using the ``$r_{n,v}$" column in their
Table 1),
Tancredi et al. (2006; hereafter T06),
Weiler et al. (2011; hereafter W11), and
Snodgrass et al. (2011; hereafter S11). 
The compilations represent a combination of radii from original data
and from reviews
of the literature, and so are not entirely independent from each other, 
but it is useful to make comparisons across a wide range of published results.
The compiled 
radii themselves are derived mostly from photometry
of apparently-bare nuclei, measurements of point-sources embedded
in faint comae, and extrapolations of activity to high $r$ --- i.e.,
from a range of methods. Among the radii that we are comparing
to SEPPCoN, they are almost entirely derived from visible-wavelength
data, and so our SEPPCoN results are
independent snapshots. 

We found 59 SEPPCoN comets that overlap with at least one of those
four works. The comparisons are shown in Fig. 11, where we show
the ratios of our radii to the radii from each older work. 
Some statistics about the ratios (excluding
the upper-limits) are listed in Table 8.
We note that the ratios do tend to straddle unity. 
We also note that there is no significant difference
in the means between the ratios from comets showing dust
in the SEPPCoN data, and the ratios from comets that do not.
This is reflected in the significance of the Student-$t$
statistic for such a comparison, given in the last column
of Table 8. One possible interpretation of this is 
that the nuclei
that we had to extract from dust comae are unlikely
to be significantly larger than those that were
bare. This is reassuring since it would mean the
dust removal process was successful.
Admittedly a more conservative interpretation might be that 
it is equally hard to extract robust nucleus photometry 
from comets with dust comae in both the infrared and visible
wavelengths. 

\margrel{Table 8}

Another representation of the ratios is shown in
the top four panels of Fig. 12, which has the histograms
of the data in Fig. 11, divided into 0.25-unit wide bins.
The white histograms include all the detected SEPPCoN comets,
and the grey histograms include only comets that showed no
discernible dust. Such comets have only slightly lower ratios
on average.

All this perhaps suggests
that there is broad consistency
between the SEPPCoN and the visible-wavelength
measurements -- and in particular that (a) the dust accounting we
(and others) have had to employ is mostly being used consistently,
and (b) a geometric albedo of $0.04$ 
is a reasonable assumption for visible-wavelength studies.

\margrel{Fig. 11}

However a little more digging reveals an interesting effect.
To demonstrate this, we start by postulating what the
expected scatter of the points in Fig. 11
around unity should be. Suppose all nuclei were prolate-ellipsoids
with semiaxes $a > b = c$, and that they all spin around
one of their short axes. If a spin axis were 
perpendicular to the observer's line of sight, a nucleus
would show a variable 
effective radius that 
oscillates between
$b$ and $\sqrt{ab}$ as it spins. For a rotation
axis that is not perpendicular, the bounds will be different,
but the effective radius will still vary (as long as
the view is not pole-on). 
Let $S$ be the cross section of the nucleus;
both L04 and S11 discuss how $S$ varies for a nucleus
viewed with aspect angle $\varepsilon$ (i.e. the angle
between the line-of-sight and the spin axis):
$$\leftline{\noindent$\displaystyle 
S = \pi a b^2 \sqrt{ {{\cos^2\varepsilon + 
                       \sin^2\varepsilon \cos^2\phi}\over{b^2}} + 
                     {{\sin^2\varepsilon \sin^2\phi}\over{a^2}} },                    
$\hfill(1)}$$
where $\phi$ is the sub-observer longitude, varying uniformly
from 0 to 360$^\circ$ over the course of a rotation period.
If we let $\axr$ be the axial ratio, $a/b$, then Eq. 1 simplifies to
$$\leftline{\noindent$\displaystyle
S = \pi b^2 \sqrt{\axr^2 + (1-\axr^2)\sin^2\varepsilon \sin^2\phi}.
$\hfill(2)}$$
The effective radius, $R_{ss}$, one would measure
from taking a snapshot of such a rotating nucleus at some time
would be $R_{ss}\equiv\sqrt{S/\pi}$, so
$$\leftline{\noindent$\displaystyle
R_{ss} = b\ \root 4 \of {\axr^2 + (1-\axr^2)\sin^2\varepsilon \sin^2\phi}.
$\hfill(3)}$$
If we were to take two snapshots of such a nucleus and
get two radii, $R_1$ and $R_2$,
the ratio of those snapshot radii would be 
$$\leftline{\noindent$\displaystyle
{{R_1}\over{R_2}} = \root 4 \of 
{{\axr^2 + (1-\axr^2)\sin^2\varepsilon \sin^2\phi_1}\over
 {\axr^2 + (1-\axr^2)\sin^2\varepsilon \sin^2\phi_2}}$$
$\hfill(4)}$$
and would only depend on $\axr$, $\varepsilon$, and the specific
longitudes -- i.e., not on the individual dimensions
themselves. Thus, an analysis of snapshot radii ratios 
can be done without having to worry about what the true
radii actually are, and so it is easy to determine
if the widths of the histograms in the top half of Fig. 12
are simply manifestations of looking at non-spherical nuclei.

\margrel{Fig. 12}

Unfortunately, it seems the easy answer is not the
correct one. This is shown in the bottom half of Fig. 12,
showing histograms of expected radii ratios from an
ensemble of 100,000 virtual nuclei. Each nucleus was given a
value of $\cos\varepsilon$ between $-1$ and 1, uniformly
distributed, and two values of $\phi$ between
0 and 360$^\circ$, also uniformly distributed. We then
assumed a common axial ratio for all 100,000, just for ease;
the value of the axial ratio is given in the upper right corner
of each panel in the bottom half of Fig. 12. The application
of Eq. 4 yields 100,000 radii ratios, which we plotted
with bins 0.01-units wide. While
larger axial ratios result in wider distributions, the
radii ratios are always strongly peaked near unity regardless,
and look nothing like the observed histograms in the top of Fig. 12.

We note that this is not the first time that snapshot
radii have been shown to be (in principle) excellent estimates of
``true" effective radii.
Weissman and Lowry (2003), L04, and S11 all discuss
how snapshot radii are related to the true effective radii,
and show that on average a snapshot radius 
is usually close (within a few percent)
to the nucleus's true effective radius. 
What we show in the bottom of Fig. 12 is simply a different manifestation 
of the same conclusion. 

The widths of the observed histograms at the top of Fig. 12
remain to be explained. There are several possible
explanations for this; 
perhaps the radii (ours and/or the earlier ones)
are simply wrong. There are certainly systematic effects
that could be influencing the radii: perhaps
the statistics are fooling us
and $\eta$ in actuality really does vary greatly across nuclei; 
perhaps incorrect phase laws in the mid-IR and/or in the visible
are being applied; 
perhaps dust
in the visible and mid-IR has not been properly accounted for,
and the photometry includes lingering dust or an overzealous
subtraction of dust; perhaps
cometary albedos have a greater range than is thought;
perhaps
nuclei that appear bare have significant, variable amounts
of dust within the seeing disk.
It is possible that a combination of these effects
is at work. 

The last point bears elaboration; as mentioned earlier
this phenomenon appears to be at work in the case
of comet 2P/Encke (Fern\'andez et al. 2000, 2005b).
Perhaps our dataset is suggesting that dust hiding within
the seeing disk is a more common phenomenon than
previously thought, and not limited to 2P. This would
indeed make studies of nuclei much more difficult
since it would require significantly more photometry
at multiple epochs and over a range of heliocentric distances.

One important clue however is the fact that each of
the four mean
ratios between our radii and the four compilations
(mentioned earlier this section)
is close to unity, as seen in Table 8. 
Furthermore the standard deviations are comparable. 
This suggests that
there is no single effect that is systematically
biassing the radii of any of the compilations 
(theirs and ours) in one direction. 
This is important because in general it is easier
to overestimate radii than to underestimate them. 
Yet that does not seem to be happening; the disparate
estimates of radii must have some other explanation.
In any case,
the relationship between SEPPCoN radii
and visible-wavelength photometry will be addressed
in future work.

We now discuss some individual comets whose SEPPCoN radii
are noticeably different from values reported in 
the four compilations, L04, T06, S11, and W11. We note that 
these four works did not necessarily use the
same basic assumptions to convert
visible photometry to radii, so the same magnitude report
can result in slightly different answers for the radius.

\noindent{\sl 4.3.2. 31P/Schwassmann-Wachmann 2} \par

We obtained a snapshot radius of 1.65 km, and this comet
did not have discernible dust (see Fig. 1). 
In contrast to this radius, T06, L04, and S11
all compiled a radius of 3.0 to 3.3 km,
all based on near-aphelion
CCD photometry reported by Luu and Jewitt (1992).
Luu and Jewitt did report a faint detection of coma in a sum
of images, but otherwise saw only a star-like point-source
in individual images. Perhaps this is a case of 31P
having dust in the seeing disk near aphelion. 

\noindent{\sl 4.3.3. 37P/Forbes} \par

We obtained a snapshot radius of 1.23 km. The accounting
for the dust was somewhat more complicated than usual
due to the tail (see Fig. 1).
This radius is in contrast to the 0.78 km compiled by
S11 and 0.81 km compiled by
W11. Both reports derive from the analysis 
presented
by Lamy et al. (2009) of their HST snapshots. Coma 
subtraction was applied
to these images as well and the nucleus was ``detected with
good contrast." It is of course possible that we
have underestimated the dust contribution.  

\noindent{\sl 4.3.4. 43P/Wolf-Harrington} \par

We obtained an upper limit of 1.01 km, though
we point out that there was a bright
star whose first Airy ring
was close to the expected position of 43P in our IRS
data (see Fig. 1), complicating the estimate of an upper limit.
Nonetheless, our radius is smaller than earlier results:
L04 list 1.8 km, T06 list 2.1 km, and
W11 list 3.4 km. S11
give an upper limit of 2.4 km, and in fact 
Snodgrass et al. (2006) discuss the case of 43P in detail,
arguing that it has unresolved coma in the known visible-wavelength
observations out to almost $r=4.9$ AU. Thus, all previously-published 
radii should be treated as upper limits. Our observations
occurred when the comet was over 5.3 AU, so our 
small radius upper-limit
is consistent with the hypothesis of unresolved coma affecting
the earlier results.

\noindent{\sl 4.3.5. 50P/Arend} \par

We obtained a snapshot radius of 1.49 km, with not much dust to remove
and a high-contrast point-source. Our four comparison 
works all compile about 0.95 km, all based on the work
of Lamy et al. (2004) using HST data when
the comet was at $r=2.4$ AU. The nucleus there was
reported to have ``moderate contrast" but the coma subtraction
fits are excellent. Our observations of course have
lower resolution so perhaps there is dust residing
in the seeing disk closer to aphelion.

\noindent{\sl 4.3.6. 69P/Taylor} \par

Our snapshot radius is 0.87 km. Dust removal in the IRS blue
image was straightforward, but was somewhat more complicated
in the red image (see Fig. 1). Nonetheless the comet appeared
fainter than expected in our imaging.
The radius is much smaller than previously
reported: W11 list 3.60 km and S11 list 3.88 km,
both based on the work of
Lowry et al. (1999), who reported an unresolved point-source
when observed at $r=4.9$ AU. 
T06 list 2.1 km, and they and L04 remark that the trend
of photometry with $r$ suggests the comet is still active
out to 4 to 5 AU. (And indeed our observations showing
dust occurred at $r=4.2$ AU.)
Perhaps this is another case of a comet showing
dust in the seeing disk near aphelion. 

\noindent{\sl 4.3.7. 79P/du Toit-Hartley} \par

This comet appeared fainter than expected also, and
was not a comet with discernible dust. Our
snapshot radius is 0.70 km, compared
to 1.4 to 1.5 km compiled by all four of our comparison works,
all based on the work of Lowry et al. (1999), who
found an unresolved point-source comet when observing
it at 4.74 AU, near aphelion. However they do also mention that
``the dominant flux source" -- i.e. nucleus vs. dust -- ``is uncertain."
Admittedly, the same could certainly be said for the
IRS imaging at 4.37 AU. 

\noindent{\sl 4.3.8. 118P/Shoemaker-Levy 4} \par

Here is yet another case where our
snapshot radius is smaller than expected. Some dust was detected
in our IRS imaging (see Fig. 1) and in particular
it was difficult to remove the dust from the blue image.
The point source in the head of both images was definitely
fainter than predicted, however. We obtained
a radius of 1.30 km. L04, W11, and S11 all list
radii of 2.4 to 2.6 km based on the 
work of Lowry et al. (2003), whose snapshots
showed an unresolved point-source when observed at $r=4.7$ AU.
Our observations occurred at $r=4.9$ AU, only slightly farther
but essentially at aphelion. Perhaps that was enough for the
comet to actually have turned off.  

\noindent{\sl 4.3.9. 120P/Mueller 1} \par

We obtained an upper limit of 0.48 km for the radius.
This is smaller than other reported values, suggesting
that we should have seen the comet in our IRS imaging.
L04 and W04 compiled 1.5 km, based on work by
Lowry et al. (1999) observing the comet at $r=3.1$ AU.
S11 compiled 0.77 km based on more recent observations
by Snodgrass et al. (2008) at higher $r$, 3.9 AU,
suggesting that the comet might have still been active at 3.1 AU.
Our SEPPCoN observations occurred at higher $r$ still,
4.8 AU, so perhaps this comet has dust in the
seeing disk until very close to aphelion.

\noindent{\sl 4.3.10. 121P/Shoemaker-Holt 2} \par

This comet was brighter than expected, with our finding
of a snapshot radius of 3.87 km. The comet showed dust in
our IRS imaging (see Fig. 1), but the point-source had high contrast
against it (with the Airy rings visible), and so the dust
removal was relatively straightforward. In comparison,
L04 compiled 1.6 km, T06 compiled 2.0 km, and W11 compiled
1.6 km, based on the work by Lowry et al. (2003)
that indicated a point-source comet when observed near $r=5$ AU.
S11 lists 3.34 km however, much closer to our value, and
is based on more recent data from Snodgrass et al. (2008)
that also show a point-source in the head of the comet
but with a faint tail next to it. These data were taken when
$r\approx4$ AU, near the value for the IRS data (4.3 AU). 
Perhaps the comet keeps dust in the seeing-disk but
fades rapidly once it turns off close to 5 AU.

\noindent{\sl 4.3.11. 243P/2003 S2 (NEAT)} \par

We obtained an upper limit of 0.55 km, which is smaller
than the 1.5 km compiled by S11 from
observations by Mazzotta Epifani et al. (2008). However,
Mazzotta Epifani et al. (2008) do state that an estimate of
coma contamination in their (albeit very deep and
stellar-looking) comet image could yield a radius as small as 0.8 km.

\noindent{\sl 4.3.12. Sky Survey Comets from W11} \par

Among the 32 comets that overlap between our survey and
the compilation by W11, there are 15 comets for which
W11 derived radii using photometry of archival sky survey observations.
These points are on the right half of W11's
panel in Fig. 11, and most of them are very clearly offset
from the other 17 comets and from the unity line. 
For 10 of those 15, the ratios 
are below 0.5, i.e. W11's radius is more than twice as large
as ours. For seven of those the W11 is radius is more
than four times as large.
This suggests perhaps a systematic effect is in play. For
example maybe the sky survey observations are not as clean of coma
as would be indicated from the radial profiles, although
W11 did try to account for such an effect. As mentioned
in \S4.3.1, future work will try to understand the large
differences in radii of these comets.

\medskip
\noindent{\sl 4.4. Radii and Size Distribution} \par

The size distribution of the JFCs depends on the properties
of the source region and the
evolutionary processes that have changed the radii
of the nuclei from their initial values. We can
use our 89 new radii to estimate this distribution.
To augment these 89, we now add in the nine other radii
we mentioned in \S 4.2.
Thus we are extracting a size distribution from 98 JFC nuclei.
The cumulative size distribution (CSD) for these 98 comets
is shown with the solid line in Fig. 13. The CSD is moderately
steep in the larger sizes, trending almost as the inverse-square of
radius, 
but it has a break in slope at smaller radii. 
The turnover 
is primarily due to our relatively poor sampling of that size regime.
However, Meech et al. (2004) modeled the expected sampling bias
one would expect from the discovery of comets and found that
such a turnover in the CSD could be explained by an actual dearth
of sub-kilometer JFCs. According
to Meech et al., many more sub-kilometer
JFCs should have been discovered
by now.  In other words, while the slope of
the CSD in Fig. 13 gets shallower at smaller sizes partly because of sampling, 
it apparently would also do the same thing if there were
not as many small JFCs out there to be
discovered in the first place. We will discuss this further in the next section. 

\margrel{Fig. 13}

We have also plotted on Fig. 13 a curve that only includes
the 54 SEPPCoN comets that had no discernible dust. 
We added the nine literature comets to make the total 63.
The figure shows that there is no significant difference
between the 63- and 98-comet CSDs. This is gratifying
since (as we have mentioned in other contexts earlier
in this paper) it suggests that the radii from
SEPPCoN comets that did appear with discernible
dust should not be wildly off, and thus their
inclusion in the CSD is not biasing our conclusions about the CSD.

The CSD approximates a power-law, and as is common we will characterize
the distribution with a power-law slope $\beta$, defined
as $N(>R_N) \propto R_N^{\beta}$, where $N$ is
the cumulative number of comets. Best fitting $\beta$ values to the
98-comet CSD are given in Table 9, where we present the slopes using  
a variety of radius-ranges to do the
fitting. Choosing an appropriate
range over which to fit the observed CSD is an uncertain process; on
the small end, it is difficult to quantify just where the
CSD starts to turnover, and
on the large end, the small-number statistics are problematic and
one does not want to give too much weight to the few largest
comets. These realities motivated us to try different choices
for the range of radii. A
better way to approach the problem is to follow the procedures
of Meech et al. (2004) or of Snodgrass et al. (2011) and account
for some of observational uncertainties and biases explicitly. We address
these issues in future work, so here we report for completeness
slopes fitted to many ranges. The median $\beta$ among
the 20 ranges we tried is $-1.92$, with a standard deviation
of $0.23$.

\margrel{Table 9}

We note that if we had {\sl not} converted the IRS
radii to a common value of $\eta$ (as described in \S 4.2), but instead
just used the radii in Table 4, we would have found
that the resulting 98-comet CSD would be shallower. Using the same
radii bounds as given in Table 9, we find that the
median $\beta$ of such
a CSD would be $-1.72$ with a standard deviation of $0.17$. 
This is largely due to the fact that (a) there are more high-$\eta$
IRS comets than low-$\eta$ IRS comets, and (b) the range of fitted
$\eta$ values seems to extend farther above the mean than it
does below it (so since higher $\eta$
means higher $R_N$, we have more of a boost to the CSD
at the large-radius end
than at the small-radius end). The validity of
this CSD over the one shown in Fig. 13 would rest on
whether or not one assumes that JFCs have a common beaming parameter.
Even though nearly all of our IRS comets have $\eta$ values consistent
with a single value near unity, suggesting that this might
be a common property (\S4.1), it is of course possible that
this is the wrong conclusion, and that there is more variety
to the beaming parameter. A more detailed analysis of 
this aspect of the CSD will be addressed in a future paper. 

Figure 13 also shows the observed CSD if we restrict the sample
by perihelion distance. One can argue that we are more likely
to have a complete census of the JFC population at the smaller
radii if we only consider JFCs that approach to within 1.5 or 2.0
AU of the Sun, since such comets are more likely to be discovered.
Perhaps such a perihelion-restricted CSD is less likely
to suffer from discovery biases. 
However Fig. 13 shows that
the shapes of the CSDs are not radically different, at least
away from the very largest end of the CSDs where there are only
one or two comets. This is confirmed quantitatively with
Kolmogorov-Smirnov tests. The test returns
a statistic with significance near 100\% for a comparison
of the 98-comet
distribution and the $q<3$ AU distribution,
a significance of 99\% for the $q<2.5$ AU distribution,
a significance of 70\% for the $q<2.0$ AU distribution,
and
a significance of 22\% for the $q<1.5$ AU distribution.
Tests using other limiting values of $q$ and tests that use
a limiting $q$ as a lower limit yield similarly high probabilities. 
Thus there is good confirmation that a $q$-limited subset
of the radii are no different than the entire 98-comet CSD.

This result is also evident from the $\beta$ values
for these $q$-limited CSDs listed in Table 9.
The typical slopes
of each CSD are similar regardless of where we cut off the perihelion.
For the CSD restricted to $q<3$ AU, the median $\beta$ is $-1.96$;
for $q<2.5$ AU, it is $-1.85$; and 
for $q<2$ AU, it is $-1.84$.
(We did not fit $\beta$ for the $q<1.5$ AU CSD since there
are too few comets left.)
There are some outliers among the slopes in Table 9, but
the similarity of the medians suggests 
that there is not a significant perihelion
bias in the overall, 98-comet CSD, at least among the
sample of comets we selected to study here. 

Closer inspection of Fig. 13 reveals an interesting
feature in the 98-comet CSD: it has a 
small bump at radii from 3 to 6 km. 
Such waviness is seen in the CSD of main-belt asteroids,
and is related to the asteroids' strength 
(e.g. Jedicke and Metcalfe 1998, O'Brien and Greenberg 2003).
If the JFC bump is real and not an artifact,
one exciting hypothesis is that
it is a remnant feature of the primordial
size distribution of the JFC's progenitor SDOs.
If the SDOs are the source
of today's JFCs and they originally had a peak in their CSD
at 3-6 km radii, subsequent evolution 
(e.g., collisions, activity, fragmentation) 
would produce smaller objects that 
would fill out the CSD but also could
have left a bump at the radii where we see
one in Fig. 13. 
In this interpretation, the CSD
is giving us clues about the original properties of the population.
However corroborating this hypothesis will require
surveying many more JFCs to gain confidence in the
reality of the bump. In particular we will need
to be sure that most of the large nuclei have
been discovered (cf. \S4.2), since we can only identify a bump
in the context of the shape of the CSD that surrounds it.
Furthermore, it would be useful
to identify such a peak in the SDO CSD, which
means finding SDOs with
R-band magnitudes near $\sim$28-29, a quite challenging task.

We also note that the perihelion-restricted CSDs in Fig. 13 appear to have
more waviness than the all-comet CSD does. We
surmise that this is due to the particular radii
of recently-discovered comets, but a better census of nucleus sizes
(among both JFCs that are known and those waiting to be discovered)
could certainly address whether the waviness is real.

\medskip
\noindent{\sl 4.5. Implications of the CSD} \par

We show in Fig. 14 
a comparison of our CSD to those of L04, T06, S11, W11, and also to
that of
Meech et al. (2004). For S11, we excluded upper limits in 
their Table 3 to make this plot. For L04, we only used
values of $r_{n,v}$ in their Table 1. For T06, we only
used entries with quality code 3 or better from their Table 1.
For W11, we only used the 46 comets in their Table 5
and excluded the sky-survey comets in their Table 4 (cf. \S4.3.12).
The top set of curves shows that our CSD is broadly similar
to these others. This is somewhat easier to see in the
bottom set of curves, where we have greyed out the individual
points for clarity and placed simple power-law fits over
a restricted (but common to all) range of radii -- 2 to 5 km -- to aid the eye.

\margrel{Fig. 14}

Comparing our CSD slope to other {\sl reported} slopes
we find that our $\beta$ is similar to that found
by L04 ($\beta=-1.9\pm0.3$ for $R_N$=1.6 to 15 km), S11 
($\beta=-1.92\pm0.20$, for $R_N\!\ge$1.25 km), and W11
($\beta=-1.76\pm0.45$,for $R_N\!\ge$2.0 km), 
as well as that reported
by Weissman et al. (2009; $\beta=-1.86\pm0.15$). In contrast,
T06 reported a steeper result, $\beta=-2.7\pm0.3$ ($R_N$=1.7 to 4.5 km). 
We note the differing ranges of radii that were applied by
these workers to derive these slopes. In some cases, slopes over multiple
ranges were published (cf. our Table 9); W11 did this, as did
Meech et al. (2004), who report
$\beta=-1.45\pm0.05$ for $R_N$=1 to 10 km but
$\beta=-1.91\pm0.06$ for $R_N$=2 to 5 km.

We also used Kolmogorov-Smirnov tests to compare our radii
with those five other compilations and quantify how well
they match. The likelihoods are quite high that all radius
distributions are drawn from the same distribution.
For our radii and those of Meech et al. (2004), the statistic
has a significance of 41\%;
for our radii and those of S11, it is 79\%;
for our radii and those of W11, it is 73\%;
for our radii and those of L04, it is 67\%; and
for our radii and those of T06, it is 22\%.

We note that all these other works
for the
most part make use of visible-wavelength
data. Thus our CSD is almost entirely independent of these earlier
results. The similarity in power-law
slope does indicate that there is likely no strong albedo
trend with radius among the JFCs. In other words, the assumption
of a common geometric albedo to all JFCs that is so frequently invoked
when interpreting visible-wavelength photometry appears to be valid. 
This is consistent with the arguments presented by Lamy et al. (2004),
though it remains to be seen what the 
measured spread of albedos across
JFCs actually is.

Also shown in Fig. 14 is model `Pb' from Meech et al. (2004),
the predicted CSD that one would measure given
an intrinsic CSD that is subject to discovery biases.
This model has a base
{\sl differential} size distribution (not CSD) with
power-law slope of $-3.5$, but multiplied
by a three-piece continuous function that was valued at 0 for $R_N\le0.3$ km,
was valued at unity for $R_N\ge2.0$ km, and was a line in between.
Hence the intrinsic size distribution is truncated entirely
below $0.3$ km and suppressed
to varying degrees below $2.0$ km. By forcing
the JFCs to have no very small members and relatively fewer 
small (kilometer-scale) members
than would otherwise be expected, and by convolving the intrinsic
distribution with a model of activity and
of discovery biases, Meech et al. (2004)
were able to reproduce their observed CSD.
As Fig. 14 demonstrates, our CSD has a similar shape
to that of Meech et al. (2004), suggesting that a similar
model with truncation at small radii (perhaps just with
different truncation parameters) could potentially reproduce
the overall shape of our CSD (excluding the bump mentioned
in \S4.4).
However it is important to once again
note that our sample criteria tended to remove relatively small
nuclei from our observations, and it is unclear if this bias is the
same in mid-IR observations as it could be at visible wavelengths.

If the Meech et al. (2004) intrinsic distribution is
correct, it suggests that the JFC population is 
collisionally relaxed (Dohnanyi 1969) but also suffers
from some processes that remove the smaller members and
flatten the true and observed distributions. These could
be, for example, regular cometary activity and fragmentation.
The effects of these processes on the CSD have been
discussed previously by, e.g., Lowry et al. (2003)
and Weissman and Lowry (2003).
In particular, fragmentation is a promising phenomenon since
it has been directly observed in many JFCs 
(Boehnhardt 2004, Chen and Jewitt 1994, Fern\'andez 2009),
and once in a Centaur
(Bauer et al. 2008). Given the observed rates
of fragmentation (Chen and Jewitt 1994), an individual
JFC likely suffers tens or hundreds of fragmentation events
over its active lifetime.

Interestingly, studies of fragment populations in individual
comets suggest that the fragmentation process itself 
leads to shallow size distributions. For example,
Fuse et al. (2007) took a visible-wavelength snapshot of Fragment B
of comet 73P/Schwassmann-Wachmann 3 and found many tiny
(meter-scale) subfragments that had come off of Fragment B
in the previous few days. The CSD of
the fragments had a power-law slope of around $-1.1$. 
Reach et al. (2007) studied several of 73P's fragments in
a wider-field Spitzer mosaic, fragments that had probably been
released $\sim$10 years in the past. The ``large" fragments
(roughly $>$150 m in radius) had a CSD slope of $-1.6$,
while the ``small" fragments (roughly $<$150 m) had a slope
of around $-0.8$. This is an especially interesting
result since it suggests that perhaps the effects of fragmentation itself
are size-dependent. 
Fern\'andez (2009) derived a power-law slope of $-1.3$ for the 
CSD of the fragments released
by comet 57P/du Toit-Neujmin-Delporte over the course of
(apparently) several months in 2002. These fragments
had diameters of roughly hectometer-scale.
We note that all these $\beta$ are even shallower than what has been derived
here for the JFC nuclei as an ensemble, 
suggesting that
fragmentation could play a significant role in controlling
the cometary size distribution. If fragmentation events
supply a shallow distribution of smaller JFCs into the inner
Solar System (as the fragments get perturbed
onto different orbits, as happened for 42P and 53P [Kres\'ak et al. 1984]), 
then the overall JFC CSD should reflect that.

We should note that if the bump mentioned
in \S4.4 is real, then perhaps the input population's
CSD is not  
a perfect collisionally-relaxed power-law. In that case,
the overall JFC CSD upon which the fragmentation 
process operates would not be a collisionally-relaxed
power-law either. Evidently this would complicate finding
a model for the intrinsic CSD
that would uniquely match the observed CSD. 

Traditionally the trans-Neptunian region -- and in
particular, the scattered disk -- has been
thought to be the primary, original source region for the
JFCs we see today (e.g. Duncan et al. 2004). 
A comparison of the size distributions among these populations
would thus be worthwhile. One significant issue however 
is the fact that 
there is minimal size matching between JFCs and 
known TNOs; few known TNOs, and no known SDOs, are small enough. 
Therefore estimates of the TNO size distribution
at the kilometer scale -- which would
be necessary if one is to make a direct comparison -- are still uncertain.
Centaurs are closer in size to the JFCs (cf. P/2004 A1 in Table 5)
but the number with known sizes matching the JFCs' sizes is still too limited.

One deep survey of the trans-Neptunian region sensitive to TNOs
with somewhat JFC-like sizes is that of Bernstein et al. (2004), which used
the Hubble Space Telescope (HST) and would
have been able to find TNOs at about $\sim$29th magnitude in R band. 
For reference,
an object at opposition that is 34 AU from the Sun and
has 4.0\% geometric albedo would have an R band magnitude of 29.0
if the radius were 5.0 km. Thus the Bernstein et al. survey
was looking for objects similar but still mostly larger than the bulk of the JFCs.
The survey detected about 1.5 orders of magnitude
fewer objects than they would expect based on the measured
luminosity function of TNOs at larger ($\sim$100 km) sizes.
In other words the survey implies there is a break in the
TNO luminosity function and so a distinct underabundance of 
JFC-sized objects currently in the trans-Neptunian region.
While there is still some uncertainty in the large-TNO
luminosity function and size distribution (as addressed
in detail by e.g. Petit et al. (2008)), the Bernstein
et al. survey would suggest that there are too few objects
to explain the current size of the JFC population, as Bernstein et al. (2004)
themselves note as do Volk and Malhotra (2008) in a more
detailed analysis. Essentially, the size distribution function
of $\sim$1-km TNOs, as extrapolated from the Bernstein et al. detections
of larger objects, is too flat. 

More recently, Schlichting et al. (2012) have reported
an analysis of small TNOs that were found
via a survey of stars observed by the Fine Guidance
Sensors aboard HST. The stars' photometry was searched
for signs of occultations by intervening TNOs. 
Their two detected objects have sub-kilometer
radii and so are objects that really do
overlap the size of many JFCs. The
distribution at these sub-kilometer scales,
as implied by this survey and its detections, is still
shallower than that implied by the size distribution
of larger TNOs -- so they confirm the break in the size 
distribution -- but it is not as shallow as implied
by the Bernstein et al. (2004) survey. As a result, 
Schlichting et al. (2012) conclude that there are 
in fact sufficient number of objects to explain the
JFC population as estimated by Volk and Malhotra (2008).

Furthermore, Schlichting et al. (2012) report
that the exponent of the differential size
distribution's power-law is around $-3.4$ to $-4.0$,
depending on the ecliptic latitude distribution. 
If true, it means there would be significant flattening
of the size distribution (likely as a result of physical
evolution) as these small TNOs become JFCs. 
Interestingly, 
Shankman et al. (2013) have recently argued for a ``divot" in
the TNO size distribution, where the number of
objects in a size bin suddenly drops but then rises again with
a different power-law. This would explain the observational
results and likewise provide for enough objects to 
explain the JFC population. 
  
In any case, since the observational biases for JFCs
are so different from those for Centaurs and TNOs
(as mentioned by Meech et al. (2004)), direct comparisons 
of CSDs can be problematic. But certainly more discoveries
of TNOs closer in size to the JFCs are necessary. Petit
et al. (2008) point out that understanding the apparent break
in the TNO size distribution is going to at least require
a survey covering several square-degrees of sky that
can reliably discover objects of 28th magnitude, an
ambitious project to be sure.

\bigskip
\noindent{\bf 5. Summary} \par

SEPPCoN has detected 89 nuclei of JFCs with the Spitzer
Space Telescope, making SEPPCoN the largest photometric
survey so far of JFCs
in the mid-infrared. Of the sample, 54 of the comets
appeared bare, and 35 appeared with discernible dust.
For comets in the latter group, a photometric extraction
technique was used to measure the nucleus's brightness alone.
We assumed that the point-source was identical to the nucleus. 
We conclude the following:

1. Of our 57 nuclei detected at two wavelengths, 55 are
consistent with the hypothesis that the nuclei all have
the same beaming parameter. We find that the mean beaming
parameter is $1.03\pm0.11$. Only two of the 57 comets (3.4\%),
89P and P/2005 JQ$_5$, have beaming parameters inconsistent with
this value, and the latter is consistent with having a dust-dominated
head with minimal nucleus. 

2. By virtue of this beaming parameter's value,  
we conclude that nearly all JFC
nuclei have low thermal inertia with little night-side
emission, and little anisotropy in
their day-side thermal emission. This conclusion is even more robust
considering that virtually all observations occurred when the comets were
4 to 5 AU from the Sun, i.e., when the nuclei should be relatively cold
and it should be easier for any thermal memory to be retained
onto the night side. 

3. The beaming parameter is uncorrelated with any observational
or orbital parameter in our sample, indicating that it is probably
intrinsic to the nuclei.

4. The census of the JFC population is apparently incomplete
even in the inner part of the Solar System and even up to
fairly large radii, $\sim$3 km. Many of the recently-discovered
comets in our sample (i.e. in the four years before
the survey was conducted)
have radii of 2 km and larger and so contribute
to the large end of the size distribution. While it is not
surprising that this is true for comets with perihelia out beyond
2.5-3 AU, we conclude that there must still be large,
undiscovered JFCs that approach to 1.5-2 AU from the Sun.

5. Combining our 89 new radii with nine radii from the literature
yields a total sample of 98 JFCs. The resulting CSD 
follows a power-law with slope of about $\beta=-1.9$. Varying
the range in radius over which the CSD is fit yields a standard
deviation in $\beta$ of about $\pm0.2$. 

6. The CSD slope is similar even if we restrict our sample
to those comets with perihelia under 2, 2.5, or 3 AU. This
indicates that even though many large
nuclei in the JFC population may be missing, our sample currently
shows no extra bias with respect to perihelion distance.    

7. Our CSD appears to show a bump near radii of 3 to 6 km,
suggesting an excess of JFCs with those sizes. A more
complete sampling of JFC radii will be required to establish
the reality of this bump, which could have interesting
implications for the origin and/or evolution of the nuclei.

8. The best fit power-law slope is consistent with results
taken in the visible-wavelengths, and suggests that
there is no broad dependence of albedo on radius among the JFC population;
i.e., that the assumption of a single geometric albedo for all nuclei
is reasonable. 

9. In particular, the power-law slope is reasonably close
to the best fitting model of Meech et al. (2004), suggesting
that our observed CSD is consistent with an intrinsic population
that is truncated at sub-kilometer scales.

\bigskip
\noindent{\bf Acknowledgements} \par
We are grateful for the recommendations and suggestions
to this manuscript made by Olivier Hainaut and one
anonymous referee. This work is based on observations made with the 
Spitzer Space Telescope, which is operated by the 
Jet Propulsion Laboratory, California Institute of 
Technology under a contract with NASA. Support for 
this work was provided by NASA through an award 
issued by JPL/Caltech, RSA \#1289123, to YRF and HC. Support for
this work also came from NASA through grant NNX-09AB44G
to YRF, SCL, KJM, and JMB;
from the National Science Foundation through grant 
AST-0808004 to YRF, SCL, and KJM; 
and from the European Union Seventh
Framework Programme (FP7/2007-2013) under grant agreement no. 268421 to CS.

\bigskip
\noindent{\bf References} \par

\def\jrnl#1{\parindent=0pc \hangindent 1.2cm \hangafter 1  {#1}\par}

\jrnl{A'Hearn, M.F., Campins, H., Schleicher, D.G., Millis, R.L., 1989.
The nucleus of comet P/Tempel 2. Astrophys. J. 347, 1155--1166.}

\jrnl{A'Hearn, M.F., Millis, R.L., Schleicher, D.G., Osip, D.J.,
Birch, P.V., 1995.
The ensemble properties of comets: Results from 
narrowband photometry of 85 comets, 1976-1992.
Icarus 118, 223--270.}

\jrnl{A'Hearn, M.F., et al., 2005.
Deep Impact: Excavating comet Tempel 1.
Science, 310, 258--264.}

\jrnl{A'Hearn, M.F., et al. 2011. EPOXI at comet
Hartley 2. Science 332, 1396--1400.}

\jrnl{A'Hearn, M.F., et al. 2012. Cometary volatiles
and the origin of comets.
Astrophys. J. 758, 29.}

\jrnl{Alvarez-Candal, A., Licandro, J., 2006.
The size distribution of asteroids in cometary orbits
and related populations.
Astron. \& Astrophys. 458, 1007--1011.}

\jrnl{Bauer, J.M., et al., 2008.
The large-grained dust coma of 174P/Echeclus.
Publ. Astron. Soc. Pacific 120, 393--404.}

\jrnl{Bernstein, G.M., et al., 2004. 
The size distribution of trans-Neptunian bodies.
Astron. J. 128, 1364--1390.}

\jrnl{Boehnhardt, H., 2004. Split comets.
In: Festou, M.C., Keller, H.U., Weaver, H.A. (Eds.),
Comets II. Univ. of Arizona Press, Tucson, pp. 301--316.}

\jrnl{Brown, R. H., 1985. 
Ellipsoidal geometry in asteroid thermal models -- 
The standard radiometric model. Icarus 64, 53--63.}

\jrnl{Brownlee, D.E., et al., 2004. Surface of young
Jupiter family comet 81 P/Wild 2: View from the Stardust spacecraft.
Science 304, 1764--1769.}

\jrnl{Campins, H., A'Hearn, M.F., McFadden, L.A., 1987.
The bare nucleus of comet Neujmin 1. 
Astrophys. J. 316, 847--857.}

\jrnl{Campins, H., Osip, D.J., Rieke, G.H., Rieke, M.J.,
1995. Estimates of the radius and albedo of 
comet-asteroid transition object 4015 Wilson-Harrington 
based on infrared observations. Planet. \& Space Sci. 43, 733--736.}

\jrnl{Chen, J., Jewitt, D., 1994.
On the rate at which comets split.
Icarus 108, 265--271.}

\jrnl{Delb\'o, M., Dell'oro, A., Harris, A.W.,
Mottola, S., Mueller, M., 2007.
Thermal inertia of near-Earth asteroids and 
implications for the magnitude of the Yarkovsky effect.
Icarus 190, 236--249.}

\jrnl{Dohnanyi, J.S., 1969. Collisional models of asteroids 
and their debris. 
J. Geophys. Res. 74, 2531--2554.}

\jrnl{Duncan, M., Levison, H., Dones, L. 2004.
Dynamical evolution of ecliptic comets. In: Festou, M.C., 
Keller, H.U., Weaver, H.A. (Eds.), Comets II. 
Univ. of Arizona Press,
Tucson, pp 193--204.}

\jrnl{Duxbury, T.C., Newburn, R.L., Brownlee, D.E. 2004.
Comet 81P/Wild 2 size, shape, and orientation.
J. Geophys. Res. 109, E1202.}

\jrnl{Ernst, C.M., Schultz, P.H., 2007.
Evolution of the Deep Impact flash: Implications 
for the nucleus surface based on
laboratory experiments. Icarus 190, 334Ð344 (2007)}

\jrnl{Farinella, P., Davis, D. R., 1996.
Short-period comets: Primordial bodies or collisional fragments?
Science 273, 938--941.}

\jrnl{Fern\'andez, J.A., Morbidelli, A., 2006.
The population of faint Jupiter family comets near the Earth.
Icarus 185, 211--222.}

\jrnl{Fern\'andez, J.A., Sosa, A., 2012.
Magnitude and size distribution of long-period comets
in Earth-crossing or approaching orbits.
Mon. Not. R. Astron. Soc. 423, 1674--1690.}

\jrnl{Fern\'andez, Y.R., 1999. Physical properties of cometary nuclei.
Thesis, University of Maryland, College Park.}

\jrnl{Fern\'andez, Y.R., 2009.
That's the way the comet crumbles: Splitting
Jupiter-family comets. Planet. \& Space Sci., 57, 1218--1227.}

\jrnl{Fern\'andez, Y.R., et al., 2000.
Physical properties of the nucleus of comet 2P/Encke.
Icarus 147, 145--160.}

\jrnl{Fern\'andez, Y.R., Meech, K. J., Lisse, C. M.,
A'Hearn, M. F., Pittichov\'a, J., Belton, M. J. S. 2003.
The nucleus of Deep Impact target comet 9P/Tempel 1.
Icarus 164, 481--491.}

\jrnl{Fern\'andez, Y.R., Jewitt, D.C., Sheppard, S.S., 2005a.
Albedos of asteroids in comet-like orbits.
Astron. J. 130, 308--318.}

\jrnl{Fern\'andez, Y.R., Lowry, S.C., Weissman, P.R.,
Mueller, B.E.A., Samarasinha, N.H., Belton, M.J.S.,
Meech, K.J. 2005b.
New near-aphelion light curves of comet 2P/Encke.
Icarus 175, 194--214.}

\jrnl{Fern\'andez, Y.R., et al., 2006.
Comet 162P/Siding Spring: A surprisingly large nucleus.
Astron. J. 132, 1354--1360.}

\jrnl{Fern\'andez, Y.R., Jewitt, D., Ziffer, J.E., 2009.
Albedos of small Jovian Trojans. Astron. J. 138, 240--250.} 

\jrnl{Fuentes, C.I., George, M.W., Holman, M.J., 2009.
A Subaru pencil-beam search for m$_R\sim$27 trans-Neptunian
bodies. Astrophys. J. 696, 91--95.}

\jrnl{Fuse, T., Yamamoto, N., Kinoshita, D., Furusawa, H.,
Watanabe, J., 2007. 
Observations of fragments split from nucleus B of comet
73P/Schwassmann-Wachmann 3 with Subaru Telescope.
Publ. Astron. Soc. Japan 59, 381--386.}

\jrnl{Groussin, O., Lamy, P., Jorda, L., Toth, I., 2004.
The nuclei of comets 126P/IRAS and 103P/Hartley 2.
Astron. \& Astrophys. 419, 375--383.}

\jrnl{Groussin, O., et al., 2007.
Surface temperature of the nucleus of comet 9P/Tempel 1.
Icarus 187, 16--25.}

\jrnl{Groussin, O., et al., 2009.
The size and thermal properties of the nucleus
of comet 22P/Kopff. Icarus 199, 568--570.}

\jrnl{Groussin, O., et al. 2013.
The temperature, thermal inertia, roughness and color
of the nuclei of comet 103P/Hartley 2 and 9P/Tempel 1.
Icarus 222, 580--594.}

\jrnl{Harmon, J.K., Nolan, M.C., 2005.
Radar observations of comet 2P/Encke during the 2003 apparition.
Icarus 176, 175--183.}

\jrnl{Harmon, J.K., Nolan, M.C., Margot, J.L., Campbell, D.B.,
Benner, L.A.M., Giorgini, J.D. 2006.
Radar observations of comet P/2005 JQ5 (Catalina).
Icarus 184, 285--288.}

\jrnl{Harris, A.W., 1998. A thermal model for near-Earth
asteroids. Icarus 131, 291--301.}

\jrnl{Harris, A.W., et al., 2011.
ExploreNEOs. II. The accuracy of the warm Spitzer
near-Earth object survey.
Astron. J. 141, 75.}

\jrnl{Houck, J.R., et al., 2004.
The Infrared Spectrograph (IRS) on the Spitzer Space Telescope.
Astrophys. J. Suppl. Ser. 154, 18--24.}

\jrnl{Howell, E.S., Vervack, R.J., Nolan, M.C.,
Magri, C., Fern\'andez, Y.R., Taylor, P.A., Rivkin, A.S. 2012.
Combining thermal and radar observations of near-Earth asteroids.
In: Asteroids, Comets, Meteors 2012, Proceedings of
the conference held May 16-20, 2012, in Niigata, Japan.
LPI Contribution No. 1667. LPI, Houston, id. 6273 [abstract].}

\jrnl{Jedicke, R., Metcalfe, T.S., 1998. The orbital and absolute 
magnitude distributions of main-belt asteroids. Icarus 131, 245--260.}

\jrnl{Jewitt, D. 1991.
Cometary photometry. In: Newburn, R. L., Neugebauer, M., 
Rahe, J. (Eds.), Comets in the Post-Halley Era.
Kluwer, Dordrecht, pp. 19--65.} 

\jrnl{Kelley, M.S., et al., 2013. The persistent activity of Jupiter-family 
comets at 3-7 AU.
Icarus 225, 475--494.}

\jrnl{Kenyon, S.J., Bromley, B.C., O'Brien, D.P.,
Davis, D.R., 2008. Formation and collisional evolution
of Kuiper Belt objects. In: Barucci, M.A., Boehnhardt, H.,
Cruikshank, D.P., Morbidelli, A. (Eds.),
The Solar System
Beyond Neptune. Univ. of Arizona Press, Tucson,
pp. 293--313.} 

\jrnl{Kres\'ak, L., Carusi, A., Perozzi, E., Valsecchi, G. 1984.
Periodic comets Neujmin 3 and van Biesbroeck.
Intl. Astron. Union Circ. 3940.}

\jrnl{Lagerros, J.S.V., 1996. Thermal physics of asteroids. I.
Effects of shape, heat conduction and beaming.
Astron. \& Astrophys. 310, 1011--1020.}

\jrnl{Lamy, P.L., et al., 1996. Observations of comet P/Faye
1991 XXI with the Planetary Camera of the
Hubble Space Telescope.
Icarus 119, 370--384.}

\jrnl{Lamy, P.L., Toth, I., Jorda, L.,
Weaver, H.A., A'Hearn, M., 1998a.
The nucleus and inner coma of comet 46P/Wirtanen.
Astron. \& Astrophys. 335, L25--L29.}

\jrnl{Lamy, P.L., Toth, I., Weaver, H. A., 1998b.
Hubble Space Telescope observations of the nucleus and inner coma
of comet 19P/1904 Y2 (Borrelly).
Astron. \& Astrophys. 337, 945--954.}

\jrnl{Lamy, P.L., Toth, I.,
Fern\'andez, Y.R., Weaver, H.A., 2004. The sizes, shapes, albedos,
and colors of cometary nuclei. In: Festou, M.C., 
Keller, H.U., Weaver, H.A. (Eds.), Comets II. 
Univ. of Arizona Press,
Tucson, pp 223--264.}

\jrnl{Lamy, P.L., et al., 2008. Spitzer Space Telescope observations of
the nucleus of comet 67P/Churyumov-Gerasimenko. 
Astron. \& Astrophys. 489, 777--785.}

\jrnl{Lamy, P.L., Toth, I., Weaver, H.A.,
A'Hearn, M.F., Jorda, L., 2009.
Properties of the nuclei and comae of 13 ecliptic comets from Hubble Space
Telescope snapshot observations.
Astron. \& Astrophys. 508, 1045--1056.}

\jrnl{Lebofsky, L.A., et al., 1986.
A refined `standard' thermal model for asteroids
based on observations of 1 Ceres and 2 Pallas.
Icarus 68, 239--251.}

\jrnl{Lebofsky, L.A., Spencer, J.R., 1989.
Radiometry and a thermal modeling of asteroids.
In: Binzel, R.P., Gehrels, T., Shapley Matthews, M. (Eds.),
Asteroids II. U. Ariz. Press, Tucson, pp. 128--147.}

\jrnl{Levison, H.F., 1996.
Comet taxonomy. In: Rettig, T., Hahn, J.M. (Eds.),
Completing the Inventory of
the Solar System. Astronomical Society of the Pacific,
San Francisco, pp. 173--191.}

\jrnl{Licandro, J., et al., 2009.
Spitzer observations of the asteroid-comet transition
object and potential spacecraft target 107P (4015)
Wilson-Harrington.
Astron. \& Astrophys. 507, 1667--1670.}

\jrnl{Lisse, C.M., et al., 1999.
The nucleus of comet Hyakutake (C/1996 B2).
Icarus 140, 189--204.}

\jrnl{Lisse, C.M., et al., 2005. Rotationally
resolved 8-35 micron Spitzer Space Telescope observations
of the nucleus of comet 9P/Tempel 1.
Astrophys. J. 625, L139--L142.}

\jrnl{Lisse, C.M., et al., 2009. Spitzer Space
Telescope observations of the nucleus of comet 103P/Hartley 2.
Publ. Astron. Soc. Pacific 121, 968--975.}

\jrnl{Lowry, S.C., Fitzsimmons, A., Cartwright, I.M.,
Williams, I.P., 1999.
CCD photometry of distant comets.
Astron. \& Astrophys. 349, 649--659.}

\jrnl{Lowry, S.C., Fitzsimmons, A., Collander-Brown, S., 2003.
CCD photometry of distant comets. III. Ensemble properties
of Jupiter-family comets. Astron. \& Astrophys. 397, 329--343.}

\jrnl{Lowry, S.C., Fitzsimmons, A., Lamy, P., Weissman, P., 2008.
Kuiper Belt Objects in the planetary region: The Jupiter-family
comets. In: Barucci, M.A., Boehnhardt, H.,
Cruikshank, D.P., Morbidelli, A. (Eds.),
The Solar System Beyond Neptune.
Univ. of Arizona Press, Tucson, pp. 397--410.}

\jrnl{Luu, J.X., Jewitt, D.C., 1992.
Near-aphelion CCD photometry of comet P/Schwassmann-Wachmann 2.
Astron. J. 104, 2243--2249.}

\jrnl{Mainzer, A., et al., 2011. 
Thermal model calibration for minor planets observed
with WISE/NEOWISE.
Astrophys. J. 736, 100.}

\jrnl{Mazzotta Epifani, E., et al., 2008.
The distant activity of short period comets II. 
Mon. Not. R. Astron. Soc. 390 265--280.}

\jrnl{Meech, K.J., Hainaut, O.R., Marsden, B.G., 2004.
Comet nucleus size distribution from HST and Keck
telescopes. Icarus 170, 463--491.}

\jrnl{Meech, K.J., and Svore\v n, J., 2004.
Using cometary activity to trace the physical
and chemical evolution of cometary nuclei.
In: Festou, M.C., 
Keller, H.U., Weaver, H.A. (Eds.), 
Comets II. Univ. of Arizona Press, Tucson, pp. 317--335.}

\jrnl{Millis, R. L., A'Hearn M. F., Campins H., 1988. An
investigation of the nucleus and coma of comet P/Arend-Rigaux.
Astrophys. J., 324, 1194¿1209.}

\jrnl{O'Brien, D.P., Greenberg, R., 2003. 
Steady-state size distributions for collisional populations: 
Analytical solution with size-dependent strength. 
Icarus 164, 334--345.}

\jrnl{Petit, J.M., Kavellars, J.J., Gladman, B.,
Loredo, T., 2008.
Size distribution of multikilometer transneptunian objects.
In: Barucci, M.A.,
Boehnhardt, H., Cruikshank, D.P., 
Morbidelli, A. (Eds.), 
The Solar System Beyond Neptune. Univ. of Arizona
Press, Tucson, pp. 71--87.}

\jrnl{Press, W.H., Teukolsky, S.A., Vetterling, W.T.,
Flannery, B.P., 1992.
Numerical recipes in FORTRAN: The art of scientific computing,
2nd Ed. Cambridge Univ. Press, Cambridge.}

\jrnl{Reach, W.T., Kelley, M.S., Sykes, M.V., 2007.
A survey of debris trails from short-period comets.
Icarus 191, 298--322.}

\jrnl{Richardson, J.E., Melosh, H.J., Lisse, C.M.,
Carcich, B., 2007.
A ballistics analysis of the Deep Impact ejecta plume:
Determining comet Tempel 1's gravity, mass, and density.
Icarus 190, 357--390.}

\jrnl{Rieke, G.H., et al., 2004.
The Multiband Imaging Photometry for Spitzer (MIPS).
Astrophys. J. Suppl. Ser. 154, 25--29.}

\jrnl{Schlichting, H.E., et al., 2012.
Measuring the abundance of sub-kilometer-sized
Kuiper Belt objects using stellar occultations.
Astrophys. J. 761, 150.}

\jrnl{Shankman, C., Gladman, B. J., Kaib, N., 
Kavelaars, J. J., Petit, J. M., 2013.
A possible divot in the size distribution of the Kuiper 
Belt's scattering objects.
Astrophys. J. 764, L2.}

\jrnl{Snodgrass, C., Fitzsimmons, A., Lowry, S.C.,
Weissman, P.R., 2011. The size distribution of Jupiter family
comet nuclei. Mon. Not. R. Astron. Soc. 414, 458--469.}

\jrnl{Snodgrass, C., Lowry, S.C., Fitzsimmons, A.,
2006. Photometry of cometary nuclei: Rotation rates,
colours, and a comparison with Kuiper Belt Objects.
Mon. Not. R. Astron. Soc. 373, 1590--1602.}

\jrnl{Snodgrass, C., Lowry, S.C., Fitzsimmons, A.,
2008. Optical observations of 23 distant Jupiter family comets,
including 36P/Whipple at multiple phase angles.
Mon. Not. R. Astron. Soc. 385, 737--756.}

\jrnl{Soderblom L.A., et al., 2002. Observations
of comet 19P/Borrelly by the Miniature Integrated Camera and
Spectrometer aboard Deep Space 1. Science 296, 1087--1091.}

\jrnl{Sunshine, J.M., et al., 2006. Exposed water ice
deposits on the surface of comet 9P/Tempel 1. Science 311, 1453--1455.}

\jrnl{Stansberry, J.A., et al., 2004.
Spitzer observations of the dust coma and
nucleus of 29P/ Schwass\-mann-Wach\-mann 1.
Astrophys. J. Suppl. Ser. 154, 463--468.}

\jrnl{Tancredi, G., Fern\'andez, J.A., Rickman, H.,
Licandro, J., 2006. Nuclear magnitudes and the size
distribution of Jupiter family comets.
Icarus 182, 527--549.}

\jrnl{Volk, K., Malhotra, R., 2008. The scattered disk
as the source of the Jupiter Family comets.
Astrophys. J. 687, 714--725.}

\jrnl{Weiler, M., Rauer, H., 
Sterken, C., 2011. 
Cometary nuclear magnitudes from sky survey observations.
Icarus 212, 351--366.}

\jrnl{Weissman, P.R., Lowry, S.C., 2003.
The size distribution of Jupiter-family cometary
nuclei. Lunar Planet. Sci. XXXIV, \#2003 (abstract).}

\jrnl{Werner, M., et al., 2004.
The Spitzer Space Telescope Mission.
Astrophys. J. Suppl. Ser. 154, 1--9.}

\jrnl{Whitman, K., Morbidelli, A., Jedicke, R., 2006.
The size frequency distribution of dormant Jupiter family comets.
Icarus 183, 101--114.}

\vfill
\eject



\vbox{%
      \halign{\hfil #P\negthinspace&#\hfil \tabskip = 0.5 em &
              # \hfil \tabskip = 0.3 em &
              \hfil #  &
              \hfil #  &
              \hfil #  &
              \hfil #  &
              \hfil #  \cr
\multispan8{\bf Table 1 \hfil} \cr
\multispan8{\vtop{\hsize=5.8in{\noindent
Observations and Target Geometry. 
Table columns are:
``Comet Desig." = comet's designation, giving permanent
number, temporary designation (if the comet
was known by that at the time of observation),
or both (if a single-apparition comet has been recovered 
since the observations); 
``I.\&C." = instrument
and Spitzer campaign number;
``UT at start" = UT in YYYY-MM-DD HH:MM:SS.S
format at the start of the AOR; ``Dur." = duration of AOR in minutes;
``$r$ $\pm$" = heliocentric distance in AU with +(--)
signifying post(pre)-perihelion; ``$\Delta$" = Spitzer-centric
distance in AU; and ``$\alpha$" = Spitzer-centric phase
angle in degrees. \hfil}}} \cr
\noalign{\vskip5pt \hrule\hrule \vskip5pt}
\multispan2{\hfil Comet Desig. \hfil} & 
	I.\&C. & \omit\hfil UT at start \hfil & \omit\hfil Dur. \hfil & 
	$r$ $\pm$ & $\Delta$ & $\alpha$ \cr 
\noalign{\vskip5pt \hrule \vskip5pt}
  6&             & IRS  38 &  2007-02-13 22:24:59.9 &  31.82 & 4.386-- & 3.849 & 11.9 \cr 
  7&             & IRS  40 &  2007-04-19 14:39:10.3 &  11.63 & 4.340-- & 4.032 & 13.1 \cr 
 11&             & MIPS 34 &  2006-09-01 14:44:54.3 &  41.99 & 4.400-- & 4.042 & 13.1 \cr 
 11&             & MIPS 34 &  2006-09-02 01:01:44.6 &  41.99 & 4.398-- & 4.048 & 13.1 \cr 
 14&             & IRS  39 &  2007-03-20 17:49:43.7 &  13.47 & 4.515-- & 4.527 & 12.7 \cr 
 15&             & MIPS 39 &  2007-02-28 02:54:36.8 &  36.86 & 4.397-- & 4.315 & 13.1 \cr 
 15&             & MIPS 39 &  2007-02-28 07:59:25.9 &  36.86 & 4.396-- & 4.311 & 13.1 \cr 
 16&             & MIPS 40 &  2007-04-10 08:57:13.0 &  49.34 & 3.398-- & 3.367 & 17.0 \cr 
 16&             & MIPS 40 &  2007-04-10 13:23:37.5 &  49.34 & 3.397-- & 3.363 & 17.0 \cr 
 22&             & IRS  40 &  2007-04-19 13:32:00.2 &  63.00 & 4.866-- & 4.381 & 10.9 \cr 
 31&             & IRS  36 &  2006-11-09 10:29:14.1 &  11.63 & 5.016-- & 4.594 & 11.1 \cr 
 32&             & IRS  33 &  2006-08-01 03:05:00.7 &  90.50 & 4.141+  & 3.763 & 13.8 \cr 
 33&             & IRS  36 &  2006-11-09 16:53:26.9 &  46.48 & 4.395-- & 4.172 & 13.4 \cr 
 37&             & IRS  39 &  2007-03-11 23:11:26.8 & 123.52 & 4.313+  & 4.193 & 13.4 \cr 
 43&             & IRS  40 &  2007-05-02 20:10:34.4 &  74.00 & 5.353+  & 4.984 & 10.4 \cr 
 47&             & IRS  39 &  2007-03-21 10:55:23.7 &   9.80 & 4.319-- & 4.317 & 13.3 \cr 
 48&             & IRS  36 &  2006-11-11 18:40:39.7 &   9.80 & 4.412+  & 4.323 & 13.4 \cr 
 50&             & IRS  35 &  2006-10-23 17:28:50.4 & 110.68 & 3.478-- & 3.307 & 17.1 \cr 
 51&-A           & MIPS 40 &  2007-04-06 18:07:47.5 &  45.67 & 3.768-- & 3.353 & 14.7 \cr 
 51&-A           & MIPS 40 &  2007-04-06 22:46:33.7 &  45.67 & 3.767-- & 3.349 & 14.7 \cr 
 54&             & IRS  39 &  2007-03-11 02:07:46.0 & 107.02 & 5.130-- & 4.684 & 10.5 \cr 
 56&             & IRS  37 &  2006-12-23 13:43:59.8 & 130.93 & 5.075+  & 5.034 & 11.5 \cr 
 57&-A           & IRS  41 &  2007-06-15 08:32:14.3 & 127.18 & 4.110-- & 3.919 & 14.2 \cr 
 62&             & MIPS 35 &  2006-10-02 12:10:15.2 &  53.37 & 4.715+  & 4.646 & 12.5 \cr 
 62&             & MIPS 35 &  2006-10-03 01:31:59.0 &  53.37 & 4.717+  & 4.656 & 12.5 \cr 
 68&             & IRS  38 &  2007-02-12 09:40:34.4 &  28.15 & 5.469-- & 5.277 & 10.5 \cr 
 69&             & IRS  33 &  2006-07-29 04:01:52.9 & 134.60 & 4.251+  & 3.663 & 12.2 \cr 
 74&             & IRS  38 &  2007-01-28 16:56:47.0 &  13.47 & 4.420-- & 4.009 & 12.5 \cr 
 77&             & IRS  38 &  2007-02-10 07:06:19.6 &  61.17 & 4.589-- & 4.178 & 12.0 \cr 
 78&             & IRS  39 &  2007-03-10 01:56:45.4 &  40.99 & 4.983+  & 4.456 & 10.3 \cr 
 79&             & IRS  34 &  2006-09-17 23:07:46.3 &  44.65 & 4.374-- & 4.082 & 13.4 \cr 
\noalign{\vskip 5pt\hrule}
}}

\vbox{%
      \halign{\hfil #P\negthinspace&#\hfil \tabskip = 0.5 em &
              # \hfil \tabskip = 0.3 em &
              \hfil #  &
              \hfil #  &
              \hfil #  &
              \hfil #  &
              \hfil #  \cr
\multispan8{Table 1 (cont'd) \hfil} \cr
\noalign{\vskip5pt \hrule\hrule \vskip5pt}
\multispan2{\hfil Comet Desig. \hfil} & 
	I.\&C. & \omit\hfil UT at start \hfil & \omit\hfil Dur. \hfil & 
	$r$ $\pm$ & $\Delta$ & $\alpha$ \cr 
\noalign{\vskip5pt \hrule \vskip5pt}        
 89&             & IRS  40 &  2007-05-01 09:13:59.3 &  81.34 & 4.758-- & 4.176 & 10.5 \cr 
 93&             & MIPS 35 &  2006-10-04 00:11:29.4 &  20.34 & 4.005-- & 3.559 & 14.0 \cr 
 93&             & MIPS 35 &  2006-10-04 02:03:46.7 &  20.34 & 4.005-- & 3.560 & 14.0 \cr 
 94&             & IRS  36 &  2006-11-14 19:05:48.3 &  77.67 & 4.789+  & 4.199 & 10.6 \cr 
101&             & IRS  40 &  2007-04-22 13:19:07.9 &  37.32 & 4.339+  & 3.760 & 11.6 \cr 
107&             & IRS  38 &  2007-02-12 07:36:55.2 &  17.14 & 4.078+  & 3.519 & 12.7 \cr 
113&             & IRS  35 &  2006-10-22 14:35:11.4 & 108.85 & 3.855-- & 3.547 & 15.2 \cr 
118&             & IRS  34 &  2006-09-17 11:37:10.0 &  18.97 & 4.946+  & 4.578 & 11.5 \cr 
119&             & IRS  38 &  2007-02-12 07:56:34.9 &  13.47 & 4.212+  & 3.672 & 12.4 \cr 
120&             & IRS  38 &  2007-02-12 06:05:17.9 &  88.67 & 4.798+  & 4.246 & 10.7 \cr 
121&             & IRS  33 &  2006-08-01 02:33:15.2 &  28.15 & 4.351+  & 3.966 & 13.1 \cr 
123&             & IRS  35 &  2006-10-21 21:32:12.6 & 136.43 & 5.344+  & 4.905 & 10.4 \cr 
124&             & IRS  34 &  2006-09-16 18:45:13.8 & 121.68 & 4.233-- & 3.762 & 13.1 \cr 
127&             & IRS  39 &  2007-03-08 16:24:16.4 &  42.82 & 4.588-- & 4.429 & 12.6 \cr 
129&             & IRS  40 &  2007-04-19 00:37:44.8 &  11.63 & 4.013+  & 3.671 & 14.1 \cr 
130&             & IRS  36 &  2006-11-14 10:12:40.0 &  50.15 & 4.452+  & 4.198 & 13.1 \cr 
131&             & IRS  38 &  2007-02-12 08:51:35.1 &  44.65 & 4.417+  & 3.964 & 12.3 \cr 
132&             & IRS  41 &  2007-06-07 18:44:05.3 & 129.02 & 3.987+  & 3.797 & 14.7 \cr 
137&             & IRS  39 &  2007-03-16 13:09:35.1 &  26.30 & 5.474-- & 5.450 & 10.5 \cr 
138&             & MIPS 38 &  2007-01-22 08:46:03.1 &  36.86 & 4.207+  & 4.214 & 13.7 \cr 
138&             & MIPS 38 &  2007-01-22 19:38:38.0 &  36.86 & 4.209+  & 4.209 & 13.7 \cr 
139&             & MIPS 36 &  2006-11-02 07:16:15.2 &   7.51 & 4.097-- & 3.572 & 13.2 \cr 
139&             & MIPS 36 &  2006-11-02 11:00:52.0 &   7.51 & 4.097-- & 3.574 & 13.2 \cr 
141&             & IRS  39 &  2007-03-25 03:18:48.5 & 136.45 & 5.112+  & 4.539 &  9.8 \cr 
143&             & IRS  39 &  2007-03-09 16:14:09.4 &   9.80 & 4.988-- & 4.740 & 11.4 \cr 
144&             & MIPS 42 &  2007-07-09 21:58:01.2 &  49.34 & 4.606-- & 4.162 & 12.0 \cr 
144&             & MIPS 42 &  2007-07-10 01:32:37.3 &  49.34 & 4.605-- & 4.158 & 12.0 \cr 
146&             & IRS  33 &  2006-08-06 18:12:08.5 & 132.68 & 5.084-- & 4.856 & 11.5 \cr 
148&             & MIPS 36 &  2006-11-02 10:10:13.2 &  36.86 & 4.307-- & 3.890 & 13.1 \cr 
148&             & MIPS 36 &  2006-11-02 15:24:37.7 &  36.86 & 4.306-- & 3.892 & 13.1 \cr 
149&             & IRS  38 &  2007-02-09 02:19:40.4 & 118.02 & 5.552-- & 5.460 & 10.4 \cr 
152&             & IRS  34 &  2006-09-17 18:38:21.7 & 129.12 & 5.698+  & 5.580 & 10.4 \cr 
159&             & IRS  38 &  2007-02-13 08:41:54.7 & 114.35 & 6.121+  & 5.737 &  9.0 \cr 
160&             & IRS  37 &  2006-12-19 16:03:31.7 & 132.77 & 4.990+  & 4.828 & 11.7 \cr 
162&             & IRS  39 &  2007-03-17 21:48:23.8 & 134.62 & 4.820+  & 4.269 & 10.6 \cr 
163&             & IRS  33 &  2006-08-06 16:36:16.4 &  92.33 & 4.089+  & 3.947 & 14.5 \cr 
168&             & MIPS 41 &  2007-05-23 02:01:22.8 &  41.99 & 4.486+  & 4.090 & 12.4 \cr 
168&             & MIPS 41 &  2007-05-23 14:05:18.3 &  41.99 & 4.488+  & 4.099 & 12.4 \cr 
\noalign{\vskip 5pt\hrule}
}}

\vbox{%
      \halign{\hfil #P\negthinspace&#\hfil \tabskip = 0.5 em &
              # \hfil \tabskip = 0.3 em &
              \hfil #  &
              \hfil #  &
              \hfil #  &
              \hfil #  &
              \hfil #  \cr
\multispan8{Table 1 (cont'd) \hfil} \cr
\noalign{\vskip5pt \hrule\hrule \vskip5pt}
\multispan2{\hfil Comet Desig. \hfil} & 
	I.\&C. & \omit\hfil UT at start \hfil & \omit\hfil Dur. \hfil & 
	$r$ $\pm$ & $\Delta$ & $\alpha$ \cr 
\noalign{\vskip5pt \hrule \vskip5pt}         
169&             & MIPS 39 &  2007-03-01 14:03:41.1 &  36.86 & 4.293+  & 4.013 & 13.3 \cr 
169&             & MIPS 39 &  2007-03-01 18:31:54.1 &  36.86 & 4.294+  & 4.011 & 13.3 \cr 
171&             & IRS  39 &  2007-03-16 10:45:15.5 & 140.12 & 4.158+  & 4.087 & 13.9 \cr 
172&             & IRS  35 &  2006-10-18 20:31:32.1 &  28.13 & 4.250-- & 4.055 & 13.9 \cr 
173&             & IRS  35 &  2006-10-24 09:11:36.4 &  31.82 & 4.817-- & 4.317 & 11.2 \cr 
197& /2003 KV$_2$& MIPS 36 &  2006-11-04 17:12:43.1 &  22.91 & 4.208-- & 4.019 & 14.0 \cr 
197& /2003 KV$_2$& MIPS 36 &  2006-11-05 00:07:12.3 &  22.91 & 4.207-- & 4.014 & 14.0 \cr 
203& /1999 WJ$_7$& MIPS 35 &  2006-10-05 15:10:54.9 &  12.65 & 5.646-- & 5.097 &  9.2 \cr 
203& /1999 WJ$_7$& MIPS 35 &  2006-10-06 01:44:01.5 &  12.65 & 5.645-- & 5.103 &  9.3 \cr 
213& /2005 R2    & IRS  36 &  2006-11-13 12:48:10.4 &  24.47 & 4.051+  & 3.706 & 14.2 \cr 
215& /2002 O8    & MIPS 38 &  2007-01-08 17:44:29.1 &   7.51 & 4.765-- & 4.235 & 11.0 \cr 
215& /2002 O8    & MIPS 38 &  2007-01-09 03:53:22.0 &   7.51 & 4.765-- & 4.229 & 10.9 \cr 
216& /2001 CV$_8$& MIPS 38 &  2007-01-01 12:48:02.7 &  12.65 & 4.411-- & 3.948 & 12.4 \cr 
216& /2001 CV$_8$& MIPS 38 &  2007-01-01 15:21:43.6 &  12.65 & 4.411-- & 3.949 & 12.4 \cr 
219& /2002 LZ$_{11}$ & 
		   MIPS 41 &  2007-06-07 04:33:40.8 &  17.78 & 4.781-- & 4.223 & 10.8 \cr 
219& /2002 LZ$_{11}$ & 
		   MIPS 41 &  2007-06-07 07:06:24.3 &  17.78 & 4.781-- & 4.224 & 10.8 \cr 
221& /2002 JN$_{16}$ & 
		   MIPS 40 &  2007-04-09 23:45:17.8 &  36.86 & 4.404-- & 3.829 & 11.4 \cr 
221& /2002 JN$_{16}$ & 
		   MIPS 40 &  2007-04-10 02:18:57.8 &  36.86 & 4.403-- & 3.830 & 11.5 \cr 
223& /2002 S1    & MIPS 40 &  2007-04-05 23:50:51.1 &  36.86 & 5.695-- & 5.144 &  8.8 \cr 
223& /2002 S1    & MIPS 40 &  2007-04-06 08:15:30.1 &  36.86 & 5.694-- & 5.139 &  8.8 \cr 
228& /2001 YX$_{127}$ & 
		   IRS  35 &  2006-10-22 11:13:26.9 &  31.82 & 4.844+  & 4.439 & 11.6 \cr 
240& /2002 X2    & IRS  40 &  2007-04-21 22:48:14.9 &  88.67 & 5.540-- & 5.265 & 10.2 \cr 
243& /2003 S2    & IRS  38 &  2007-02-13 23:24:41.7 & 132.77 & 5.178+  & 4.846 & 10.9 \cr 
244& /2000 Y3    & MIPS 35 &  2006-10-03 02:28:12.6 &  26.59 & 5.976-- & 5.895 &  9.9 \cr 
244& /2000 Y3    & MIPS 35 &  2006-10-03 10:30:13.7 &  26.59 & 5.975-- & 5.900 &  9.9 \cr 
246& /2004 F3    & IRS  37 &  2006-12-17 23:18:30.2 &  11.63 & 4.280+  & 3.707 & 12.1 \cr 
256& /2003 HT$_{15}$ & 
		   MIPS 37 &  2006-12-06 12:30:12.3 &  31.72 & 6.212+  & 5.660 &  8.2 \cr 
256& /2003 HT$_{15}$ & 
		   MIPS 37 &  2006-12-06 19:57:46.7 &  31.72 & 6.213+  & 5.664 &  8.3 \cr 
260& /2005 K3    & MIPS 38 &  2007-01-01 03:29:24.1 &  31.72 & 4.202+  & 3.946 & 13.8 \cr 
260& /2005 K3    & MIPS 38 &  2007-01-01 07:24:09.7 &  31.72 & 4.203+  & 3.944 & 13.8 \cr 
   & /1998 VS$_{24}$ & 
		   MIPS 36 &  2006-11-02 07:27:01.2 &   7.51 & 4.173-- & 3.781 & 13.6 \cr 
   & /1998 VS$_{24}$ & 
		   MIPS 36 &  2006-11-02 10:50:05.4 &   7.51 & 4.173-- & 3.783 & 13.6 \cr 
   & /2001 R6    & MIPS 40 &  2007-04-05 10:50:32.6 &  45.67 & 5.743-- & 5.178 &  8.7 \cr 
   & /2001 R6    & MIPS 40 &  2007-04-05 13:56:06.1 &  45.67 & 5.743-- & 5.176 &  8.7 \cr 
   & /2003 O3    & MIPS 42 &  2007-07-09 11:40:04.5 &  36.86 & 4.282-- & 4.100 & 13.7 \cr 
   & /2003 O3    & MIPS 42 &  2007-07-09 17:09:50.8 &  36.86 & 4.281-- & 4.103 & 13.7 \cr 
   & /2003 S1    & IRS  39 &  2007-03-08 10:42:49.4 &  48.32 & 5.772+  & 5.371 &  9.4 \cr 
   & /2004 A1$^a$ & 
		   MIPS 39 &  2007-03-04 20:37:18.6 &  36.86 & 6.517+  & 6.418 &  8.8 \cr 
   & /2004 A1$^a$ & 
		   MIPS 39 &  2007-03-05 01:00:51.0 &  36.86 & 6.517+  & 6.418 &  8.8 \cr 
\noalign{\vskip 5pt\hrule}
}}

\vbox{%
      \halign{\hfil #P\negthinspace&#\hfil \tabskip = 0.5 em &
              # \hfil \tabskip = 0.3 em &
              \hfil #  &
              \hfil #  &
              \hfil #  &
              \hfil #  &
              \hfil #  \cr
\multispan8{Table 1 (cont'd) \hfil} \cr
\noalign{\vskip5pt \hrule\hrule \vskip5pt}
\multispan2{\hfil Comet Desig. \hfil} & 
	I.\&C. & \omit\hfil UT at start \hfil & \omit\hfil Dur. \hfil & 
	$r$ $\pm$ & $\Delta$ & $\alpha$ \cr 
\noalign{\vskip5pt \hrule \vskip5pt} 
   & /2004 DO$_{29}$ & 
		   IRS  34 &  2006-09-18 03:38:42.4 &  99.68 & 5.550+  & 5.207 & 10.3 \cr 
   & /2004 H2    & MIPS 38 &  2007-01-01 20:34:16.9 &  12.65 & 5.457+  & 5.070 & 10.2 \cr 
   & /2004 H2    & MIPS 38 &  2007-01-02 00:06:03.0 &  12.65 & 5.457+  & 5.072 & 10.2 \cr 
   & /2004 T1    & MIPS 39 &  2007-02-28 04:21:38.2 &  31.72 & 4.914+  & 4.419 & 10.7 \cr 
   & /2004 T1    & MIPS 39 &  2007-02-28 09:33:17.3 &  31.72 & 4.915+  & 4.417 & 10.7 \cr 
   & /2004 V3    & IRS  34 &  2006-09-18 05:29:58.8 &  83.17 & 5.358+  & 4.895 & 10.3 \cr 
   & /2004 V5-A  & IRS  39 &  2007-03-10 00:42:43.1 &  48.32 & 5.811+  & 5.565 &  9.8 \cr 
   & /2004 VR$_8$& IRS  33 &  2006-08-07 18:02:45.4 &  66.67 & 3.443+  & 3.350 & 17.2 \cr 
   & /2005 GF$_8$& IRS  36 &  2006-11-12 23:26:43.4 &  11.63 & 4.164+  & 3.793 & 13.7 \cr 
   & /2005 JD$_{108}$ & 
		   MIPS 38 &  2007-01-01 20:08:55.7 &   7.51 & 4.772+  & 4.318 & 11.4 \cr 
   & /2005 JD$_{108}$ & 
		   MIPS 38 &  2007-01-01 21:19:22.2 &   7.51 & 4.772+  & 4.319 & 11.4 \cr 
   & /2005 JQ$_5$& IRS  36 &  2006-11-21 13:10:28.5 & 108.85 & 4.022+  & 3.996 & 14.6 \cr 
   & /2005 L4    & IRS  39 &  2007-03-11 22:46:06.3 &  22.63 & 4.145+  & 4.075 & 13.9 \cr 
   & /2005 Q4    & MIPS 39 &  2007-02-28 03:36:18.6 &  41.99 & 4.409+  & 3.933 & 12.1 \cr 
   & /2005 Q4    & MIPS 39 &  2007-02-28 08:47:57.7 &  41.99 & 4.411+  & 3.931 & 12.1 \cr 
   & /2005 R1    & IRS  37 &  2006-12-19 12:35:13.0 &  29.98 & 4.116+  & 3.719 & 13.7 \cr 
   & /2005 S3    & MIPS 41 &  2007-05-23 02:47:40.2 &   7.51 & 4.117+  & 3.876 & 14.0 \cr 
   & /2005 S3    & MIPS 41 &  2007-05-23 14:51:35.7 &   7.51 & 4.119+  & 3.885 & 14.0 \cr 
   & /2005 T5    & MIPS 35 &  2006-10-09 13:43:37.7 &   7.51 & 3.995+  & 3.705 & 14.7 \cr 
   & /2005 T5    & MIPS 35 &  2006-10-09 15:53:20.8 &   7.51 & 3.995+  & 3.704 & 14.7 \cr 
\omit\hfil C& /2005 W2$^b$ & 
		   MIPS 38 &  2007-01-01 02:59:55.5 &   7.51 & 4.071+  & 3.764 & 14.1 \cr 
\omit\hfil C& /2005 W2$^b$ & 
		   MIPS 38 &  2007-01-01 04:26:05.0 &   7.51 & 4.071+  & 3.764 & 14.1 \cr 
   & /2005 W3    & MIPS 37 &  2006-12-08 02:37:58.4 &   7.51 & 4.359+  & 4.349 & 13.4 \cr 
   & /2005 W3    & MIPS 37 &  2006-12-08 04:33:05.3 &   7.51 & 4.360+  & 4.348 & 13.4 \cr 
   & /2005 XA$_{54}$ & 
		   MIPS 41 &  2007-06-04 15:57:41.9 &  41.99 & 4.423+  & 3.858 & 11.7 \cr 
   & /2005 XA$_{54}$ & 
		   MIPS 41 &  2007-06-04 23:09:21.8 &  41.99 & 4.425+  & 3.856 & 11.6 \cr 
   & /2005 Y2    & MIPS 38 &  2007-01-19 22:46:26.2 &  10.08 & 5.410+  & 5.060 & 10.4 \cr 
   & /2005 Y2    & MIPS 38 &  2007-01-20 05:00:15.2 &  10.08 & 5.411+  & 5.065 & 10.4 \cr 
\noalign{\vskip 5pt\hrule\vskip 5pt}
\multispan8{\vtop{\hsize=5.6in{$^a$ P/2004 A1 is also
a Centaur but has $2<T_J<3$. \hfil}}}\cr
\multispan8{\vtop{\hsize=5.6in{$^b$ C/2005 W2
is possibly a Halley-family comet but has $2<T_J<3$.
\hfil}}} \cr
}}

\vfill
\eject

\vbox{%
      \halign{\hfil #P\negthinspace&#\hfil \tabskip = 0.3 em &
               # \hfil \tabskip = 0.3 em &
              \hfil # \tabskip = 2.0 em &
              \hfil #P\negthinspace\tabskip = 0 em&#\hfil \tabskip = 0.3 em &
              \hfil #  &
              \hfil #  \cr
\multispan8{\bf Table 2 \hfil} \cr
\multispan8{\vtop{\hsize=6.3in{\noindent Nucleus contribution to comet flux
within a radius 3-pixel circular aperture.
Comets observed with IRS are listed first, followed by comets observed with
MIPS.  Table columns are:
``Comet Desig." = comet's designation, as in Table 1;
``$f_A$" = fraction of comet's flux from the nucleus
at blue PU wavelength (for IRS observations)
or at first epoch (for MIPS observations);
``$f_B$" = fraction at red PU wavelength (for IRS observations)
or at second epoch (for MIPS observations).
The 35 comets that showed discernible dust emission
are indicated with underscored \unb{ratios}
(see \S3.1). \hfil}}} \cr 
\noalign{\vskip5pt \hrule\hrule \vskip5pt}
\multispan2{\hfil Comet Desig. \hfil} & \omit{\hfil$f_{A}$\hfil} & 
                                        \omit{\hfil$f_{B}$\hfil} &
\multispan2{\hfil Comet Desig. \hfil} & \omit{\hfil$f_{A}$\hfil} & 
                                        \omit{\hfil$f_{B}$\hfil} \cr
\noalign{\vskip5pt \hrule \vskip10pt}
\multispan8{{\sl IRS}\hfil } \cr
  6&                  &\unb{$0.99\pm0.01$} & \unb{$0.98\pm0.01$} &
127&                  &    {$1.14\pm0.10$} &     {$1.11\pm0.10$} \cr
  7&                  &    {$1.01\pm0.02$} &     {$1.00\pm0.02$} &
129&                  &\unb{$0.35\pm0.04$} & \unb{$0.26\pm0.03$} \cr
 14&                  &    {$1.01\pm0.02$} &     {$1.00\pm0.01$} &
130&                  &    {$0.99\pm0.01$} &     {$0.95\pm0.01$} \cr
 22&                  &\unb{$0.71\pm0.03$} & \unb{$0.63\pm0.05$} &
131&                  &    {$1.11\pm0.04$} &     {$1.01\pm0.04$} \cr
 31&                  &    {$1.07\pm0.09$} &     {$1.11\pm0.09$} &
132&                  &    {$0.94\pm0.03$} &     {$1.00\pm0.04$} \cr
 32&                  &\unb{$0.18\pm0.03$} & \unb{$0.25\pm0.02$} &
137&                  &    {$1.03\pm0.02$} &     {$0.99\pm0.01$} \cr
 33&                  &    {$0.95\pm0.05$} &     {$1.06\pm0.04$} &
143&                  &    {$1.00\pm0.01$} &     {$1.01\pm0.01$} \cr
 37&                  &\unb{$0.61\pm0.10$} & \unb{$0.48\pm0.04$} &
146&                  &    {$1.08\pm0.11$} &     {$1.00\pm0.08$} \cr
 47&                  &    {$1.00\pm0.01$} &     {$0.98\pm0.01$} &
149&                  &    {$0.91\pm0.04$} &     {$0.99\pm0.04$} \cr
 48&                  &\unb{$0.78\pm0.08$} & \unb{$0.89\pm0.08$} &
152&                  &\unb{$0.33\pm0.07$} & \unb{$0.34\pm0.11$} \cr
 50&                  &\unb{$0.97\pm0.01$} & \unb{$0.91\pm0.02$} &
159&                  &\unb{$0.35\pm0.12$} & \unb{$0.55\pm0.10$} \cr
 56&                  &\unb{$0.90\pm0.14$} & \unb{$0.99\pm0.03$} &
160&                  &    {$1.16\pm0.09$} &     {$0.89\pm0.05$} \cr
 57&                  &    {$1.00\pm0.03$} &     {$1.02\pm0.03$} &
162&                  &    {$1.00\pm0.01$} &     {$1.00\pm0.01$} \cr
 68&                  &    {$0.99\pm0.03$} &     {$1.03\pm0.02$} &
163&                  &    {$1.01\pm0.02$} &     {$1.00\pm0.02$} \cr
 69&                  &\unb{$0.82\pm0.02$} & \unb{$0.47\pm0.06$} & 
171&                  &\unb{$0.68\pm0.04$} & \unb{$0.68\pm0.07$} \cr
 74&                  &\unb{$0.33\pm0.04$} & \unb{$0.41\pm0.03$} & 
172&                  &    {$1.01\pm0.01$} &     {$1.01\pm0.01$} \cr
 77&                  &    {$0.96\pm0.02$} &     {$0.96\pm0.01$} & 
173&                  &\unb{$0.39\pm0.08$} & \unb{$0.51\pm0.08$} \cr
 78&                  &\unb{$0.51\pm0.10$} & \unb{$0.46\pm0.09$} & 
213&       /2005 R2   &\unb{$0.10\pm0.01$} & \unb{$0.12\pm0.02$} \cr
 79&                  &    {$0.87\pm0.09$} &     {$0.94\pm0.10$} & 
228&       /2001 YX127&    {$0.77\pm0.11$} &     {$0.88\pm0.06$} \cr
 89&                  &    {$1.00\pm0.02$} &     {$0.88\pm0.03$} & 
246&       /2004 F3   &\unb{$0.27\pm0.03$} & \unb{$0.25\pm0.03$} \cr
 94&                  &    {$1.01\pm0.01$} &     {$0.97\pm0.01$} & 
   &       /2004 DO29 &    {$1.02\pm0.10$} &     {$0.99\pm0.08$} \cr
101&                  &\unb{$0.28\pm0.06$} & \unb{$0.26\pm0.05$} & 
   &       /2004 V3   &    {$1.07\pm0.09$} &     {$1.17\pm0.08$} \cr
107&                  &    {$1.04\pm0.02$} &     {$1.02\pm0.02$} & 
   &       /2004 V5-A &\unb{$0.13\pm0.05$} & \unb{$0.15\pm0.05$} \cr
113&                  &    {$1.02\pm0.01$} &     {$1.01\pm0.01$} & 
   &       /2004 VR8  &\unb{$0.83\pm0.11$} & \unb{$0.84\pm0.02$} \cr
118&                  &\unb{$0.23\pm0.06$} & \unb{$0.63\pm0.05$} & 
   &       /2005 GF8  &    {$0.96\pm0.01$} &     {$0.98\pm0.01$} \cr
119&                  &\unb{$0.26\pm0.04$} & \unb{$0.24\pm0.08$} & 
   &       /2005 JQ5  &    {$0.88\pm0.07$} &     {$0.90\pm0.05$} \cr
121&                  &\unb{$0.99\pm0.01$} & \unb{$1.02\pm0.01$} & 
   &       /2005 L4   &    {$1.02\pm0.02$} &     {$1.06\pm0.02$} \cr
123&                  &    {$0.97\pm0.01$} &     {$1.00\pm0.01$} & 
   &       /2005 R1   &    {$1.00\pm0.02$} &     {$1.00\pm0.02$} \cr
124&                  &    {$1.02\pm0.01$} &     {$0.99\pm0.01$} \cr
\noalign{\vskip 5pt\hrule\vskip 5pt}
}}

\vfill
\eject

\vbox{%
      \halign{\hfil #P\negthinspace&#\hfil \tabskip = 0.3 em &
               # \hfil \tabskip = 0.3 em &
              \hfil # \tabskip = 2.0 em &
              \hfil #P\negthinspace\tabskip = 0 em&#\hfil \tabskip = 0.3 em &
              \hfil #  &
              \hfil #  \cr
\multispan8{\bf Table 2 (cont'd) \hfil} \cr
\noalign{\vskip5pt \hrule\hrule \vskip5pt}
\multispan2{\hfil Comet Desig. \hfil} & \omit{\hfil$f_{A}$\hfil} & 
                                        \omit{\hfil$f_{B}$\hfil} &
\multispan2{\hfil Comet Desig. \hfil} & \omit{\hfil$f_{A}$\hfil} & 
                                        \omit{\hfil$f_{B}$\hfil} \cr
\noalign{\vskip5pt \hrule \vskip10pt}
\multispan8{{\sl MIPS}\hfil } \cr
 11&                  &    {$1.03\pm0.11$} &     {$1.31\pm0.13$} &
221&       /2002 JN16 &    {$0.96\pm0.03$} &     {$0.92\pm0.02$} \cr
 15&                  &    {$0.93\pm0.05$} &     {$0.94\pm0.06$} &
223&       /2002 S1   &    {$1.01\pm0.01$} &     {$1.01\pm0.01$} \cr
 16&                  &\unb{$0.61\pm0.09$} & \unb{$0.47\pm0.06$} &
256&       /2003 HT15 &    {$0.95\pm0.19$} &     {$0.76\pm0.18$} \cr
 51&                  &    {$0.79\pm0.09$} &     {$0.94\pm0.09$} &
260&       /2005 K3   &\unb{$0.98\pm0.01$} & \unb{$0.97\pm0.02$} \cr
 62&                  &\unb{$0.62\pm0.19$} & \unb{$0.33\pm0.10$} &
   &       /2001 R6   &    {$1.07\pm0.18$} &     {$1.32\pm0.18$} \cr
 93&                  &    {$1.01\pm0.01$} &     {$0.98\pm0.01$} & 
   &       /2003 O3   &    {$0.95\pm0.10$} &     {$1.07\pm0.15$} \cr
138&                  &    {$1.33\pm0.13$} &     {$1.05\pm0.09$} & 
   &       /2004 A1   &\unb{$0.23\pm0.02$} & \unb{$0.22\pm0.03$} \cr
139&                  &    {$0.89\pm0.05$} &     {$0.91\pm0.06$} & 
   &       /2004 H2   &\unb{$0.45\pm0.05$} & \unb{$0.46\pm0.15$} \cr
144&                  &\unb{$1.19\pm0.06$} & \unb{$1.09\pm0.05$} & 
   &       /2005 JD108&\unb{$0.34\pm0.04$} & \unb{$0.30\pm0.03$} \cr
148&                  &    {$1.14\pm0.02$} &     {$1.12\pm0.02$} & 
   &       /2005 Q4   &    {$1.11\pm0.01$} &     {$1.08\pm0.01$} \cr
168&                  &    {$0.76\pm0.12$} &     {$0.97\pm0.16$} &
   &       /2005 S3   &    {$0.96\pm0.05$} &     {$0.96\pm0.13$} \cr
169&                  &    {$1.00\pm0.01$} &     {$1.03\pm0.01$} &
   &       /2005 T5   &\unb{$0.76\pm0.08$} & \unb{$0.77\pm0.07$} \cr
197&       /2003 KV2  &    {$0.90\pm0.07$} &     {$0.89\pm0.06$} &
   &       /2005 W2   &\unb{$0.18\pm0.06$} & \unb{$1.24\pm0.44$} \cr
215&       /2002 O8   &    {$0.86\pm0.07$} &     {$0.83\pm0.08$} &
   &       /2005 W3   &\unb{$0.25\pm0.10$} & \unb{$0.27\pm0.11$} \cr
216&       /2001 CV8  &    {$1.85\pm0.31$} &     {$0.94\pm0.14$} &
   &       /2005 XA54 &    {$1.01\pm0.01$} &     {$0.97\pm0.01$} \cr
219&       /2001 LZ11 &\unb{$1.01\pm0.07$} & \unb{$0.94\pm0.05$} &
   &       /2005 Y2   &\unb{$0.39\pm0.04$} & \unb{$0.35\pm0.04$} \cr
\noalign{\vskip 5pt\hrule\vskip 5pt}
}}

\vfill
\eject

\vbox{%
      \halign{\hfil #P\negthinspace&#\hfil \tabskip = 0.3 em &
              # \hfil \tabskip = 0.3 em &
              \hfil # \tabskip = 2.0 em &
              \hfil #P\negthinspace\tabskip = 0 em&#\hfil \tabskip = 0.3 em &
              \hfil #  &
              \hfil #  \cr
\multispan8{\vtop{\hsize=6.5in{\noindent {\bf Table 3.}
Nucleus photometry. Entries
with ``$<$" indicate 3$\sigma$ upper limits. Comets observed
with IRS are listed first, followed by comets observed with 
MIPS.  Table columns are:
``Comet Desig." = comet's designation, as in Table 1;
``$F_A$" = flux density in mJy at blue PU wavelength (for IRS observations)
or at first epoch (for MIPS observations);
``$F_B$" = flux density in mJy at red PU wavelength (for IRS observations)
or at second epoch (for MIPS observations). 
The 35 comets that showed discernible dust emission
are indicated with underscored \unb{flux densities}
(see \S3.1). \hfil}}} \cr
\noalign{\vskip5pt \hrule\hrule \vskip5pt}
\multispan2{\hfil Comet Desig. \hfil} & $F_{A}$ (mJy) & $F_{B}$ (mJy) &
\multispan2{\hfil Comet Desig. \hfil} & $F_{A}$ (mJy) & $F_{B}$ (mJy) \cr
\noalign{\vskip5pt \hrule \vskip10pt}
\multispan8{{\sl IRS}\hfil } \cr
  6&                  & \unb{$2.22\pm0.11$} & \unb{$3.47\pm0.17$}  &
127&                  &      $ 0.25\pm0.02$  &      $ 0.35\pm0.05$  \cr
  7&                  &      $2.96\pm0.15$  &      $4.50\pm0.22$   &
129&                  & \unb{$ 0.82\pm0.08$} & \unb{$ 1.56\pm0.16$} \cr
 14&                  &      $2.55\pm0.13$  &      $4.27\pm0.21$   &
130&                  &      $ 1.64\pm0.08$  &      $ 3.13\pm0.16$  \cr
 22&                  & \unb{$1.28\pm0.10$} & \unb{$1.93\pm0.18$}  &
131&                  &      $ 0.54\pm0.04$  &      $ 0.78\pm0.06$  \cr
 31&                  &      $0.59\pm0.04$  &      $1.03\pm0.08$   &
132&                  &      $ 0.36\pm0.03$  &      $ 0.58\pm0.04$  \cr
 32&                  & \unb{$3.03\pm0.61$} & \unb{$4.64\pm0.93$}  &
137&                  &      $ 2.07\pm0.10$  &      $ 3.56\pm0.18$  \cr
 33&                  &      $0.46\pm0.02$  &      $0.86\pm0.04$   &
141&                  & \hfil ---$^a$  \hfil & \hfil ---$^a$  \hfil \cr
 37&                  & \unb{$0.57\pm0.07$} & \unb{$0.96\pm0.10$}  &
143&                  &      $ 4.73\pm0.24$  &      $ 8.34\pm0.42$  \cr
 43&                  & \hfil $< 0.27$ \hfil & \hfil $< 0.30$ \hfil  &
146&                  &      $ 0.17\pm0.02$  &      $ 0.31\pm0.03$  \cr
 47&                  &      $3.41\pm0.17$  &      $5.77\pm0.29$   &
149&                  &      $ 0.25\pm0.02$  &      $ 0.44\pm0.02$  \cr
 48&                  & \unb{$2.96\pm0.20$} & \unb{$5.02\pm0.32$}  & 
152&                  & \unb{$ 0.10\pm0.02$} & \unb{$ 0.26\pm0.08$} \cr
 50&                  & \unb{$2.07\pm0.10$} & \unb{$3.73\pm0.19$}  & 
159&                  & \unb{$ 0.12\pm0.04$} & \unb{$ 0.43\pm0.08$} \cr
 54&                  & \hfil $< 0.08$ \hfil & \hfil $< 0.12$ \hfil & 
160&                  &      $ 0.17\pm0.02$  &      $ 0.38\pm0.03$  \cr
 56&                  & \unb{$0.58\pm0.07$} & \unb{$1.22\pm0.10$}  & 
162&                  &      $13.28\pm0.66$  &      $24.89\pm1.25$  \cr
 57&-A                &      $0.46\pm0.03$  &      $0.65\pm0.09$  & 
163&                  &      $ 0.84\pm0.04$  &      $ 1.38\pm0.07$  \cr
 68&                  &      $1.02\pm0.05$  &      $1.88\pm0.09$  & 
171&                  & \unb{$ 0.71\pm0.06$} & \unb{$ 1.05\pm0.11$} \cr
 69&                  & \unb{$0.39\pm0.02$} & \unb{$0.62\pm0.08$} & 
172&                  &      $13.67\pm0.68$  &      $22.21\pm1.11$  \cr
 74&                  & \unb{$3.09\pm0.39$} & \unb{$9.98\pm0.77$} & 
173&                  & \unb{$ 3.70\pm0.74$} & \unb{$15.17\pm2.28$} \cr
 77&                  &      $0.82\pm0.06$  &      $1.63\pm0.08$  & 
213&/2005 R2          & \unb{$ 1.16\pm0.16$} & \unb{$ 3.62\pm0.62$} \cr
 78&                  & \unb{$0.41\pm0.08$} & \unb{$0.78\pm0.16$} & 
228&/2001 YX$_{127}$  &      $ 0.28\pm0.04$  &      $ 0.80\pm0.06$  \cr
 79&                  &      $0.20\pm0.02$  &      $0.30\pm0.03$  & 
240&/2002 X2          & \hfil ---$^a$ \hfil & \hfil ---$^a$ \hfil \cr
 89&                  &      $0.69\pm0.03$  &      $0.95\pm0.05$  & 
243&/2003 S2          & \hfil $< 0.05$ \hfil & \hfil $< 0.12$ \hfil \cr
 94&                  &      $1.52\pm0.12$  &      $2.62\pm0.13$  & 
246&/2004 F3          & \unb{$ 8.08\pm0.81$} & \unb{$17.86\pm2.33$} \cr
101&                  & \unb{$0.41\pm0.08$} & \unb{$0.85\pm0.16$} & 
   &/2003 S1          & \hfil $< 0.12^b$ \hfil & \hfil $< 0.22^b$ \hfil \cr
107&                  &      $1.25\pm0.06$  &      $2.20\pm0.11$  & 
   &/2004 DO$_{29}$   &      $ 0.15\pm0.01$  &      $ 0.27\pm0.02$  \cr
113&                  &      $2.00\pm0.10$  &      $3.12\pm0.16$ &  
   &/2004 V3          &      $ 0.14\pm0.01$  &      $ 0.24\pm0.02$  \cr
118&                  & \unb{$0.32\pm0.06$} & \unb{$1.03\pm0.21$} & 
   &/2004 V5-A        & \unb{$ 0.21\pm0.16$} & \unb{$ 0.57\pm0.16$} \cr
119&                  & \unb{$0.48\pm0.08$} & \unb{$1.16\pm0.39$} &
   &/2004 VR$_{8}$    & \unb{$29.48\pm2.41$} & \unb{$42.73\pm2.14$} \cr
120&                  & \hfil $< 0.06$ \hfil & \hfil $< 0.14$ \hfil &
   &/2005 GF$_{8}$    &      $ 3.49\pm0.17$  &      $ 6.46\pm0.32$  \cr
121&                  & \unb{$ 5.92\pm0.30$} & \unb{$10.85\pm0.54$} &
   &/2005 JQ$_{5}$    &      $ 0.19\pm0.01$  &      $ 0.44\pm0.02$  \cr
123&                  &      $ 0.68\pm0.03$  &      $ 1.75\pm0.09$  &
   &/2005 L4          &      $ 1.71\pm0.09$  &      $ 2.36\pm0.12$  \cr
124&                  &      $ 3.37\pm0.17$  &      $ 5.65\pm0.28$  &
   &/2005 R1          &      $ 1.50\pm0.08$  &      $ 2.35\pm0.12$  \cr
\noalign{\vskip 5pt\hrule}
}}

\vbox{%
      \halign{\hfil #P\negthinspace&#\hfil \tabskip = 0.3 em &
              # \hfil \tabskip = 0.3 em &
              \hfil # \tabskip = 2.0 em &
              \hfil #P\negthinspace\tabskip = 0 em&#\hfil \tabskip = 0.3 em &
              \hfil #  &
              \hfil #  \cr
\multispan6{Table 3 (cont'd) \hfil} \cr
\noalign{\vskip5pt \hrule\hrule \vskip5pt}
\multispan2{\hfil Comet Desig. \hfil} & $F_{A}$ (mJy) & $F_{B}$ (mJy) &
\multispan2{\hfil Comet Desig. \hfil} & $F_{A}$ (mJy) & $F_{B}$ (mJy) \cr
\noalign{\vskip5pt \hrule \vskip10pt}
\multispan8{{\sl MIPS}\hfil } \cr
 11&                 &       $ 0.22\pm0.02$ &      $ 0.23\pm0.02$ &
223&/2002 S1         &       $2.37\pm0.02$  &      $2.10\pm0.02$  \cr
 15&                 &       $ 0.52\pm0.03$ &      $ 0.49\pm0.03$ &
244&/2000 Y3         &  \omit\hfil $< 0.30$\hfil & \omit\hfil $< 0.30$ \hfil\cr
 16&                 &  \unb{$0.91\pm0.13$} & \unb{$0.64\pm0.08$} &
256&/2003 HT$_{15}$  &       $0.11\pm0.02$  &      $0.09\pm0.02$  \cr
 51&-A               &       $0.21\pm0.02$  &      $0.23\pm0.02$  &
260&/2005 K3         &  \unb{$1.86\pm0.03$} & \unb{$1.81\pm0.03$} \cr
 62&                 &  \unb{$0.16\pm0.05$} & \unb{$0.16\pm0.05$} &
   &/1998 VS$_{24}$  &  \omit\hfil $<1.02^b$\hfil & \omit\hfil $<1.02^b$\hfil\cr
 93&                 &       $5.94\pm0.03$  &      $7.90\pm0.04$  &
   &/2001 R6         &       $0.12\pm0.02$  &      $0.15\pm0.02$  \cr
138&                 &       $0.43\pm0.04$  &      $0.35\pm0.03$  &
   &/2003 O3         &       $0.25\pm0.03$  &      $0.20\pm0.03$  \cr
139&                 &       $1.70\pm0.09$  &      $2.01\pm0.12$  &
   &/2004 A1         &  \unb{$1.61\pm0.16$} & \unb{$1.45\pm0.16$} \cr
144&                 &  \unb{$0.39\pm0.02$} & \unb{$0.44\pm0.02$} &
   &/2004 H2         &  \unb{$0.43\pm0.05$} & \unb{$0.48\pm0.16$} \cr
148&                 &       $1.00\pm0.02$  &      $0.99\pm0.02$  &
   &/2004 T1         &  \omit\hfil $<0.30^b$\hfil & \omit\hfil $<0.30^b$\hfil\cr
168&                 &       $0.15\pm0.02$  &      $0.15\pm0.02$  &
   &/2005 JD$_{108}$ &  \unb{$4.33\pm0.48$} & \unb{$4.01\pm0.40$} \cr
169&                 &       $4.20\pm0.02$  &      $4.72\pm0.02$  &
   &/2005 Q4         &       $1.49\pm0.02$  &      $1.68\pm0.02$  \cr
197&/2003 KV$_{2}$   &       $0.58\pm0.05$  &      $0.66\pm0.04$  &
   &/2005 S3         &       $1.84\pm0.10$  &      $1.49\pm0.20$  \cr
203&/1999 WJ$_{7}$   &  \omit\hfil $< 0.28$ \hfil & \omit\hfil $< 0.28$ \hfil  &
   &/2005 T5         &  \unb{$1.60\pm0.16$} & \unb{$1.76\pm0.16$} \cr
215&/2002 O8         &       $0.97\pm0.08$  &      $0.81\pm0.07$  &
\omit\hfil C&/2005 W2&  \unb{$0.24\pm0.08$} & \unb{$0.24\pm0.08$} \cr
216&/2001 CV$_{8}$   &       $0.25\pm0.04$  &      $0.25\pm0.04$  &
   &/2005 W3         &  \unb{$0.80\pm0.32$} & \unb{$0.80\pm0.32$} \cr
219&/2002 LZ$_{11}$  &  \unb{$0.49\pm0.03$} & \unb{$0.58\pm0.03$} &
   &/2005 XA$_{54}$  &       $5.96\pm0.02$  &      $4.97\pm0.02$  \cr
221&/2002 JN$_{16}$  &       $0.77\pm0.02$  &      $0.83\pm0.02$  &
   &/2005 Y2         &  \unb{$8.04\pm0.80$} & \unb{$7.24\pm0.80$} \cr
\noalign{\vskip 5pt\hrule \vskip 5pt}
\multispan8{\vtop{\hsize=6.2in{\noindent $^a$ Neither 141P nor 240P were
in the field of view, hence no upper limit. \hfil}}} \cr
\multispan8{\vtop{\hsize=6.2in{\noindent $^b$ P/1998 VS$_{24}$, 
P/2003 S1, and P/2004 T1 may or may not have been in the field of view,
the upper limits given here assume they were. \hfil}}} \cr
}}

\vfill
\eject

\vbox{%
      \halign{\hfil #P\negthinspace&#\hfil \tabskip = 0.3 em &
              # \hfil \tabskip = 0.3 em &
              \hfil # \tabskip = 2.5 em &
              \hfil #P\negthinspace\tabskip = 0 em&#\hfil \tabskip = 0.3 em &
              \hfil #  &
              \hfil #  \cr
\multispan8{\bf Table 4 \hfil} \cr
\multispan8{\vtop{\hsize=5.6in{\noindent Radius and beaming parameter for
each of the 57 comets with two IRS detections.  Table columns are:
``Comet Desig." = comet's designation, as in Table 1;
``$R_N$" = effective radius in kilometers;
``$\eta$" = beaming parameter. \hfil}}} \cr
\noalign{\vskip5pt \hrule\hrule \vskip5pt}
\multispan2{\hfil Comet Desig. \hfil} & $R_N$ (km) & \omit\hfil$\eta$\hfil &
\multispan2{\hfil Comet Desig. \hfil} & $R_N$ (km) & \omit\hfil$\eta$\hfil \cr
\noalign{\vskip5pt \hrule \vskip5pt}
  6&                    & $ 1.99^{+0.33}_{-0.27}$  & $0.84^{+0.26}_{-0.21}$ &
127&                    & $ 0.62^{+0.25}_{-0.20}$  & $0.52^{+0.45}_{-0.27}$ \cr
  7&                    & $ 2.30^{+0.37}_{-0.33}$  & $0.80^{+0.25}_{-0.21}$ &
129&                    & $ 1.65^{+0.59}_{-0.43}$  & $1.71^{+1.04}_{-0.69}$ \cr
 14&                    & $ 2.87^{+0.46}_{-0.39}$  & $0.98^{+0.28}_{-0.23}$ &
130&                    & $ 2.67^{+0.45}_{-0.35}$  & $1.39^{+0.40}_{-0.29}$ \cr
 22&                    & $ 1.59^{+0.46}_{-0.36}$  & $0.60^{+0.35}_{-0.24}$ &
131&                    & $ 0.88^{+0.22}_{-0.18}$  & $0.65^{+0.34}_{-0.24}$ \cr
 31&                    & $ 1.47^{+0.37}_{-0.28}$  & $0.79^{+0.39}_{-0.27}$ &
132&                    & $ 0.89^{+0.24}_{-0.19}$  & $1.24^{+0.32}_{-0.30}$ \cr
 32&                    & $ 2.20^{+1.87}_{-1.06}$  & $0.89^{+1.62}_{-0.67}$ &
137&                    & $ 3.22^{+0.51}_{-0.44}$  & $0.71^{+0.20}_{-0.16}$ \cr
 33&                    & $ 1.36^{+0.22}_{-0.18}$  & $1.36^{+0.38}_{-0.29}$ &
143&                    & $ 4.45^{+0.72}_{-0.62}$  & $0.91^{+0.26}_{-0.21}$ \cr
 37&                    & $ 1.45^{+0.57}_{-0.42}$  & $1.36^{+0.98}_{-0.62}$ &
146&                    & $ 0.93^{+0.34}_{-0.25}$  & $0.98^{+0.64}_{-0.21}$ \cr
 47&                    & $ 3.21^{+0.51}_{-0.46}$  & $1.09^{+0.31}_{-0.27}$ &
149&                    & $ 1.25^{+0.24}_{-0.20}$  & $0.82^{+0.28}_{-0.22}$ \cr
 48&                    & $ 3.09^{+0.75}_{-0.60}$  & $1.14^{+0.51}_{-0.37}$ &
152&                    & $ 1.36^{+1.53}_{-0.77}$  & $1.19^{+2.40}_{-0.92}$ \cr
 50&                    & $ 2.17^{+0.36}_{-0.31}$  & $2.00^{+0.56}_{-0.46}$ &
159&                    & $ 2.90^{+4.53}_{-1.72}$  & $2.46^{+5.19}_{-1.63}$ \cr
 56&                    & $ 2.51^{+0.86}_{-0.63}$  & $1.60^{+0.88}_{-0.60}$ &
160&                    & $ 1.21^{+0.42}_{-0.28}$  & $1.41^{+0.83}_{-0.49}$ \cr
 57&-A                  & $ 0.78^{+0.21}_{-0.21}$  & $0.73^{+0.21}_{-0.21}$ &
162&                    & $ 7.53^{+1.23}_{-1.10}$  & $1.15^{+0.31}_{-0.27}$ \cr
 68&                    & $ 2.50^{+0.40}_{-0.34}$  & $0.86^{+0.23}_{-0.19}$ &
163&                    & $ 1.39^{+0.23}_{-0.20}$  & $1.13^{+0.34}_{-0.27}$ \cr
 69&                    & $ 0.83^{+0.26}_{-0.22}$  & $0.96^{+0.54}_{-0.40}$ &
171&                    & $ 1.06^{+0.34}_{-0.26}$  & $0.74^{+0.48}_{-0.32}$ \cr
 74&                    & $ 9.03^{+3.16}_{-2.28}$  & $4.51^{+2.05}_{-1.46}$ &
172&                    & $ 5.60^{+0.92}_{-0.81}$  & $1.00^{+0.31}_{-0.25}$ \cr
 77&                    & $ 2.01^{+0.41}_{-0.31}$  & $1.43^{+0.49}_{-0.35}$ &
173&                    & $16.18^{+11.14}_{-6.51}$ & $5.81^{+4.47}_{-2.71}$ \cr
 78&                    & $ 1.41^{+1.20}_{-0.67}$  & $1.11^{+1.66}_{-0.77}$ &
213&/2005 R2            & $ 4.84^{+2.86}_{-1.89}$  & $5.04^{+3.66}_{-2.42}$ \cr
 79&                    & $ 0.56^{+0.21}_{-0.15}$  & $0.65^{+0.55}_{-0.33}$ &
228&/2001 YX$_{127}$    & $ 2.41^{+0.91}_{-0.63}$  & $2.95^{+1.57}_{-1.05}$ \cr
 89&                    & $ 0.95^{+0.15}_{-0.11}$  & $0.48^{+0.16}_{-0.11}$ &
246&/2004 F3            & $ 6.87^{+2.89}_{-2.11}$  & $2.18^{+1.38}_{-0.95}$ \cr
 94&                    & $ 2.05^{+0.40}_{-0.33}$  & $0.85^{+0.32}_{-0.23}$ &
   &/2004 DO$_{29}$     & $ 0.90^{+0.22}_{-0.16}$  & $0.77^{+0.33}_{-0.22}$ \cr
101&                    & $ 1.40^{+1.10}_{-0.63}$  & $1.83^{+2.36}_{-1.18}$ &
   &/2004 V3            & $ 0.75^{+0.20}_{-0.15}$  & $0.73^{+0.35}_{-0.24}$ \cr
107&                    & $ 1.70^{+0.26}_{-0.24}$  & $1.35^{+0.37}_{-0.31}$ &
   &/2004 V5-A          & $ 2.28^{+2.57}_{-1.20}$  & $1.82^{+3.17}_{-1.28}$ \cr
113&                    & $ 1.77^{+0.27}_{-0.27}$  & $1.11^{+0.32}_{-0.30}$ &
   &/2004 VR$_{8}$      & $ 6.02^{+1.48}_{-1.12}$  & $1.31^{+0.68}_{-0.45}$ \cr
118&                    & $ 3.29^{+2.71}_{-1.53}$  & $3.57^{+3.61}_{-2.04}$ &
   &/2005 GF$_{8}$      & $ 3.35^{+0.57}_{-0.46}$  & $1.49^{+0.43}_{-0.33}$ \cr
119&                    & $ 1.96^{+2.13}_{-1.19}$  & $2.78^{+4.11}_{-2.14}$ &
   &/2005 JQ$_{5}$      & $ 1.25^{+0.20}_{-0.17}$  & $2.72^{+0.66}_{-0.54}$ \cr
121&                    & $ 4.49^{+0.72}_{-0.63}$  & $1.33^{+0.37}_{-0.31}$ &
   &/2005 L4            & $ 1.50^{+0.23}_{-0.22}$  & $0.65^{+0.22}_{-0.18}$ \cr
123&                    & $ 3.44^{+0.56}_{-0.45}$  & $1.95^{+0.44}_{-0.36}$ &
   &/2005 R1            & $ 1.60^{+0.24}_{-0.24}$  & $0.97^{+0.27}_{-0.26}$ \cr
124&                    & $ 2.73^{+0.44}_{-0.38}$  & $1.11^{+0.32}_{-0.26}$ \cr
\noalign{\vskip 5pt\hrule}
}}

\vfill
\eject

\vbox{%
      \halign{\hfil #P\negthinspace&#\hfil \tabskip = 0.3 em &
              # \hfil \tabskip = 0.3 em &
              \hfil # \tabskip = 2.5 em &
              \hfil #P\negthinspace\tabskip = 0 em&#\hfil \tabskip = 0.3 em &
              \hfil #  &
              \hfil #  \cr
\multispan8{\bf Table 5 \hfil} \cr
\multispan8{\vtop{\hsize=6.2in{\noindent Radii or radii limits 
for the other 43 comets in our survey. Two radii are
given for each of the 32 comets
detected by MIPS at two epochs. 
Table columns are:
``Comet Desig." = comet's designation, as in Table 1;
``$R_{N1}$" = effective radius in kilometers at first MIPS epoch
or blue IRS wavelength;
``$R_{N2}$" = effective radius in kilometers at second MIPS
epoch or red IRS wavelength. \hfil}}} \cr
\noalign{\vskip5pt \hrule\hrule \vskip5pt}
\multispan2{\hfil Comet Desig. \hfil} & $R_{N1}$ (km) & $R_{N2}$ (km) &
\multispan2{\hfil Comet Desig. \hfil} & $R_{N1}$ (km) & $R_{N2}$ (km) \cr
\noalign{\vskip5pt \hrule \vskip5pt}
\multispan8{\sl MIPS detections \hfil} \cr
\noalign{\vskip5pt}
 11&                 &  $0.57^{+0.04}_{-0.04}$ & $0.58^{+0.04}_{-0.04}$ &
221&/2002 JN$_{16}$  &  $1.01^{+0.05}_{-0.06}$ & $1.04^{+0.07}_{-0.06}$ \cr
 15&                 &  $0.93^{+0.05}_{-0.04}$ & $0.91^{+0.04}_{-0.04}$ &
223&/2002 S1         &  $3.02^{+0.18}_{-0.18}$ & $2.84^{+0.17}_{-0.18}$ \cr
 16&                 &  $0.78^{+0.06}_{-0.06}$ & $0.65^{+0.05}_{-0.06}$ &
256&/2003 HT$_{15}$  &  $0.78^{+0.09}_{-0.08}$ & $0.71^{+0.08}_{-0.09}$ \cr
 51&-A               &  $0.40^{+0.03}_{-0.02}$ & $0.42^{+0.03}_{-0.02}$ &
260&/2005 K3         &  $1.55^{+0.09}_{-0.09}$ & $1.53^{+0.09}_{-0.09}$ \cr
 62&                 &  $0.59^{+0.09}_{-0.09}$ & $0.59^{+0.09}_{-0.09}$ &
   &/2001 R6         &  $0.69^{+0.06}_{-0.06}$ & $0.77^{+0.07}_{-0.06}$ \cr
 93&                 &  $2.40^{+0.13}_{-0.13}$ & $2.77^{+0.14}_{-0.13}$ &
   &/2003 O3         &  $0.60^{+0.04}_{-0.04}$ & $0.54^{+0.04}_{-0.04}$ \cr
138&                 &  $0.80^{+0.05}_{-0.06}$ & $0.72^{+0.04}_{-0.06}$ &
   &/2004 A1         &  $3.59^{+0.27}_{-0.28}$ & $3.40^{+0.27}_{-0.28}$ \cr
139&                 &  $1.31^{+0.07}_{-0.07}$ & $1.43^{+0.08}_{-0.07}$ &
   &/2004 H2         &  $1.22^{+0.09}_{-0.10}$ & $1.29^{+0.20}_{-0.10}$ \cr
144&                 &  $0.81^{+0.04}_{-0.04}$ & $0.86^{+0.05}_{-0.04}$ &
   &/2005 JD$_{108}$ &  $2.90^{+0.21}_{-0.23}$ & $2.79^{+0.19}_{-0.23}$ \cr
148&                 &  $1.15^{+0.05}_{-0.07}$ & $1.14^{+0.06}_{-0.07}$ &
   &/2005 Q4         &  $1.44^{+0.08}_{-0.08}$ & $1.53^{+0.09}_{-0.08}$ \cr
168&                 &  $0.48^{+0.04}_{-0.03}$ & $0.48^{+0.04}_{-0.03}$ &
   &/2005 S3         &  $1.49^{+0.08}_{-0.08}$ & $1.34^{+0.11}_{-0.08}$ \cr
169&                 &  $2.41^{+0.13}_{-0.13}$ & $2.56^{+0.13}_{-0.13}$ &
   &/2005 T5         &  $1.30^{+0.08}_{-0.10}$ & $1.36^{+0.08}_{-0.10}$ \cr
197&/2003 KV$_{2}$   &  $0.88^{+0.06}_{-0.05}$ & $0.94^{+0.06}_{-0.05}$ &
\omit\hfil C&/2005 W2 & $0.52^{+0.07}_{-0.09}$ & $0.52^{+0.07}_{-0.09}$ \cr
215&/2002 O8         &  $1.34^{+0.09}_{-0.09}$ & $1.23^{+0.07}_{-0.09}$ &
   &/2005 W3         &  $1.16^{+0.22}_{-0.27}$ & $1.16^{+0.22}_{-0.27}$ \cr
216&/2001 CV$_{8}$   &  $0.59^{+0.05}_{-0.05}$ & $0.59^{+0.05}_{-0.05}$ &
   &/2005 XA$_{54}$  &  $2.83^{+0.16}_{-0.15}$ & $2.59^{+0.14}_{-0.15}$ \cr
219&/2002 LZ$_{11}$  &  $0.95^{+0.06}_{-0.05}$ & $1.04^{+0.05}_{-0.05}$ &
   &/2005 Y2         &  $5.22^{+0.37}_{-0.40}$ & $4.95^{+0.37}_{-0.40}$ \cr
\noalign{\vskip5pt }
\multispan8{\sl MIPS non-detections \hfil} \cr
\noalign{\vskip5pt}
203&/1999 WJ$_{7}$   &  \hfil $<1.02$   \hfil & \hfil $<1.02$   \hfil &
   &/1998 VS$_{24}$  &  \hfil $<1.09^b$ \hfil & \hfil $<1.10^b$ \hfil \cr
244&/2000 Y3         &  \hfil $<1.30$   \hfil & \hfil $<1.30$   \hfil &
   &/2004 T1         &  \hfil $<0.80^b$ \hfil & \hfil $<0.80^b$ \hfil \cr
\noalign{\vskip5pt }
\multispan8{\sl IRS non-detections \hfil} \cr
\noalign{\vskip5pt}
 43&                 &  \hfil $<1.34$   \hfil & \hfil $<1.01$   \hfil &
240&/2002 X2         &   \hfil ---$^a$ \hfil &  \hfil ---$^a$ \hfil \cr
 54&                 &  \hfil $<0.64$   \hfil & \hfil $<0.58$   \hfil &
243&/2003 S2         &  \hfil $<0.55$   \hfil & \hfil $<0.60$   \hfil \cr
120&                 &  \hfil $<0.48$   \hfil & \hfil $<0.53$   \hfil &
   &/2003 S1         &  \hfil $<1.08^b$ \hfil & \hfil $<1.01^b$ \hfil \cr
141&                 &   \hfil ---$^a$ \hfil  & \hfil ---$^a$ \hfil \cr 
\noalign{\vskip 5pt\hrule \vskip 5pt}
\multispan8{\vtop{\hsize=6.2in{\noindent $^a$ 141P and 240P were
not in the field of view, hence no upper limit. \hfil}}} \cr
\multispan8{\vtop{\hsize=6.2in{\noindent $^b$ P/1998 VS$_{24}$, 
P/2003 S1, and P/2004 T1 may or may not have been in the field of view,
the upper limits given here assume they were. \hfil}}} \cr
}}

\vfill
\eject

\vbox{%
      \halign{# \hfil & \tabskip = 0.3 em
              \hfil # \hfil &
              \hfil # &
              \hfil # \hfil &
              \hfil # \hfil &
              \hfil # & 
              \hfil # \cr 
\multispan7{\bf Table 6 \hfil} \cr
\multispan7{\vtop{\hsize=5.8in{\noindent Statistics from
search for trends of $\eta$ with various quantities. See
\S 4.1 for details.
Table columns are:
``Quantity" = the independent variable
that is compared to $\eta$, with $^\star$
indicating that the scatter plot is shown in Fig. 7;
``$m$" = best-fitting slope to the linear fit (with
$1\sigma$ error);
``$Z$" = after calculating the Spearman ``$\rho$," the
number of standard deviations by which $\rho$
deviates from its null-hypothesis expected value;
``$P_Z$" = the probability that the value of $Z$ would
occur by chance in an uncorrelated sample;
``$\cal N$" = the percentage of Monte Carlo runs that resulted
in $|Z| \ge 3\sigma$;
``$\bar Z$" = expected value of $Z$ based on the Monte Carlo runs; 
$P_t$ = after calculating the Student-$t$ statistic for
a comparison of means between a group of $\eta$ values with low values
of the independent quantity and a second group with high values
of the independent quantity,
the probability that that value of $t$ would occur by chance
in an uncorrelated sample. Regarding absolute magnitudes,
we note that they were drawn from both the Minor Planet Center
website and from JPL's Horizons ephemeris system, hence the
two entries.
\hfil}}}\cr
\noalign{\vskip5pt \hrule\hrule \vskip5pt}
Quantity &  $m$  &  \omit\hfil$Z$\hfil & $P_Z$ & $\cal N$ &
        \omit\hfil$\bar Z$\hfil & \omit\hfil$P_t$\hfil \cr
\noalign{\vskip5pt \hrule \vskip5pt}
heliocentric distance$^\star$       &
 $   -0.18\pm0.18     $ & $ 0.92$ & $ 0.359$ & $ 0.4\%$ & $ 0.95$ & $0.60$ \cr
Spitzer-centric distance    &
 $  -0.218\pm0.074    $ & $ 1.17$ & $ 0.244$ & $ 0.9\%$ & $ 1.05$ & $0.66$ \cr
phase angle$^\star$                 &
 $   0.089\pm0.061    $ & $-0.71$ & $ 0.478$ & $ 0.3\%$ & $-0.88$ & $0.52$ \cr
days from perihelion$^\star$        &
 $ 0.00019\pm0.00014  $ & $-2.93$ & $ 0.003$ & $ 3.4\%$ & $-1.67$ & $0.37$ \cr
days until aphelion         &
 $(4.8\pm9.9)\times10^{-5}$ & 
                          $-1.24$ & $ 0.214$ & $ 0.0\%$ & $-0.44$ & $0.43$ \cr
duration of AORs    &
 $  0.0007\pm0.0023   $ & $ 0.67$ & $ 0.506$ & $ 0.0\%$ & $ 0.47$ & $0.92$ \cr
perihelion distance$^\star$         &
 $   -0.05\pm0.15     $ & $-1.94$ & $ 0.052$ & $ 0.3\%$ & $-1.02$ & $0.79$ \cr
semimajor axis              &
 $   -0.03\pm0.11     $ & $-0.98$ & $ 0.325$ & $ 0.0\%$ & $-0.32$ & $0.90$ \cr
eccentricity                &
 $    0.34\pm0.88     $ & $ 1.74$ & $ 0.082$ & $ 0.2\%$ & $ 1.11$ & $0.69$ \cr
inclination                 &
 $   0.000\pm0.011    $ & $ 0.17$ & $ 0.862$ & $ 0.0\%$ & $ 0.21$ & $0.88$ \cr
argument of perihelion      &
 $ -0.0001\pm0.0010   $ & $ 0.81$ & $ 0.419$ & $ 0.0\%$ & $ 0.62$ & $0.99$ \cr
longitude of asc. node &
 $ 0.00014\pm0.00095  $ & $ 0.38$ & $ 0.704$ & $ 0.0\%$ & $ 0.25$ & $0.69$ \cr
Tisserand invariant         &
 $    0.05\pm0.57     $ & $-1.20$ & $ 0.231$ & $ 0.3\%$ & $-1.07$ & $0.92$ \cr
true anomaly                &
 $ -0.0012\pm0.0023   $ & $ 1.13$ & $ 0.257$ & $ 0.0\%$ & $ 0.35$ & $0.56$ \cr
MPC absolute magnitude      & 
 $  -0.011\pm0.029    $ & $ 0.93$ & $ 0.355$ & $ 0.0\%$ & $ 0.54$ & $0.60$ \cr
JPL absolute magnitude      &
 $  -0.003\pm0.036    $ & $ 1.59$ & $ 0.111$ & $ 0.1\%$ & $ 0.89$ & $0.85$ \cr
\noalign{\vskip 5pt\hrule \vskip 5pt}
}}

\vfill
\eject

\vbox{%
      \halign{\hfil #P\negthinspace&#\hfil \tabskip = 0.3 em &
              \hfil # \tabskip = 2.0 em &
              \hfil #P\negthinspace\tabskip = 0 em&#\hfil \tabskip = 0.3 em &
              \hfil # \tabskip = 2.0 em &
              \hfil #P\negthinspace\tabskip = 0 em&#\hfil \tabskip = 0.3 em &
              \hfil #  \cr
\multispan9{\bf Table 7 \hfil} \cr
\multispan9{\vtop{\hsize=6.2in{\noindent Radii  used
in the derivation of size distribution. All radii assume
$\bar\eta=1.03\pm0.11$. These include the 57 nuclei
successfully observed by IRS and the 32 by MIPS, for
a total of 89 nuclei.
Table columns are:
``Comet Desig." = comet's designation, as in Table 1;
``$R_{N}$" = effective radius in kilometers. \hfil}}} \cr
\noalign{\vskip5pt \hrule\hrule \vskip5pt}
\multispan2{\hfil Comet Desig. \hfil} & $R_{N}$ (km) & 
\multispan2{\hfil Comet Desig. \hfil} & $R_{N}$ (km) & 
\multispan2{\hfil Comet Desig. \hfil} & $R_{N}$ (km)  \cr
\noalign{\vskip5pt \hrule \vskip5pt}
  6&                   & $  2.23^{+0.13}_{-0.15}$ &
118&                   & $  1.30^{+0.19}_{-0.22}$  &
215&/2002 O8           & $  1.29^{+0.08}_{-0.08}$ \cr
  7&                   & $  2.64^{+0.17}_{-0.17}$ &
119&                   & $  0.98^{+0.08}_{-0.11}$  &
216&/2001 CV$_{8}$     & $  0.59^{+0.05}_{-0.05}$ \cr
 11&                   & $  0.58^{+0.04}_{-0.04}$ &
121&                   & $  3.87^{+0.26}_{-0.27}$  &
219&/2002 LZ$_{11}$    & $  1.00^{+0.06}_{-0.06}$ \cr
 14&                   & $  2.95^{+0.19}_{-0.19}$ &
123&                   & $  2.18^{+0.23}_{-0.23}$  &
221&/2002 JN$_{16}$    & $  1.03^{+0.06}_{-0.06}$ \cr
 15&                   & $  0.92^{+0.05}_{-0.05}$ &
124&                   & $  2.62^{+0.16}_{-0.16}$  &
223&/2002 S1           & $  2.93^{+0.18}_{-0.18}$ \cr
 16&                   & $  0.72^{+0.09}_{-0.09}$ &
127&                   & $  0.90^{+0.07}_{-0.07}$  &
228&/2001 YX$_{127}$   & $  1.23^{+0.15}_{-0.17}$ \cr
 22&                   & $  2.15^{+0.17}_{-0.17}$ &
129&                   & $  1.22^{+0.10}_{-0.10}$  &
246&/2004 F3           & $  4.20^{+0.42}_{-0.43}$ \cr
 31&                   & $  1.65^{+0.11}_{-0.12}$ &
130&                   & $  2.23^{+0.15}_{-0.17}$  &
256&/2003 HT$_{15}$    & $  0.75^{+0.08}_{-0.09}$ \cr
 32&                   & $  2.38^{+0.21}_{-0.23}$ &
131&                   & $  1.11^{+0.09}_{-0.07}$  &
260&/2005 K3           & $  1.54^{+0.09}_{-0.08}$ \cr
 33&                   & $  1.15^{+0.08}_{-0.09}$ &
132&                   & $  0.81^{+0.04}_{-0.05}$  &
   &/2001 R6           & $  0.73^{+0.07}_{-0.06}$ \cr
 37&                   & $  1.23^{+0.08}_{-0.09}$ &
137&                   & $  4.04^{+0.31}_{-0.32}$  &
   &/2003 O3           & $  0.57^{+0.04}_{-0.04}$ \cr
 47&                   & $  3.11^{+0.20}_{-0.21}$ &
138&                   & $  0.76^{+0.06}_{-0.06}$  &
   &/2004 A1           & $  3.49^{+0.27}_{-0.28}$ \cr
 48&                   & $  2.97^{+0.19}_{-0.20}$ &
139&                   & $  1.37^{+0.08}_{-0.08}$  &
   &/2004 DO$_{29}$    & $  1.07^{+0.09}_{-0.07}$ \cr
 50&                   & $  1.49^{+0.13}_{-0.13}$ &
143&                   & $  4.79^{+0.32}_{-0.33}$  &
   &/2004 H2           & $  1.23^{+0.15}_{-0.17}$ \cr
 51&-A                 & $  0.41^{+0.03}_{-0.02}$ &
144&                   & $  0.84^{+0.05}_{-0.04}$  &
   &/2004 V3           & $  0.92^{+0.08}_{-0.07}$ \cr
 56&                   & $  1.92^{+0.13}_{-0.14}$ &
146&                   & $  0.96^{+0.07}_{-0.08}$  &
   &/2004 V5-A         & $  1.66^{+0.22}_{-0.24}$ \cr
 57&-A                 & $  0.96^{+0.06}_{-0.07}$ &
148&                   & $  1.14^{+0.06}_{-0.07}$  &
   &/2004 VR$_{8}$     & $  5.41^{+0.30}_{-0.31}$ \cr
 62&                   & $  0.59^{+0.09}_{-0.09}$ &
149&                   & $  1.42^{+0.09}_{-0.10}$  &
   &/2005 GF$_{8}$     & $  2.70^{+0.19}_{-0.20}$ \cr
 68&                   & $  2.80^{+0.19}_{-0.21}$ &
152&                   & $  1.03^{+0.10}_{-0.11}$  &
   &/2005 JD$_{108}$   & $  2.84^{+0.20}_{-0.22}$ \cr
 69&                   & $  0.87^{+0.05}_{-0.07}$ &
159&                   & $  1.46^{+0.20}_{-0.23}$  &
   &/2005 JQ$_{5}$     & $  0.67^{+0.07}_{-0.08}$ \cr
 74&                   & $  3.31^{+0.60}_{-0.69}$ &
160&                   & $  0.99^{+0.09}_{-0.07}$  &
   &/2005 L4           & $  1.88^{+0.13}_{-0.14}$ \cr
 77&                   & $  1.66^{+0.11}_{-0.12}$ &
162&                   & $  7.03^{+0.47}_{-0.48}$  &
   &/2005 Q4           & $  1.48^{+0.08}_{-0.08}$ \cr
 78&                   & $  1.35^{+0.12}_{-0.14}$ &
163&                   & $  1.32^{+0.08}_{-0.08}$  &
   &/2005 R1           & $  1.65^{+0.10}_{-0.10}$ \cr
 79&                   & $  0.70^{+0.04}_{-0.04}$ &
168&                   & $  0.48^{+0.04}_{-0.03}$  &
   &/2005 S3           & $  1.41^{+0.11}_{-0.11}$ \cr
 89&                   & $  1.43^{+0.13}_{-0.13}$ &
169&                   & $  2.48^{+0.13}_{-0.14}$  &
   &/2005 T5           & $  1.33^{+0.08}_{-0.10}$ \cr
 93&                   & $  2.59^{+0.26}_{-0.26}$ &
171&                   & $  1.25^{+0.09}_{-0.08}$  &
\omit\hfil C&/2005 W2  & $  0.52^{+0.07}_{-0.09}$ \cr
 94&                   & $  2.27^{+0.13}_{-0.15}$ &
172&                   & $  5.70^{+0.36}_{-0.37}$  &
   &/2005 W3           & $  1.16^{+0.22}_{-0.27}$ \cr
101&                   & $  0.98^{+0.09}_{-0.08}$ &
173&                   & $  4.28^{+0.95}_{-1.17}$  &
   &/2005 XA$_{54}$    & $  2.71^{+0.17}_{-0.17}$ \cr
107&                   & $  1.45^{+0.09}_{-0.10}$ &
197&/2003 KV$_{2}$     & $  0.92^{+0.06}_{-0.05}$  &
   &/2005 Y2           & $  5.08^{+0.37}_{-0.40}$ \cr
113&                   & $  1.70^{+0.10}_{-0.10}$ &
213&/2005 R2           & $  1.53^{+0.24}_{-0.29}$  \cr
\noalign{\vskip 5pt\hrule \vskip 5pt}
}}

\vfill
\eject

\vbox{%
      \halign{# \hfil  \tabskip = 5 mm &
              \hfil #  &
              \hfil #  &
              \hfil #  &
              \hfil # \cr
\multispan5{\bf Table 8 \hfil} \cr
\multispan5{\vtop{\hsize=6.5in{\noindent Statistics associated
with the radii ratios in Figure 11. ``Compilation" indicates
the paper mentioned in \S4.3.1 and in Figure 11. The
middle columns give the mean $\mu_m$ and
standard deviation $\sigma$ among the $N$ radii ratios; 
the three columns pertain to all detected SEPPCoN comets that
appear in the comparison compilation, just the SEPPCoN
comets that appeared without discernible dust, and
just the SEPPCoN comets that appear with discernible
dust, respectively. The last column gives the
significance of a Student-$t$ statistic comparing the means
of the comets with and without discernible dust. \hfil}}} \cr
\noalign{\vskip5pt \hrule\hrule \vskip5pt}
Compilation & All comets & No dust & Yes dust & Student-$t$ \cr
            & ($N; \mu_m\pm\sigma$)  & ($N; \mu_m\pm\sigma$) & 
           ($N; \mu_m\pm\sigma$)  & \cr
\noalign{\vskip5pt \hrule \vskip5pt}
S11 & $25; 0.93\pm0.42$ & $13; 0.81\pm0.35$ & $12; 1.06\pm0.46$ & $14.7\%$ \cr
W11 & $29; 0.82\pm0.55$ & $17; 0.73\pm0.41$ & $12; 0.95\pm0.70$ & $29.7\%$ \cr
T06 & $37; 1.08\pm0.41$ & $19; 1.19\pm0.34$ & $18; 0.95\pm0.44$ & $ 7.2\%$ \cr
L04 & $20; 1.18\pm0.43$ & $10; 1.04\pm0.35$ & $10; 1.32\pm0.48$ & $15.5\%$ \cr
\noalign{\vskip 5pt\hrule \vskip 5pt}
}}

\vfill
\eject

\vbox{%
      \halign{# \hfil \tabskip = 5 mm &
              \hfil # \hfil \tabskip = 0 mm & 
              # \tabskip = 5 mm &
              \hfil # \hfil \tabskip = 0 mm & 
              # \tabskip = 5 mm &
              \hfil # \hfil \tabskip = 0 mm & 
              # \tabskip = 5 mm &
              \hfil # \hfil \tabskip = 0 mm & 
              #  \cr
\multispan9{\bf Table 9 \hfil} \cr
\multispan9{\vtop{\hsize=4.9in{\noindent
Power-Law Slopes to CSDs in Figure 13. 
Table columns:
``Range" = range in radii (in km) that was used for the fitting;
``all $q$" = slopes calculated including all perihelia;
``$q<x$ AU" = slopes calculated using only comets with perihelia
under $x$ AU;
``$\beta$" = power-law slope in the
function $R_N^\beta$; ``No." = number of comets 
included in the fit. \hfil}}} \cr
\noalign{\vskip5pt \hrule\hrule \vskip5pt}
Range & \multispan2{\hfil all $q$ \hfil} & \multispan2{\hfil $q<3$ AU \hfil} & 
        \multispan2{\hfil $q<2.5$ AU \hfil} & \multispan2{\hfil $q<2$ AU \hfil} \cr
      & $\beta$ & No. & $\beta $ & No. & $\beta $ & No. & $\beta $ & No. \cr
\noalign{\vskip5pt \hrule \vskip5pt}
$1.4-4.0$ & $-1.43$ & 42 & $-1.52$ & 36 & $-1.73$ & 29 & $-1.57$ & 18 \cr
$1.4-5.0$ & $-1.55$ & 47 & $-1.60$ & 40 & $-1.68$ & 31 & $-1.59$ & 20 \cr
$1.4-7.0$ & $-1.75$ & 51 & $-1.74$ & 43 & $-1.71$ & 34 & $-1.63$ & 21 \cr
$1.4-9.0$ & $-1.82$ & 52 & $-1.80$ & 44 & $-1.75$ & 35 & $-1.63$ & 22 \cr
$1.6-4.0$ & $-1.60$ & 33 & $-1.80$ & 28 & $-2.10$ & 23 & $-2.98$ & 14 \cr
$1.6-5.0$ & $-1.70$ & 38 & $-1.80$ & 32 & $-1.84$ & 25 & $-2.01$ & 16 \cr
$1.6-7.0$ & $-1.92$ & 42 & $-1.93$ & 35 & $-1.81$ & 28 & $-1.94$ & 17 \cr
$1.6-9.0$ & $-2.00$ & 43 & $-1.99$ & 36 & $-1.85$ & 29 & $-1.82$ & 18 \cr
$1.8-4.0$ & $-1.88$ & 28 & $-2.06$ & 25 & $-2.67$ & 20 & $-2.98$ & 14 \cr
$1.8-5.0$ & $-1.88$ & 33 & $-1.91$ & 29 & $-1.99$ & 22 & $-2.01$ & 16 \cr
$1.8-7.0$ & $-2.10$ & 37 & $-2.03$ & 32 & $-1.87$ & 25 & $-1.94$ & 17 \cr
$1.8-9.0$ & $-2.17$ & 38 & $-2.08$ & 33 & $-1.91$ & 26 & $-1.82$ & 18 \cr
$2.0-4.0$ & $-1.97$ & 26 & $-2.15$ & 23 & $-2.84$ & 19 & $-2.98$ & 14 \cr
$2.0-5.0$ & $-1.92$ & 31 & $-1.94$ & 27 & $-2.00$ & 21 & $-2.01$ & 16 \cr
$2.0-7.0$ & $-2.15$ & 35 & $-2.06$ & 30 & $-1.87$ & 24 & $-1.94$ & 17 \cr
$2.0-9.0$ & $-2.22$ & 36 & $-2.11$ & 31 & $-1.91$ & 25 & $-1.82$ & 18 \cr
$2.2-4.0$ & $-2.00$ & 22 & $-2.13$ & 19 & $-2.90$ & 15 & $-3.32$ & 11 \cr
$2.2-5.0$ & $-1.93$ & 27 & $-1.89$ & 23 & $-1.85$ & 17 & $-1.88$ & 13 \cr
$2.2-7.0$ & $-2.18$ & 31 & $-2.04$ & 26 & $-1.78$ & 20 & $-1.84$ & 14 \cr
$2.2-9.0$ & $-2.26$ & 32 & $-2.11$ & 27 & $-1.84$ & 21 & $-1.74$ & 15 \cr
\noalign{\vskip 5pt\hrule \vskip 5pt}
}}

\vfill
\eject


\figembedps[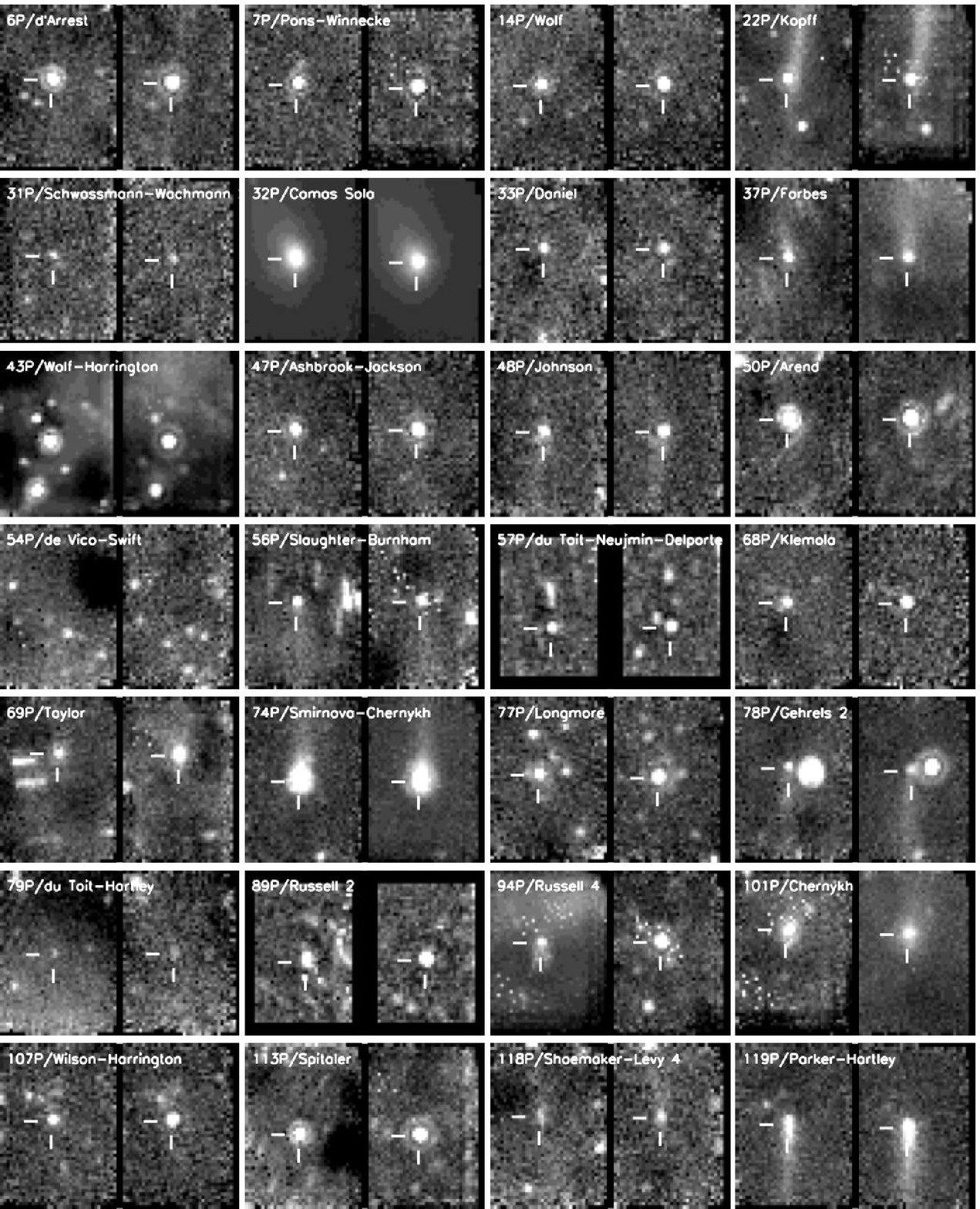,7.0 in]{Figure 1,
caption at end of figure. \hfil}
\bigskip
\eject

\figembedps[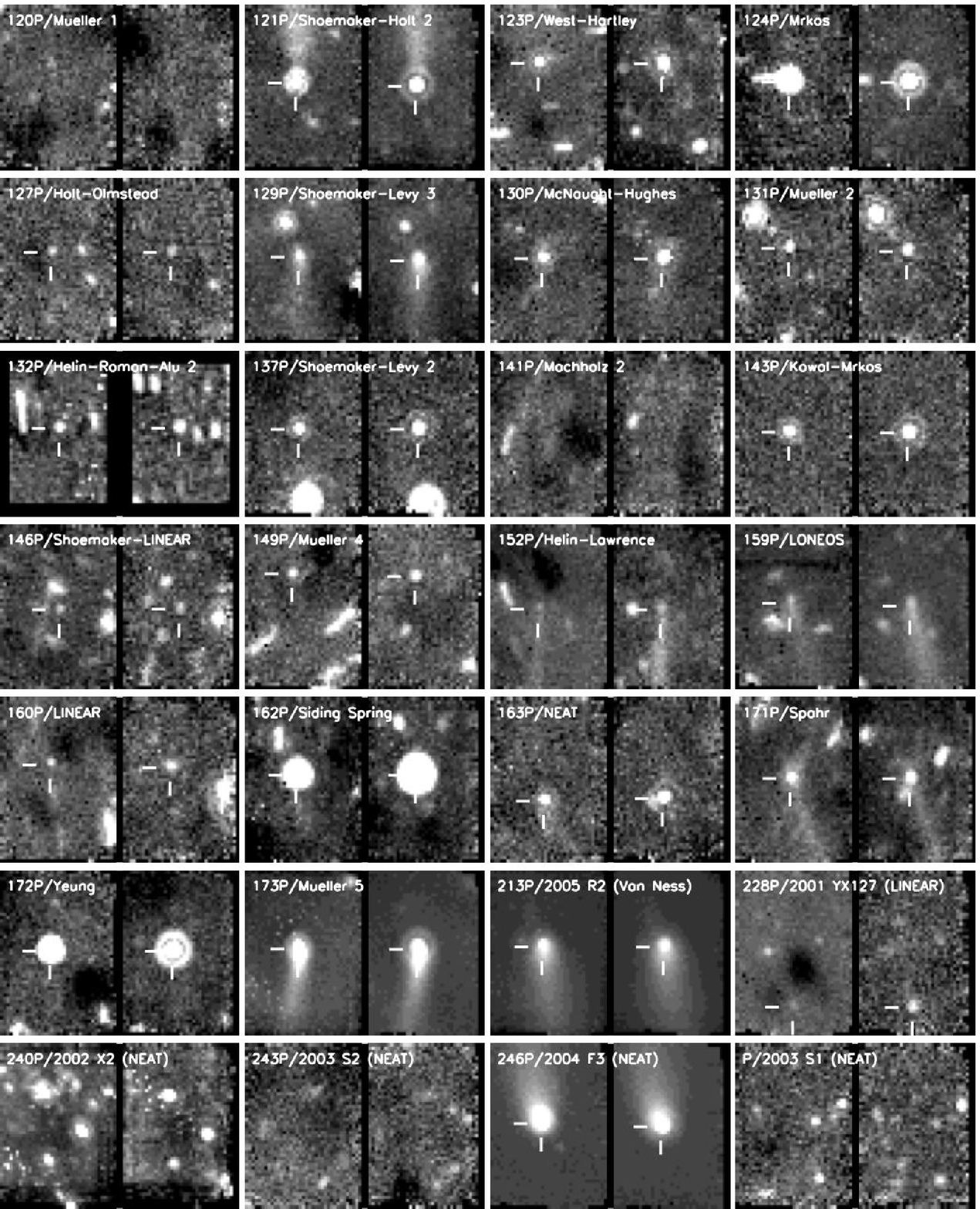,7.0 in]{Figure 1 continued,
caption at end of figure. \hfil}
\bigskip
\eject

\figembedps[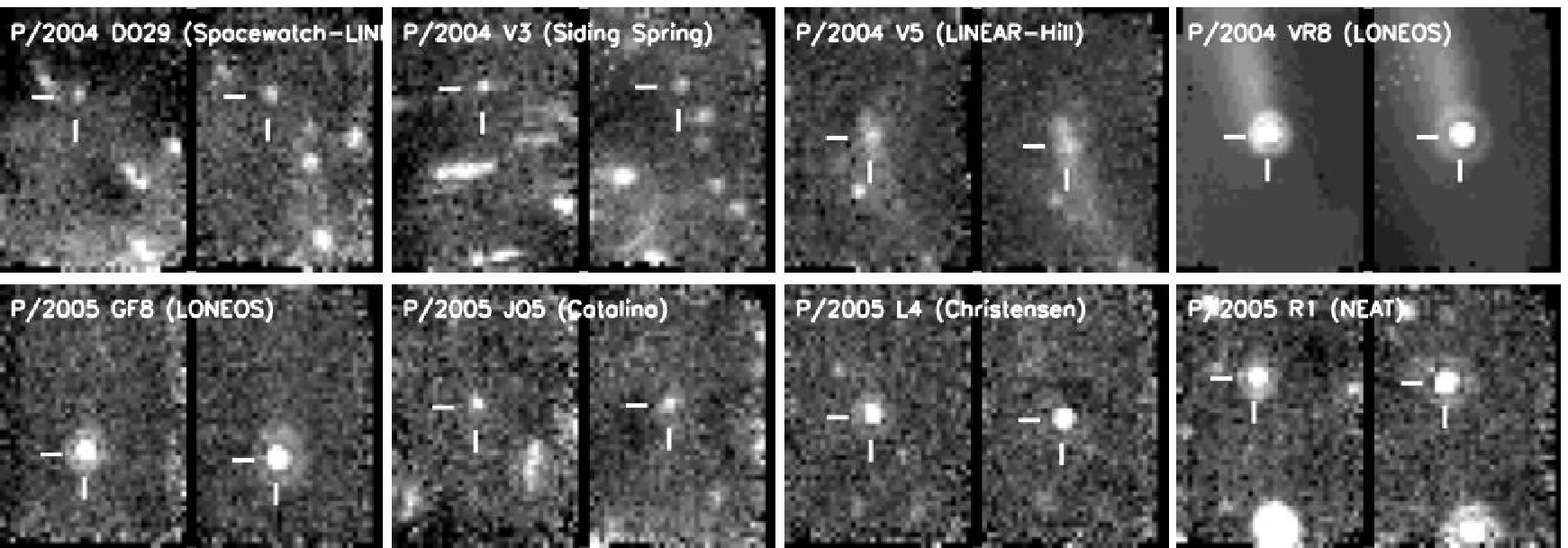,7.0 in]{Figure 1 continued.
Montage of the 64 comets in our survey observed by
Spitzer's IRS. For each comet, there are
two panels: the left panel shows the ``blue"
16 $\mu$m PU image; the right, the ``red"
22 $\mu$m PU image. 
Tick marks indicate the location of the comet,
except for the seven comets (see text) where we
had no detection. Many comets
appear as point-sources, evincing only emission
from the nucleus; some however show extended
emission from dust. Note that some images
show bright background sources as well. 
Each panel is 67-by-92 arcsec. \hfil}
\bigskip
\eject

\figembedps[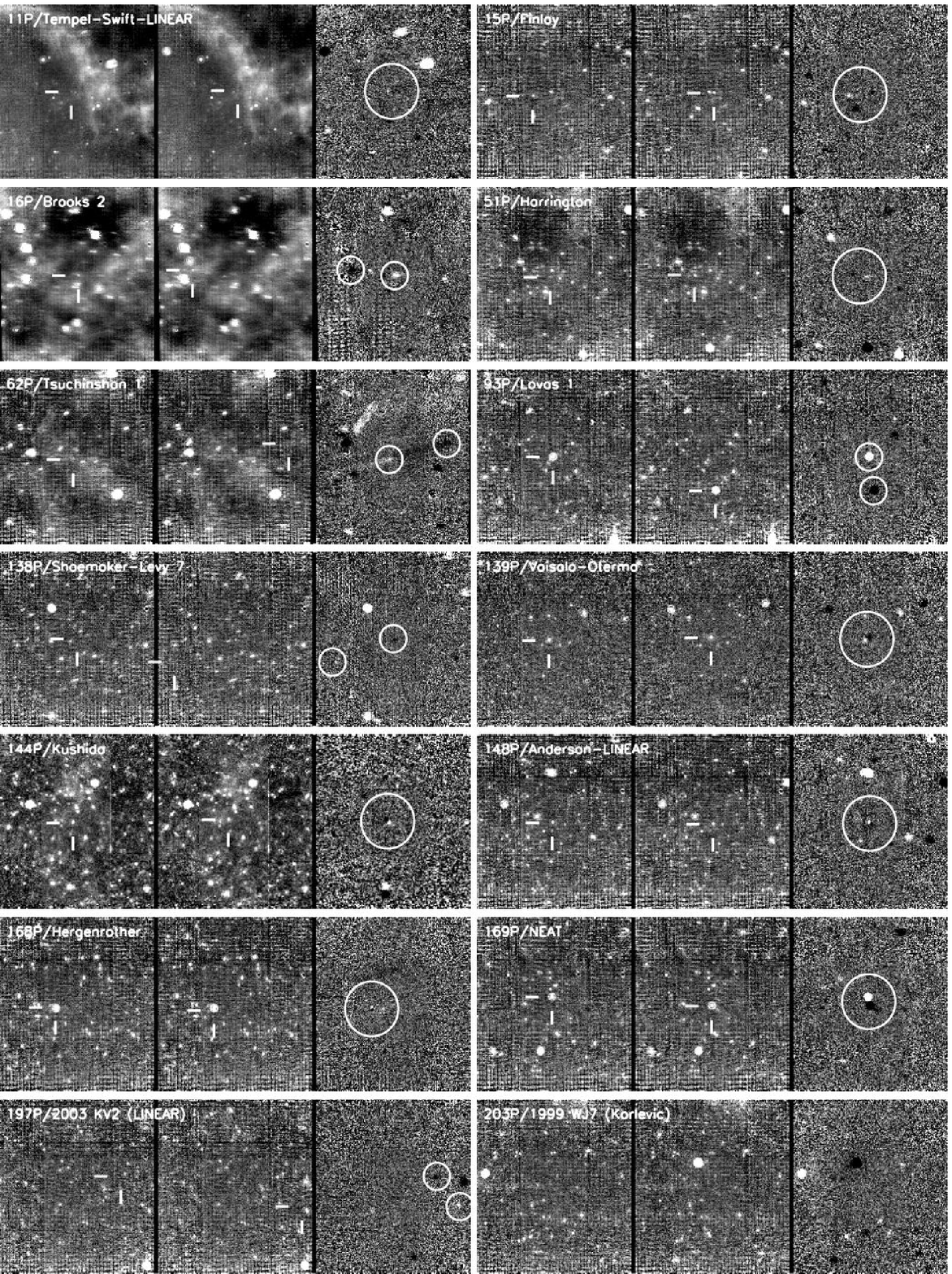,6.5 in]{Figure 2,
caption at end of figure. \hfil}
\bigskip
\eject

\figembedps[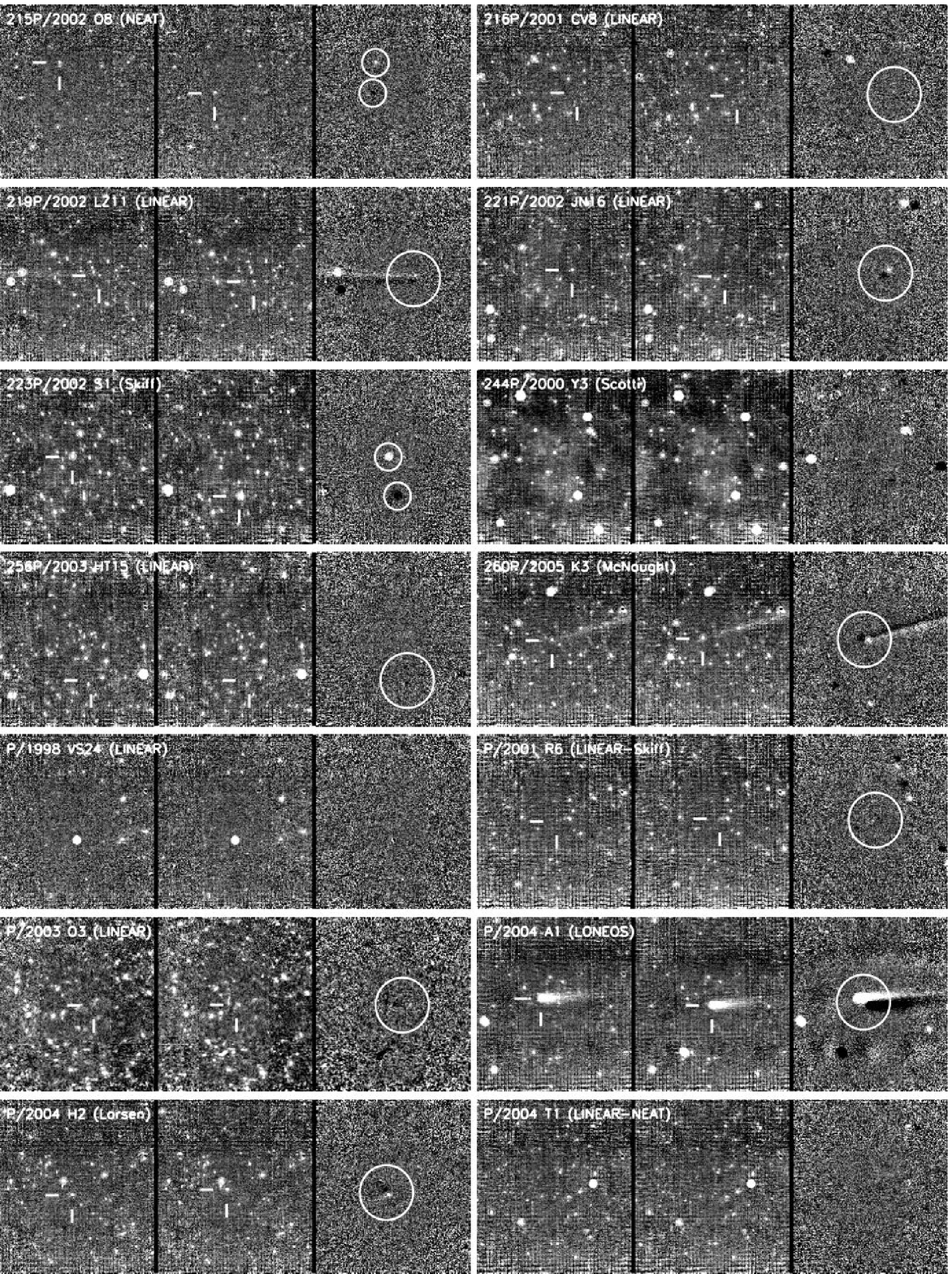,6.5 in]{Figure 2 continued,
caption at end of figure. \hfil}
\bigskip
\eject

\figembedps[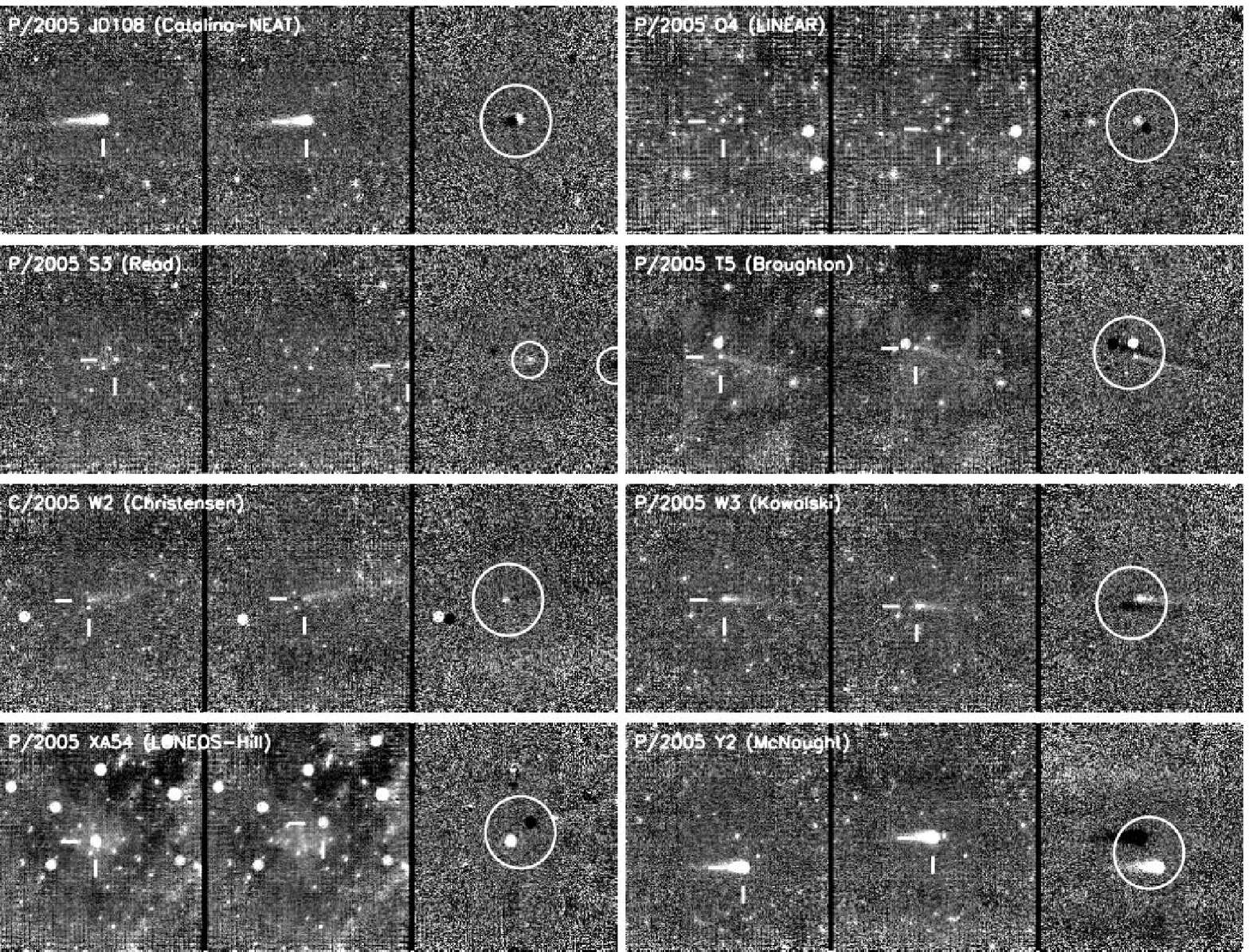,6.5 in]{Figure 2 continued. 
Montage of the 36 comets in our survey observed by
Spitzer's MIPS at 24 $\mu$m. For each comet, there are
three panels: the ``on-source" image, the
``shadow" image, and a difference
image that is the subtraction of the
two. Tick marks in the on-source and shadow images
indicate the location of the comet, though
the comets are most easily seen
in the difference images.
Circles in the difference images
indicate the location of the comet's positive
and negative images (though if the two are close, then
only one large circle is drawn around both).
No ticks or circles are drawn
for the 4 comets (see text) where we had no detection.
As in Fig. 1,
many comets appear as point-sources but
several show extended emission from dust.
Each panel is 7.3-by-7.9 arcmin. Several
panels also have serendipitously-observed
asteroids in the field. \hfil}
\bigskip
\eject

\figembedps[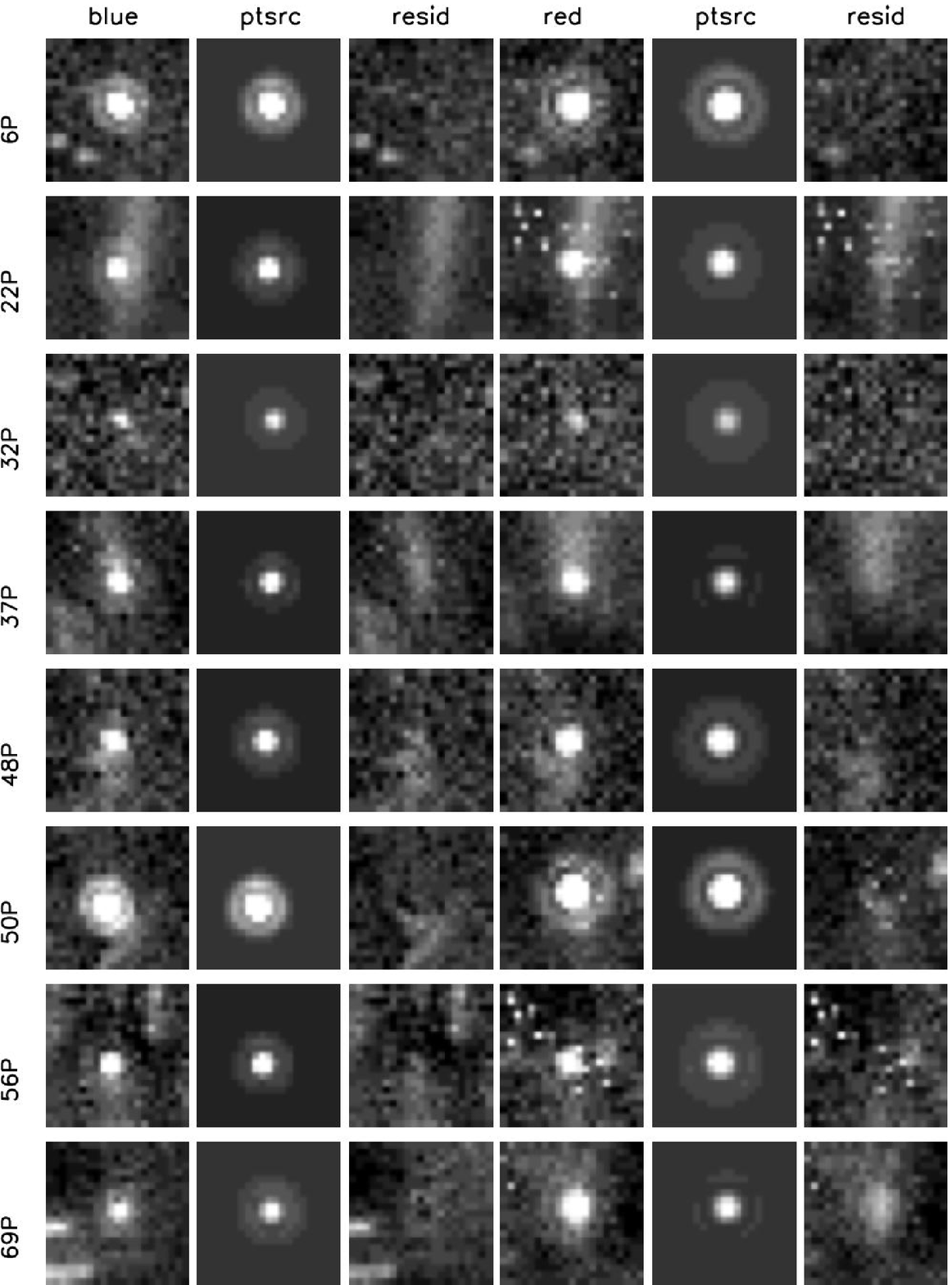,6.5 in]{Figure 3,
caption at end of figure. \hfil}
\bigskip
\eject

\figembedps[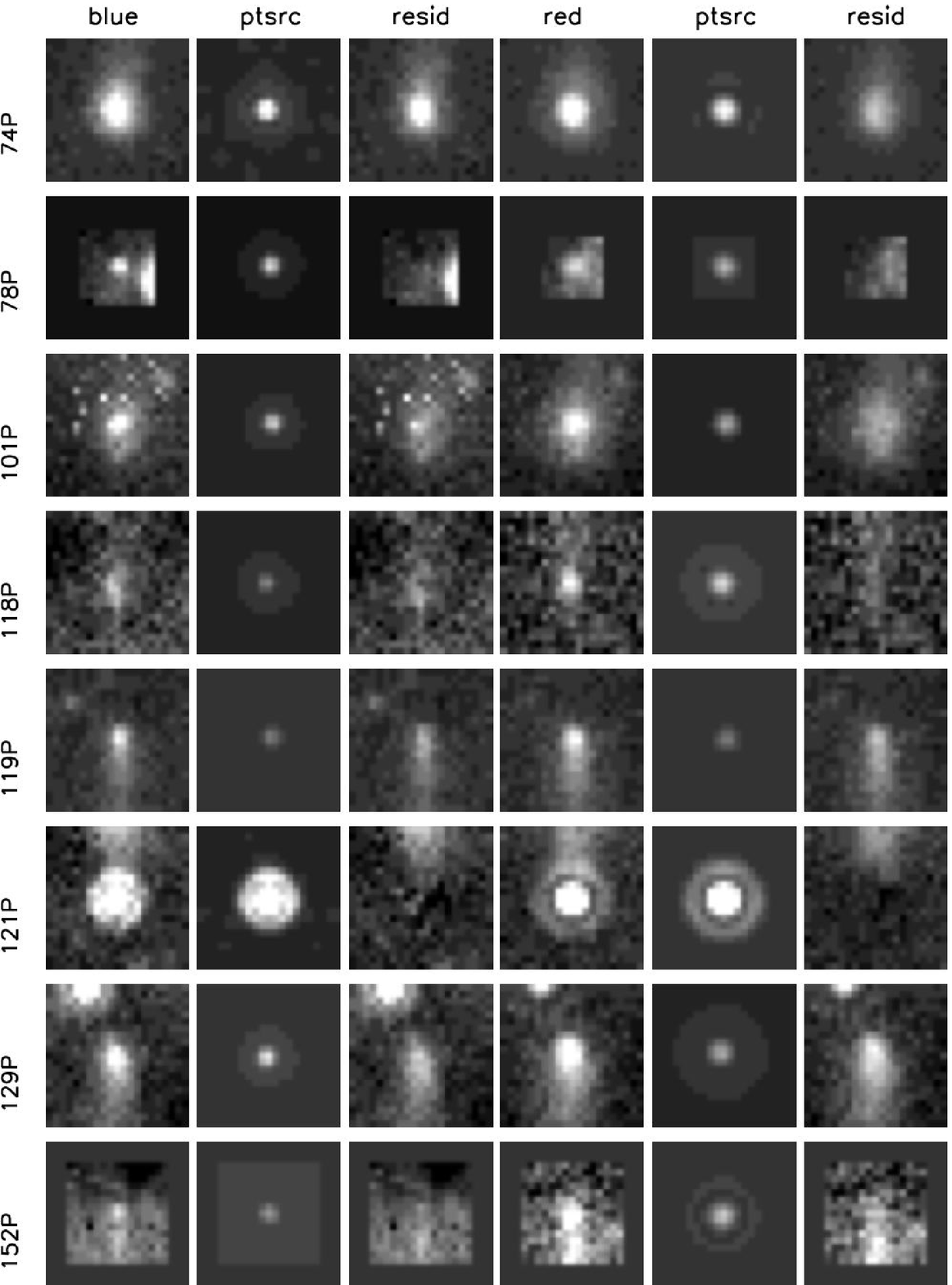,6.5 in]{Figure 3 continued,
caption at end of figure. \hfil}
\bigskip
\eject

\figembedps[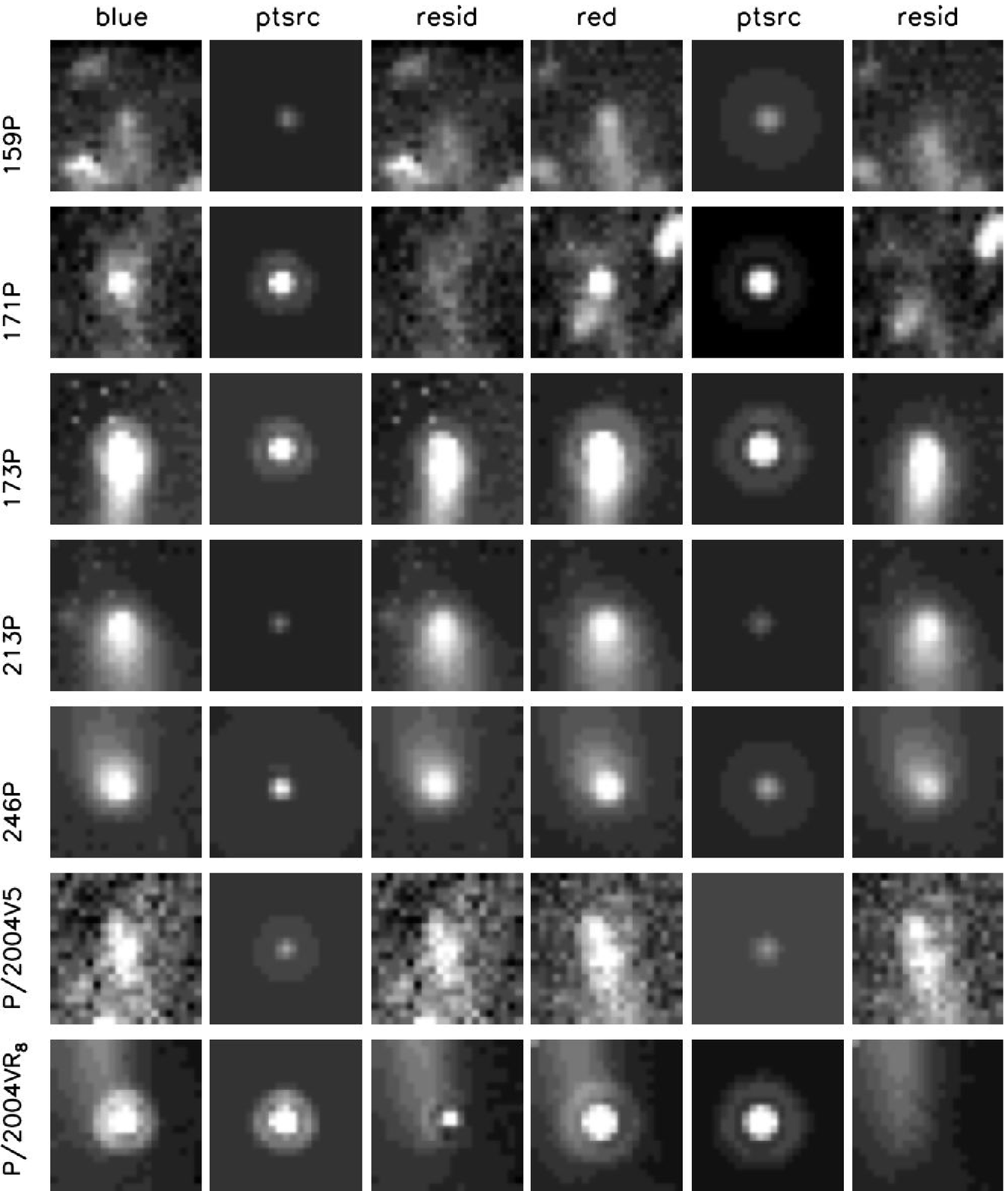,6.5 in]{Figure 3 continued.
Montage of the 23 comets observed
by IRS with extended emission from dust.
Each comet has a row of six images. Each image
shows a 21-by-21 pixel region around the comet.
The 6 images are, from left to right:
``blue" PU image, best-fitting
PSF for that image, residual after removing the PSF,
``red" PU image, best-fitting PSF for that image,
residual after removing the PSF. Note that for
some comets (e.g. 6P, 32P) the dust is not obvious. \hfil}
\bigskip
\eject

\figembedps[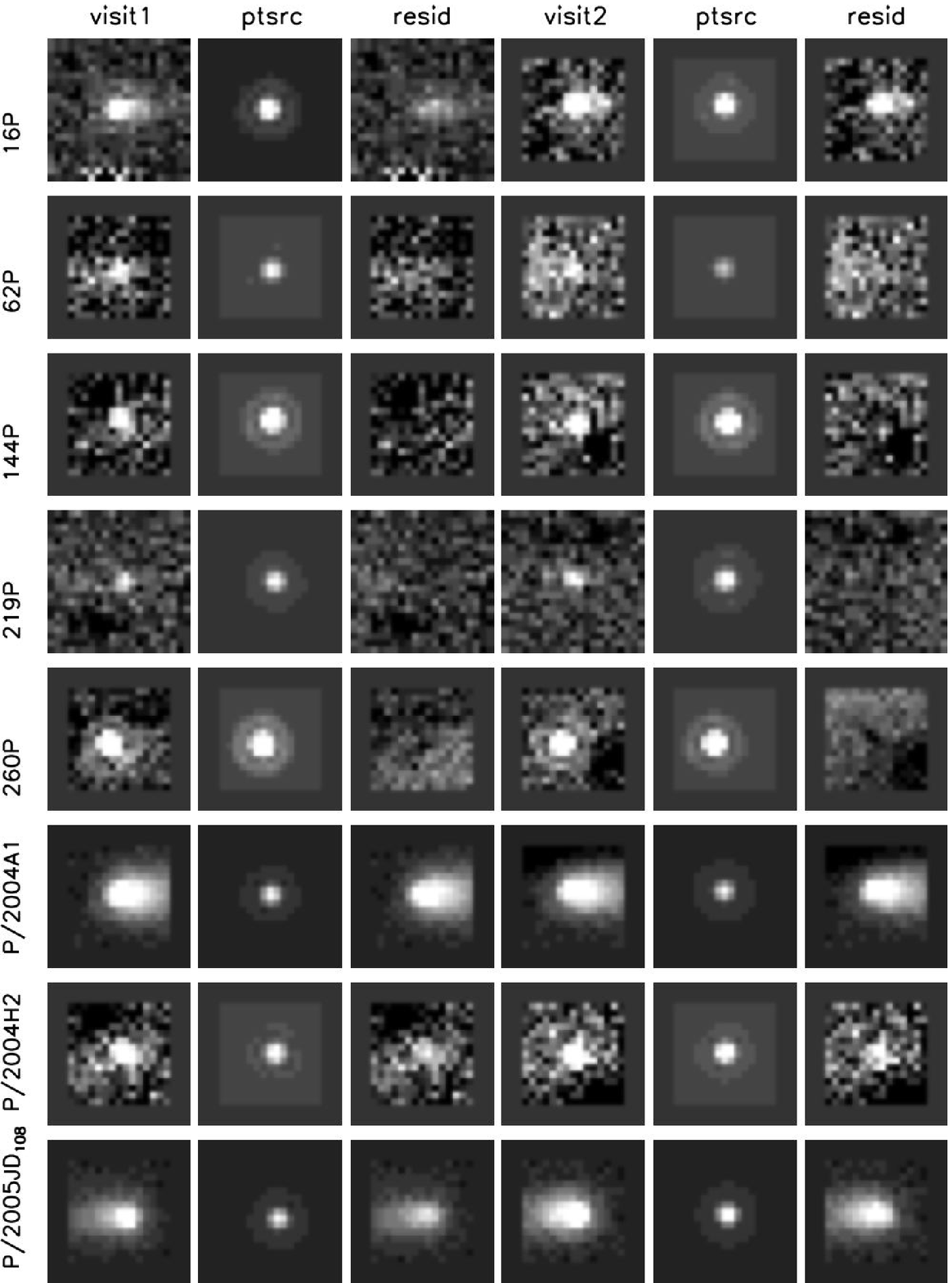,6.5 in]{Figure 4,
caption at end of figure. \hfil}
\bigskip
\eject

\figembedps[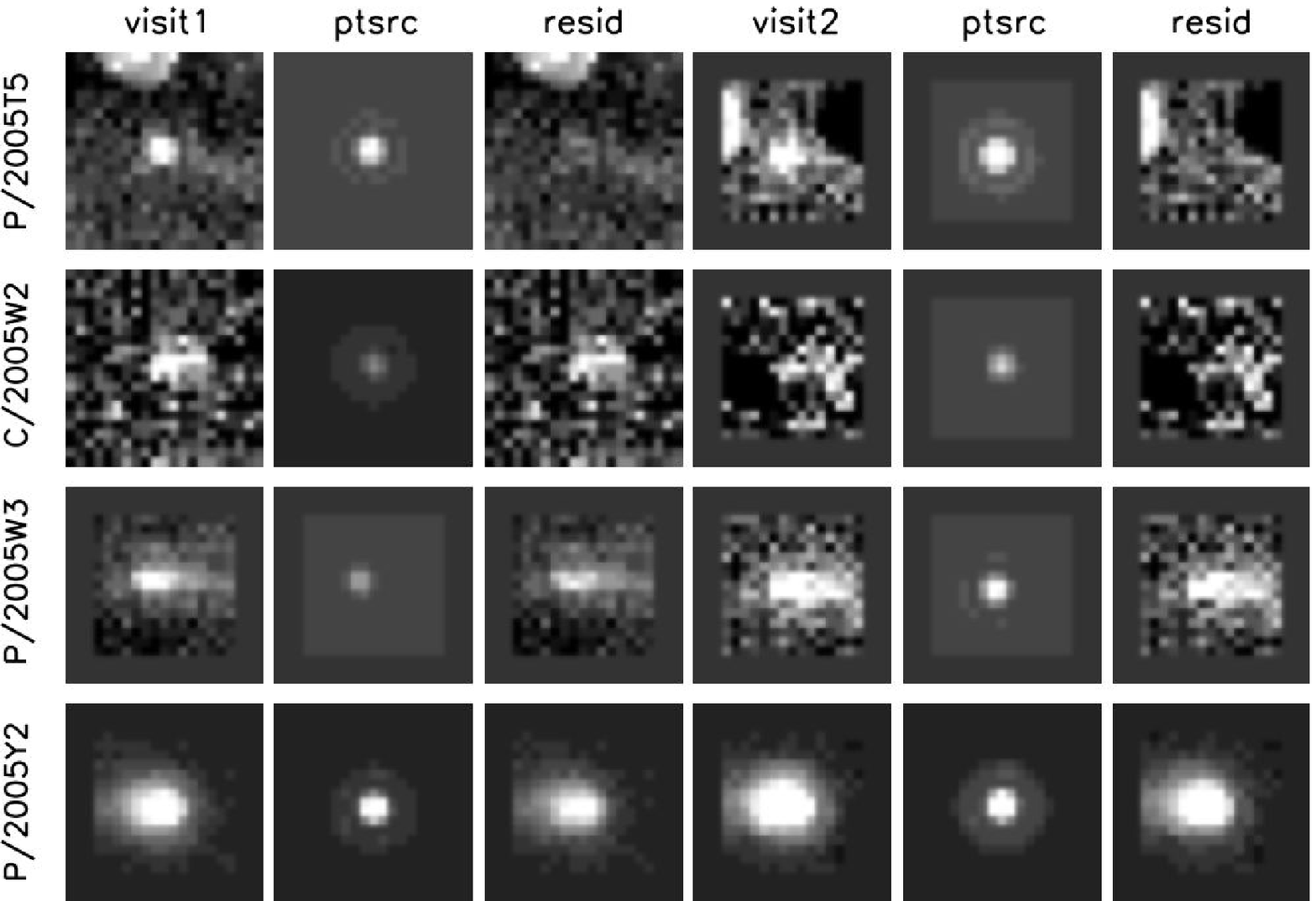,6.5 in]{Figure 4 continued.
Same as Fig. 3, except showing the
12 comets observed by MIPS with extended emission. The first and
fourth images in each row display the 
comet at each visits after shadow-subtraction with the other visit.
Note that for
some comets (e.g. 144P, 219P) the dust is not obvious. \hfil}
\bigskip
\eject

\figembedps[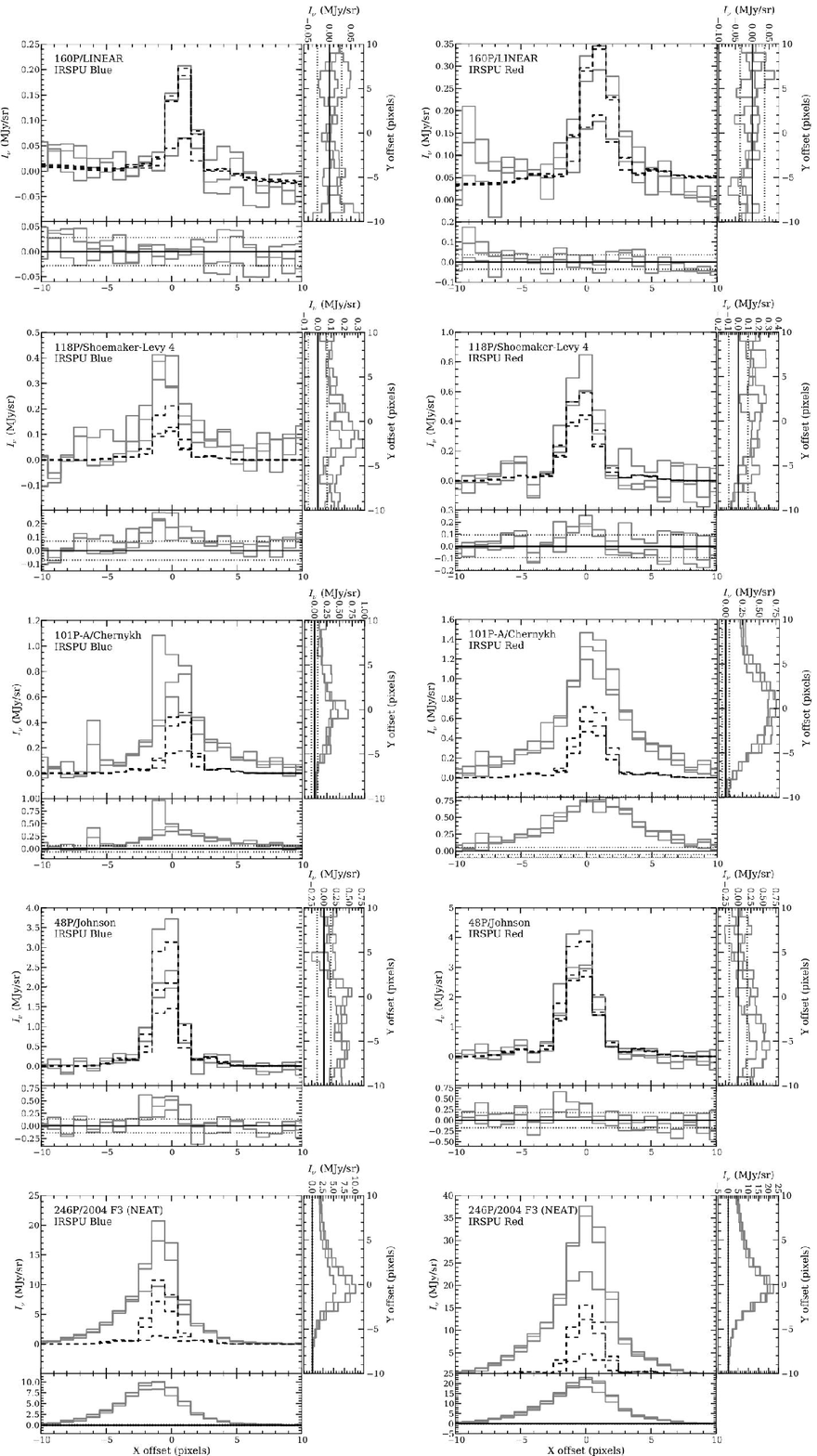,4.8 in]{Figure 5,
caption after end of figure. \hfil}
\bigskip
\eject

\figembedps[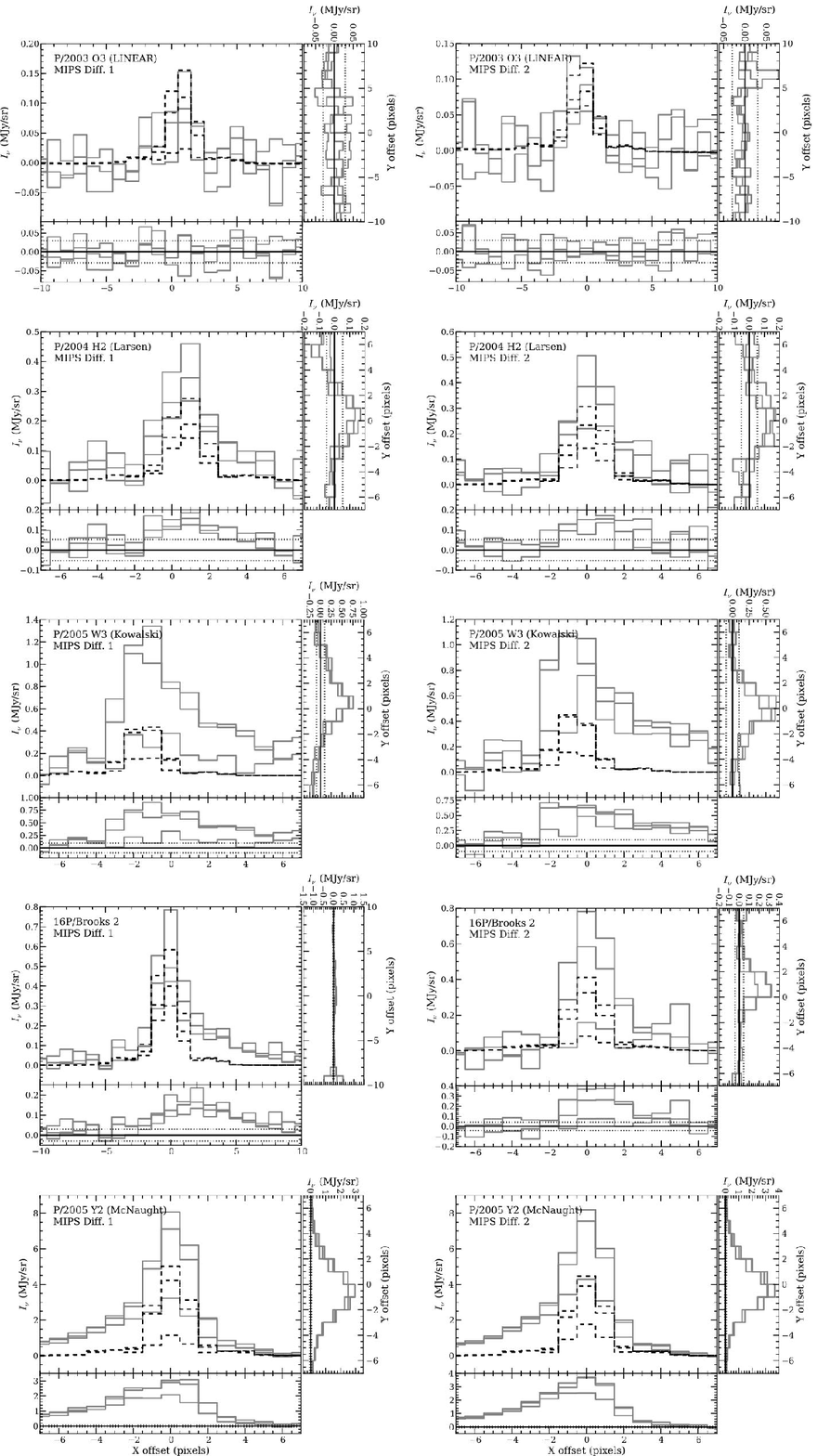,4.8 in]{Figure 5 continued,
caption after end of figure. \hfil}
\bigskip
\eject

Figure 5. Sample of surface-brightness profiles from
ten comets in our survey. The first five pairs
of panels are from five IRS-observed comets, with 
profiles
from the ``blue" and ``red" PU images (one panel for each).
The second five pairs are from five MIPS-observed comets, with
profiles from the two shadow-subtracted visits (one panel for each).
The name of the comet is given in each panel in the upper-left.
A variety of coma brightnesses and nucleus-to-coma contrasts is shown.
In each panel, there are three plots. The largest plot shows
three line cuts through the image in the x-direction; these are the
solid grey lines.  The plot also shows three line cuts 
through the best-fitting PSF
in the x-direction; these are the black dashed lines. 
The cuts come from the row that has the pixel with the comet's centroid,
the row above that, and the row below that.
The plot below the largest plot shows residuals in the line cuts
after subtracting the best-fitting PSF from the image; the solid
horizontal black line marks a residual of zero, and the horizontal
dotted lines indicated $\pm1\sigma$. 
The plot to the right of the largest plot is similar in that
it shows residuals, except it shows y-direction line cuts.

\vfill
\eject

\figembedps[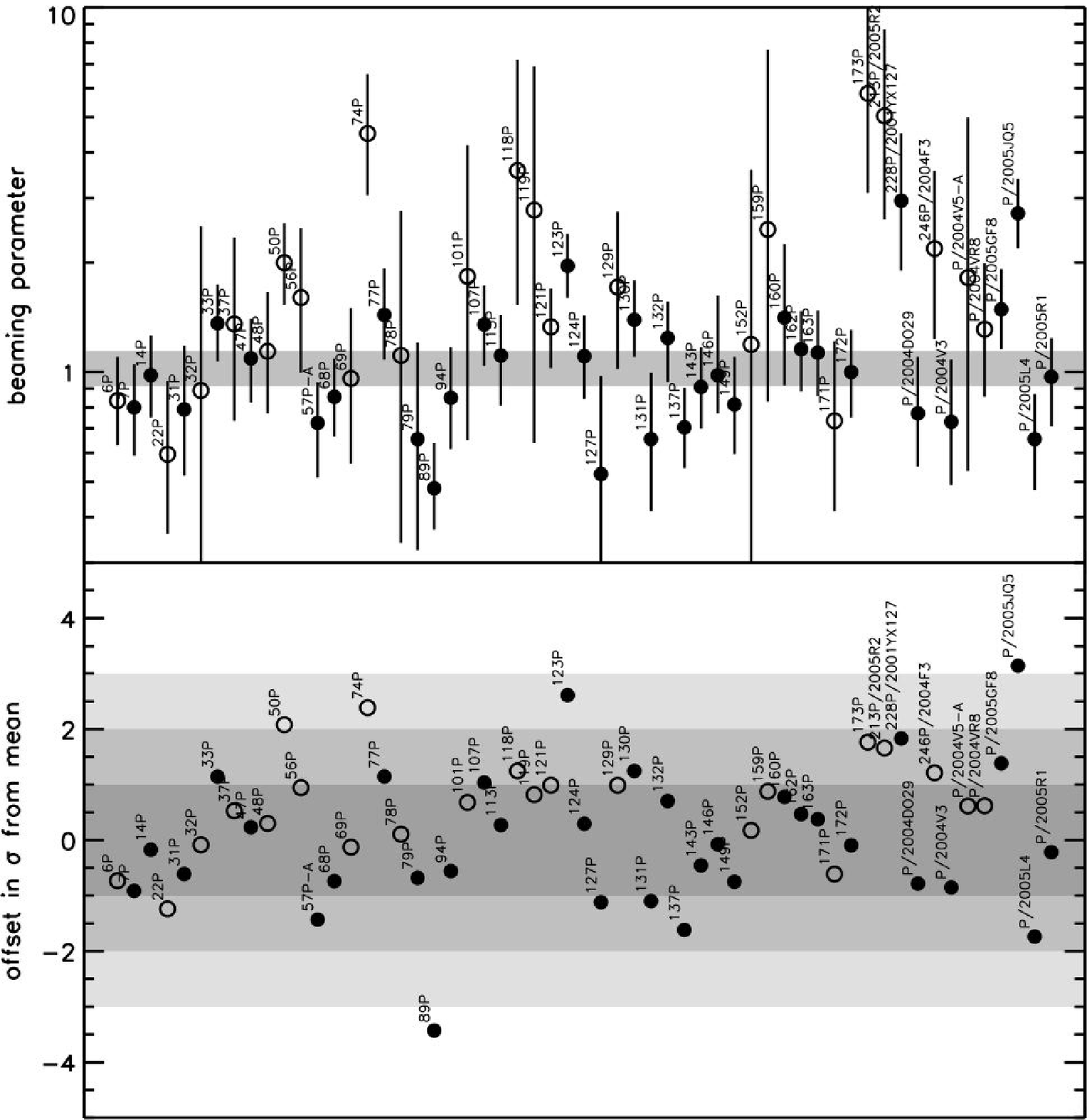,6.5 in]{Figure 6.
Graphical representation of the
57 beaming parameters reported in Table 4.
The top panel is a scatter plot of the points,
with each point labeled with its comet. 
Filled circles indicate comets for which there was
no discernible dust coma; unfilled circles are
for comets that showed some coma (i.e. those
underlined comets in Table 2). Error
bars are effectively 1$\sigma$. The grey
rectangle is vertically centered on
the mean beaming parameter, $\bar\eta = 1.03$,
and represents the 1$\sigma$ range
of the error in the mean, $\pm0.11$. 
The bottom panel shows the offset of each point
from $\bar\eta$ in terms of \underbar{its} 
\underbar{own} $\sigma$. 
The (dark,middle,light) grey rectangles represent
offsets of (1,2,3)$\sigma$. Given the wide
variety of error bars in the top panel, the bottom
panel makes it easier to see that there is good
clustering of our
measured $\eta$, and only two outliers. \hfil}
\bigskip
\eject

\figembedps[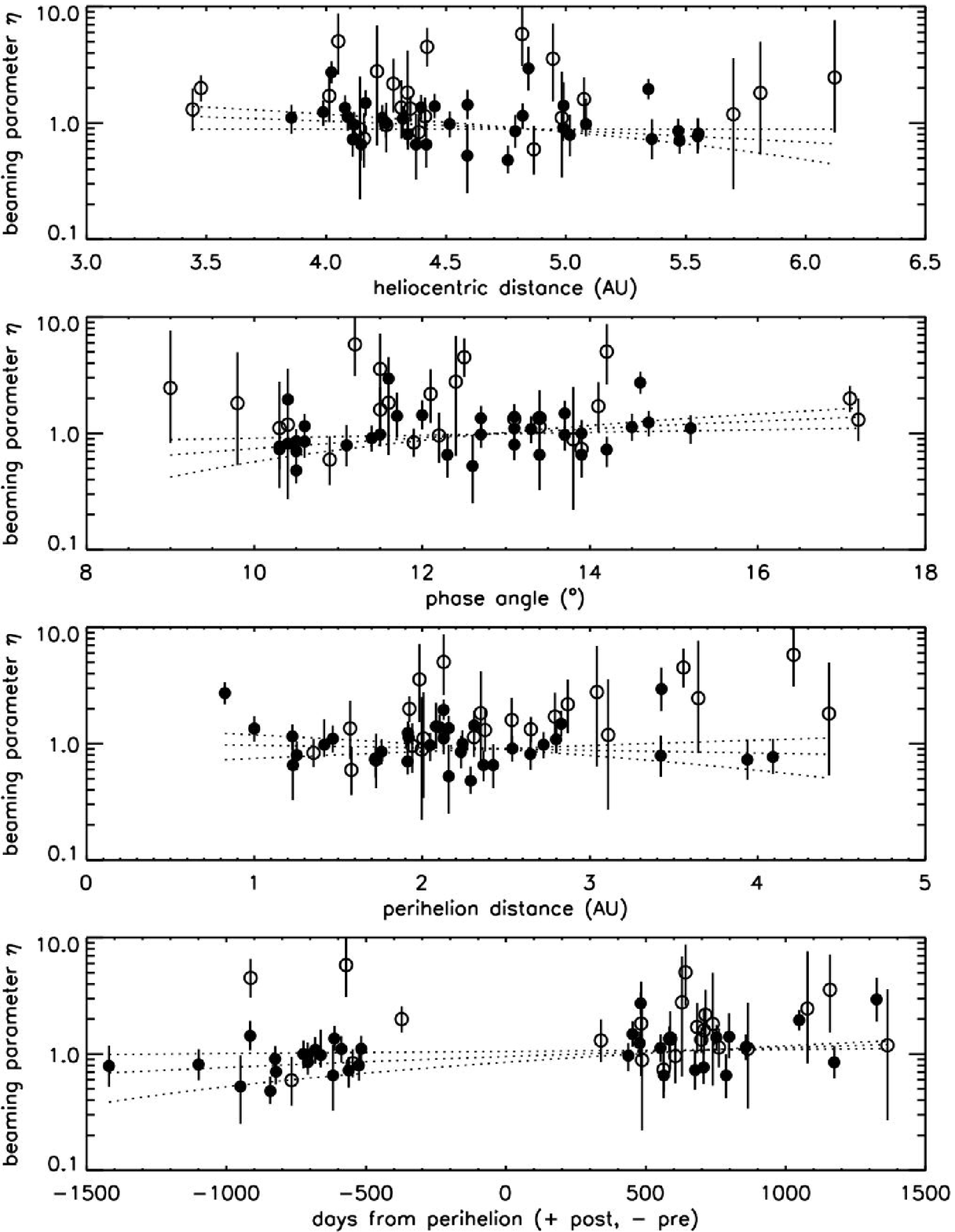,6.5 in]{Figure 7,
caption after end of figure. \hfil}
\bigskip
\eject

Figure 7. Search for trends of $\eta$ with
various significant quantities, as discussed
in \S 4.1. All 57 beaming parameters from Table 4
are plotted vs. heliocentric distance,
phase angle, perihelion distance, and days from
perihelion.
Filled circles indicate comets for which there was
no discernible dust coma; unfilled circles are
for comets that showed some coma (i.e. those
underlined comets in Table 2). The three
dotted lines in each panel represent the best-fitting
line to the data and the lines corresponding
to $\pm1\sigma$ range in slope and intercept
(see Table 6 for slopes). 
\vfill
\eject

\figembedps[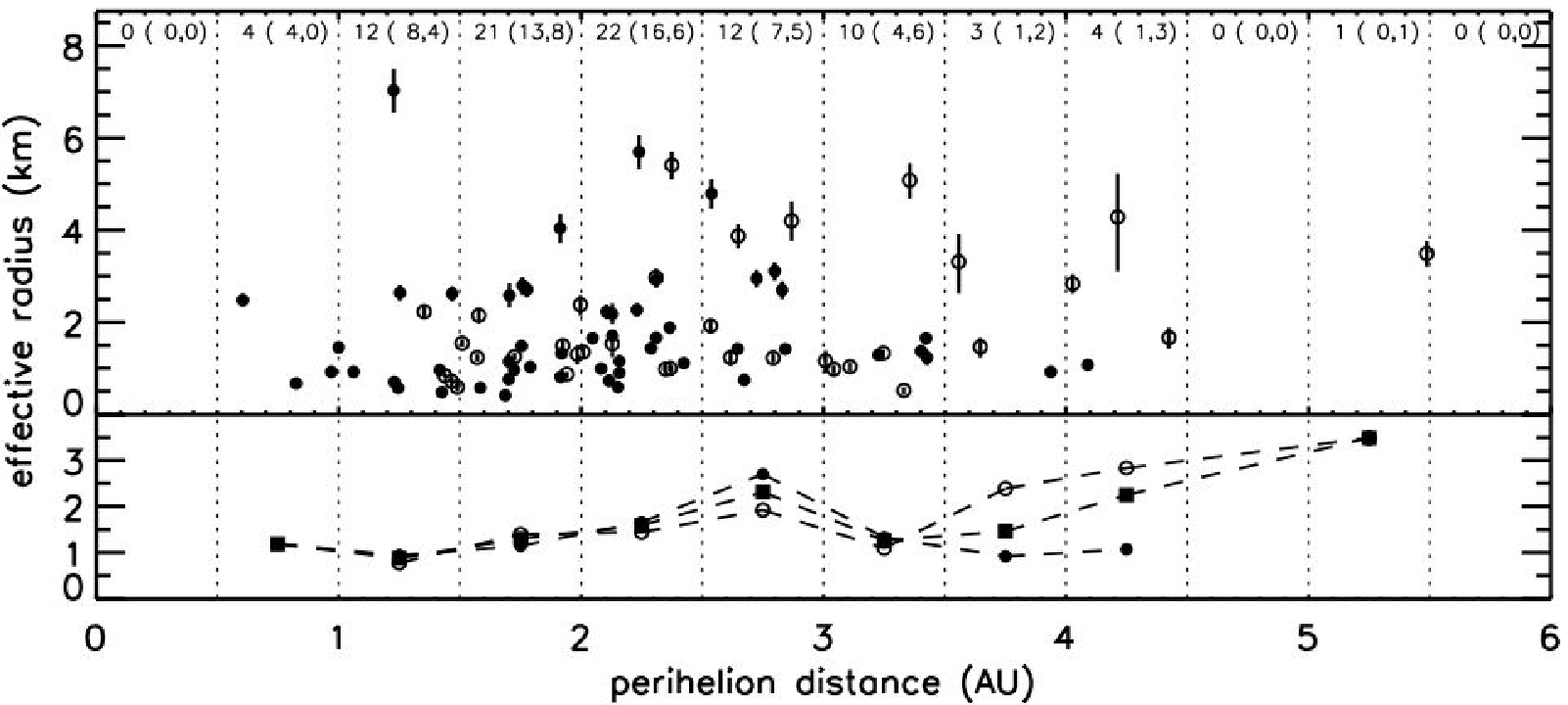,7.1 in]{Figure 8.
Top panel: scatter plot of all 89 radii vs.
each nucleus's perihelion distance. Error
bars are plotted for all points, though some
errors are smaller than the symbol used. 
Filled circles indicate comets for which there was
no discernible dust coma; unfilled circles are
for comets that showed some coma (i.e. those
underlined comets in Table 2). 
Numbers
at the top of the plot of the form ``$m(n,p)$"
indicate the number
of nuclei within each 0.5 AU-wide bin ($m$),
the number of comets without discernible dust in that bin ($n$),
and the number of comets with dust in that bin ($p$).
Bottom panel: Median values of the radii within
in each bin. Squares indicate medians of all nuclei within
each bin; filled circles, of comets without discernible dust;
open circles, of comets with dust.  \hfil}
\bigskip
\eject

\figembedps[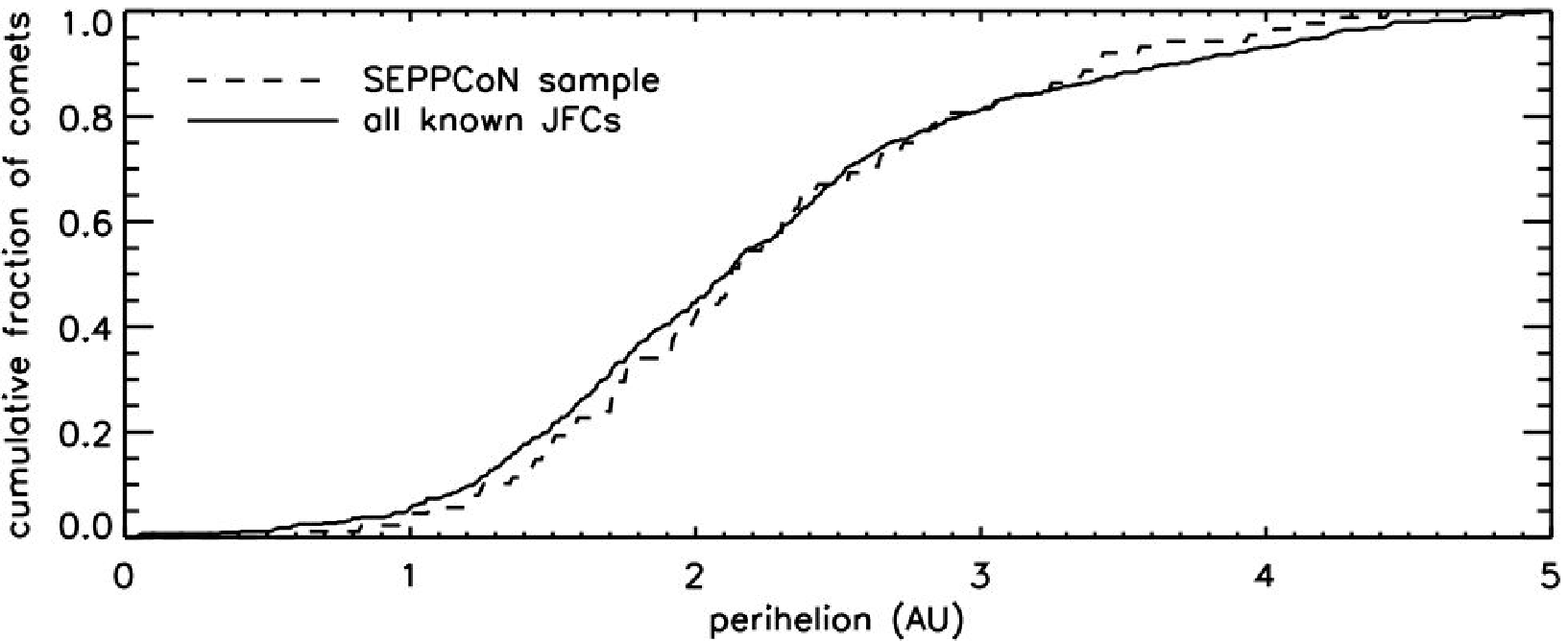,7.1 in]{Figure 9. 
Comparison of the cumulative perihelion distribution
of our 89 detected SEPPCoN JFCs (dashed curve)
and all 450 JFCs (solid curve) that were known at
time of writing. The curves match well,
indicating that our sample's perihelion bias
is no different than the overall one that is
present in the JFC discovery rate. \hfil}
\bigskip
\eject

\figembedps[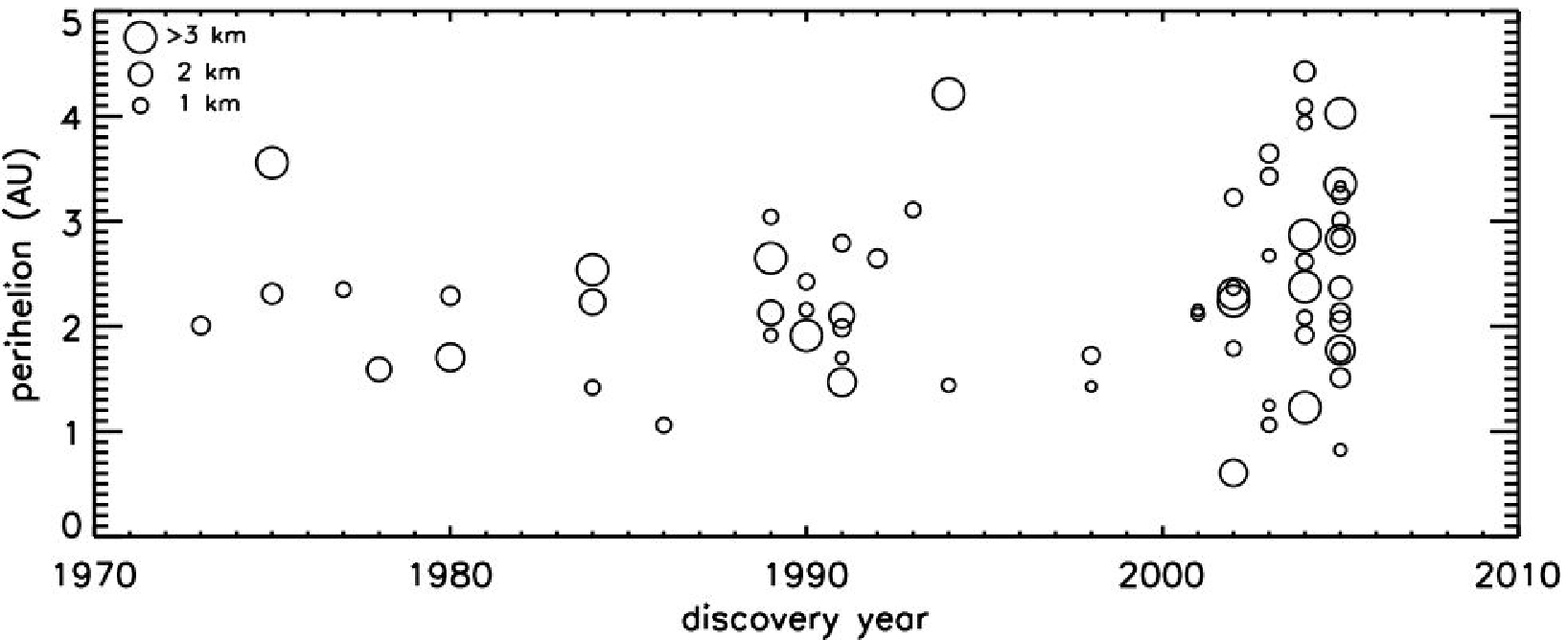,7.1 in]{Figure 10.
Scatter plot of discovery year and perihelion
distance for the 98 nuclei considered here (i.e. the 89 from
our survey plus nine reliable estimates from the literature).
For clarity we have left out the 32 such comets discovered
before 1970 and the 1 comet (P/2004 A1) with perihelion over 5 AU. 
The symbol size is tied to
the radius of the nucleus.
Many of our large nuclei ($\sim$3 km
and larger) and several such nuclei at low
perihelion were discovered only recently.  \hfil}
\bigskip
\eject

\figembedps[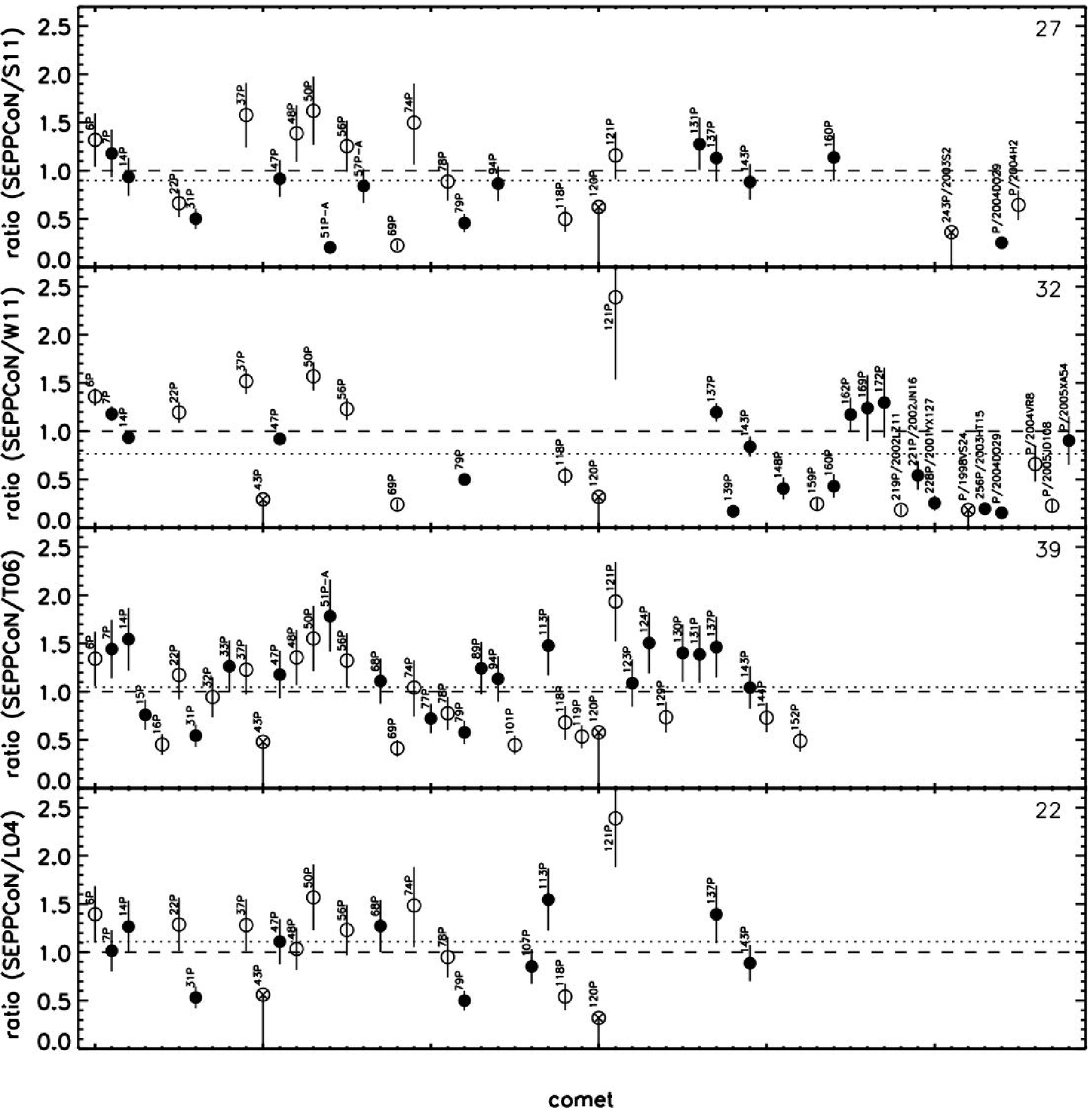,6.5 in]{Figure 11.
Plots showing the ratio of our radii in Table 7
to those reported in previously-published compilations: (from
top to bottom) Snodgrass et al. (2011; S11), 
Weiler et al. (2011; W11), Tancredi et al. (2006; T06),
Lamy et al. (2004; L04). Fifty-nine of our SEPPCoN comets
appear at least once in these works. Error bars are
propagated from an assumed 20\% error (in the case
of S11, T06, and L04) or the reported error (in the
case of W11). The same comet
is at the same abscissa value across all panels. 
Filled circles indicate comets for which there was
no discernible dust coma; unfilled circles are
for comets that showed some coma (i.e. those
underlined comets in Table 2); crossed circles
are for comets that we did not detect. Number
in the upper right of each panel indicates the number of comets
that overlap with that particular work. Horizontal dashed line
indicates unity; horizontal dotted line indicates the
average ratio of the points in that panel. \hfil}
\bigskip
\eject

\figembedps[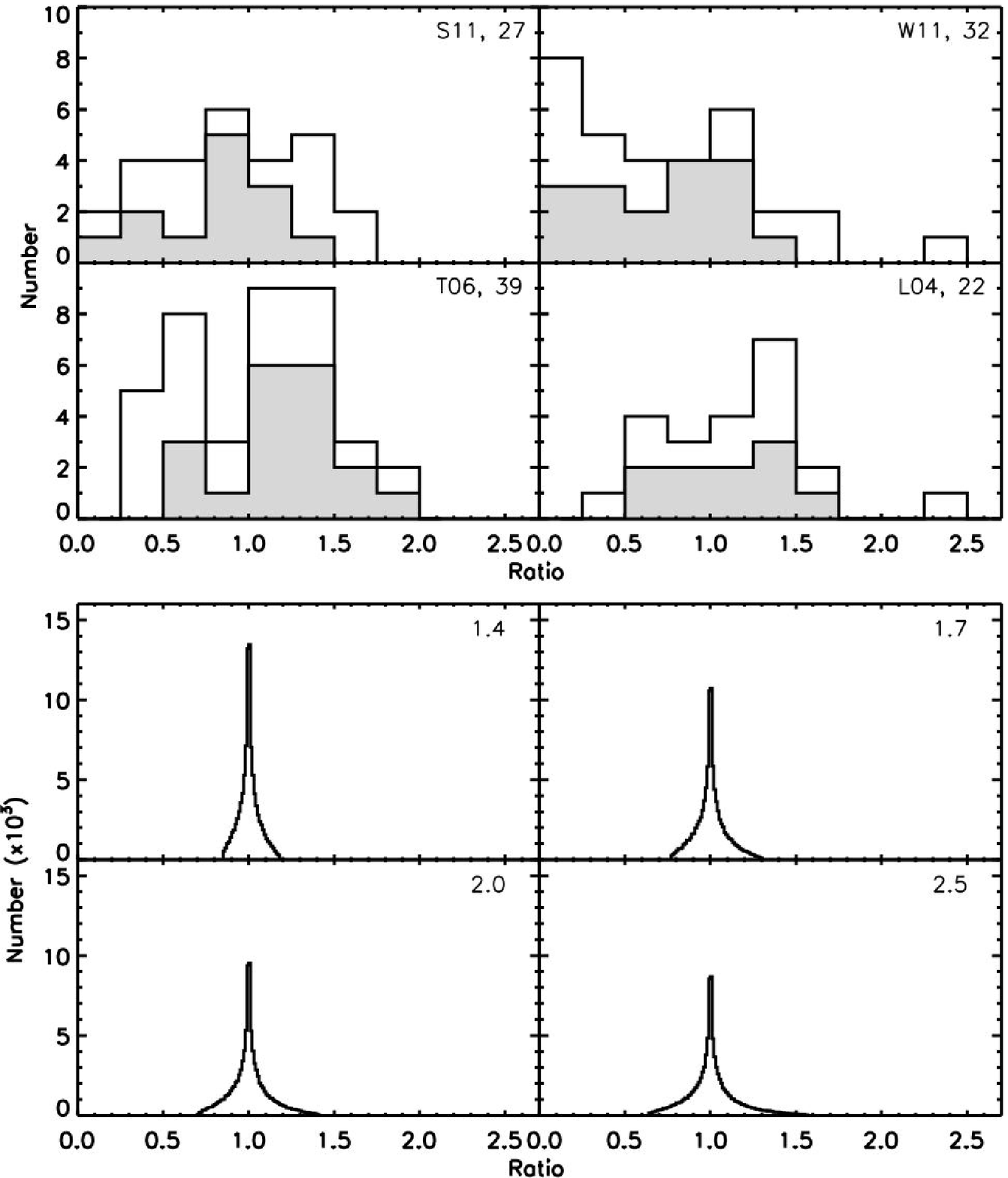,6.5 in]{Figure 12. Histograms of 
measured and expected ratios between
two snapshots of radii. Top half: four panels that give
the histograms to the panels in Fig. 11. White histogram
is for all the comets; grey histogram is for the comets
for which there was no discernible dust. Bottom half:
four panels that show the expected histograms given
the axial ratio listed in the upper right corner of each panel.
We note the differences between the character
of the histograms in the top and
bottom halves. \hfil}
\bigskip
\eject

\figembedps[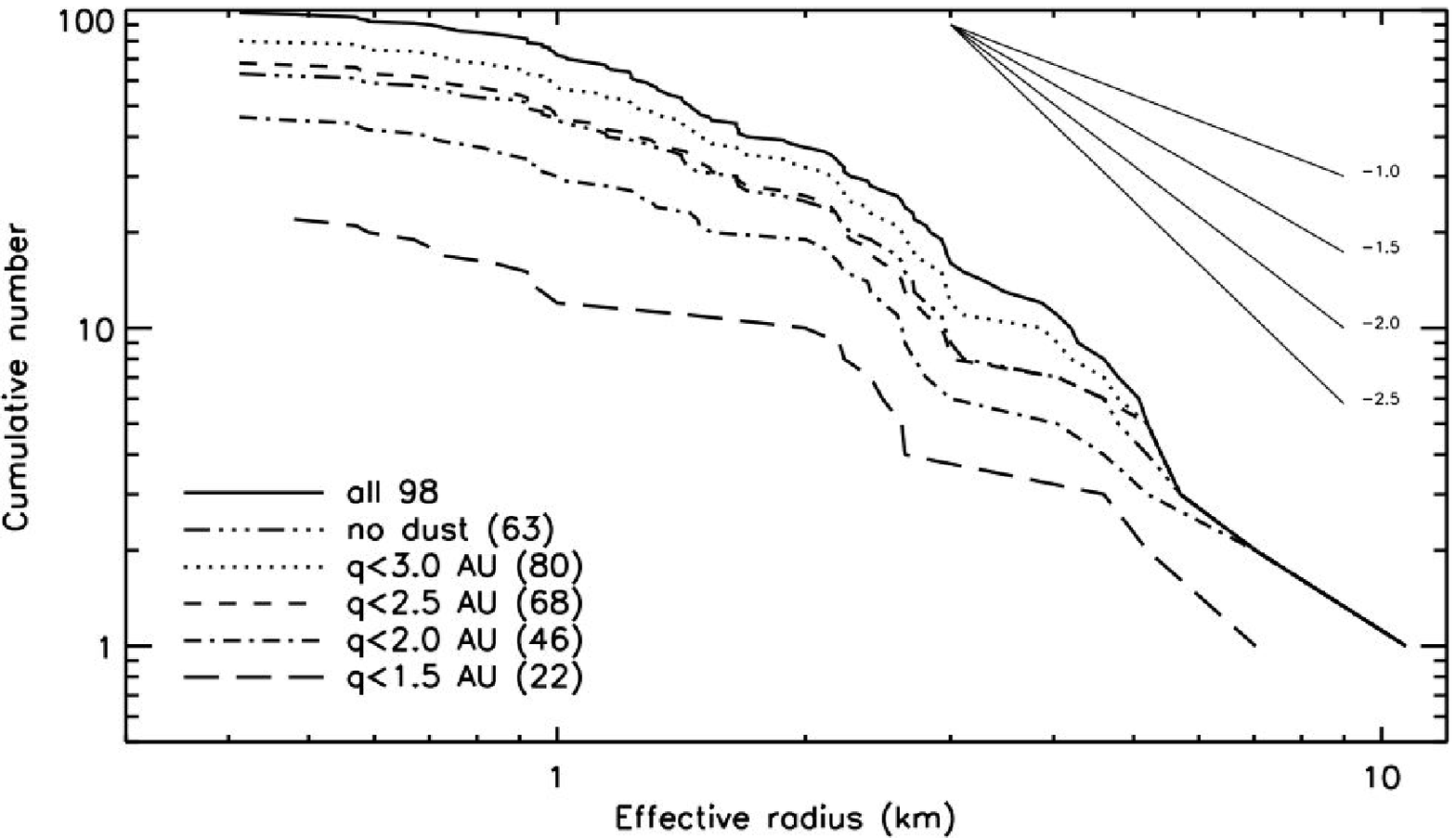,7.1 in]{Figure 13.
Observed cumulative size distributions (CSDs) of JFCs.
To create each curve, the list of radii was sorted and
the cumulative number was incremented by one
at the radius of each nucleus in that sorted list. 
Individual points are not shown for clarity. 
The overall
CSD (``all 98") includes the 89 comets reported here plus nine others from
the literature for a total of 98. Separate curves
break down the CSD by perihelion distance. 
Also shown is the CSD (``no dust") if we included only the 54 SEPPCoN
comets that did not show discernible dust plus the nine
others from the literature for a total of 63. 
Straight diagonal
lines in the upper right correspond to different (labeled)
power-law slopes. The slopes of the CSDs themselves are
given in Table 9 and \S4.4. \hfil}
\bigskip
\eject

\figembedps[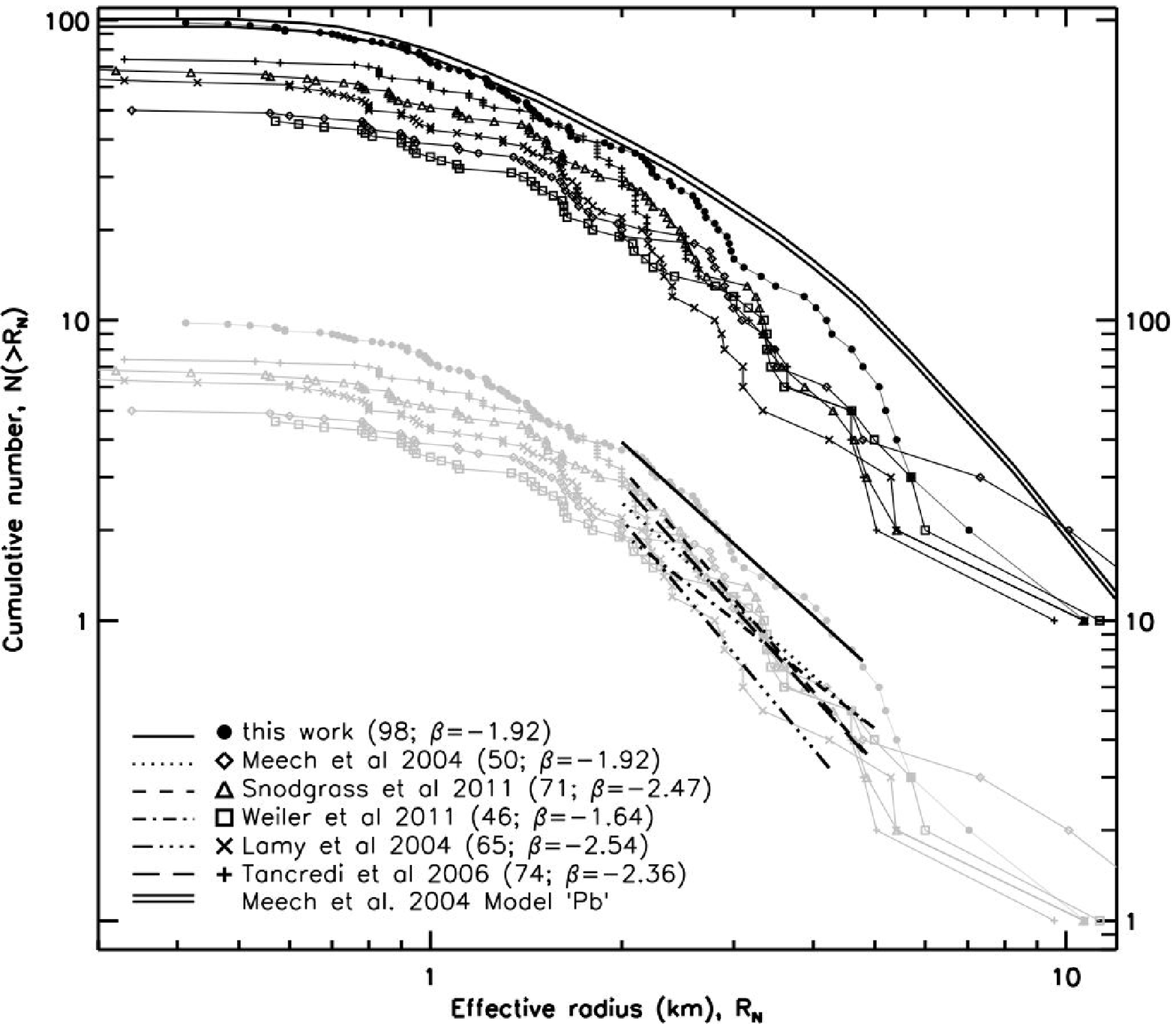,7.1 in]{Figure 14.
Comparison of our CSD with previous results. 
The top set of data curves shows six CSDs, including ours
and the ones from L04, T06, W11, S11, and Meech et al. (2004). 
Also plotted here, with
a double solid line, is model `Pb' from Meech et al. (2004),
the model that fit their CSD best. 
The bottom set of curves
shows the same CSDs in grey -- just displaced by
a factor of 10 --  underneath simple power-law
fits to each of the CSDs over the range of $2 {\rm\ km} < R_N < 5$ km.
This is simply to help lead the eye in the general trend of the
CSDs, and shows how ours has a similar overall shape. (It is
{\sl not} 
meant to show {\sl a definitive} answer on the slope of each CSD.)
The numbers in parentheses in each line of the legend indicate how many
comets were included in the plot and the slope of the power-law fit in
that 2-5 km range. We note that the ordinate's left axis should be
used with the top set of curves, and the right axis should be
used with the bottom set of curves. \hfil}
\bigskip
\eject

\bye